\documentclass[11pt,a4paper]{article}
\usepackage[english]{babel}
\usepackage[utf8]{inputenc}
\usepackage{epsf}
\usepackage{cite}
\usepackage{graphicx}
\usepackage{subcaption} % allows subfigures

\usepackage{verbatim} % allows block comments

\usepackage{amsmath}
\usepackage{amssymb}
\usepackage{mathrsfs}
\usepackage{ulem}
\usepackage{mathbbol}
\usepackage{slashed}
\usepackage{amsfonts}
\usepackage{bbding}
\usepackage{array}
\usepackage[
colorlinks=true
,urlcolor=black
,anchorcolor=black
,citecolor=black
,filecolor=black
,linkcolor=black
,menucolor=black
,linktocpage=true
,pdfproducer=medialab
,pdfa=true
]{hyperref}

\usepackage{color} % colored text
\usepackage[left=2cm,right=2cm,top=2cm,bottom=2cm]{geometry}
\usepackage{tikz}
\usetikzlibrary{shapes,arrows,positioning,automata,backgrounds,calc,er,patterns}
\usepackage[compat=1.1.0]{tikz-feynman}
\usepackage{contour}

\vspace*{1cm}
\allowdisplaybreaks
\begin{document}

\begin{center}
\vspace*{1mm}
%%%%%%%%%%%%%%%%%%%%%
\vspace{1.3cm}
{\Large\bf
Heavy neutral lepton corrections to SM boson decays: \\

\vspace*{3mm}
lepton flavour universality violation in low-scale seesaw realisations
}

\vspace*{1.2cm}
{\bf A.~Abada $^{a}$, J.~Kriewald $^{b}$, E. Pinsard $^{c}$, S.~Rosauro-Alcaraz $^{a}$ and A.~M.~Teixeira $^{c}$}

\vspace*{.5cm}
$^{a}$ P\^ole Th\'eorie, Laboratoire de Physique des 2 Infinis Irène Joliot Curie (UMR 9012), \\
CNRS/IN2P3,
15 Rue Georges Clemenceau, 91400 Orsay, France

\vspace*{.2cm}
${^b}$ Jožef Stefan Institut, Jamova Cesta 39, P. O. Box 3000, 1001 Ljubljana, Slovenia

\vspace*{.2cm}
$^{c}$ Laboratoire de Physique de Clermont (UMR 6533), CNRS/IN2P3,\\
Univ. Clermont Auvergne, 4 Av. Blaise Pascal, 63178 Aubi\`ere Cedex,
France
\end{center}

\vspace*{5mm}
\begin{abstract}
\noindent
We study lepton flavour universality violation in SM boson decays in low-scale seesaw models of neutrino mass generation, also addressing other electroweak precision observables. 
We compute the electroweak next-to-leading order corrections, 
which turn out to be important - notably in the case of the invisible decay width of the $Z$ boson, for which the corrections can be as large as the current experimental uncertainty.
As a well-motivated illustrative study case, we choose a realisation of the Inverse Seesaw mechanism, and discuss the complementary role of lepton flavour conserving, lepton flavour violating and precision observables, both in constraining and in probing such models of neutrino mass generation. 
Our findings suggest that invisible $Z$ decays are especially important, potentially at the origin of the most stringent constraints for certain regimes of the Inverse Seesaw (while complying with charge lepton flavour violation and other electroweak precision tests). We also discuss the probing power of the considered observables in view of the expected improvement in experimental precision at FCC-ee.
\end{abstract}

\newpage
\section{Introduction}
In addition to their role in well-motivated mechanisms of neutrino mass generation (for instance the type-I seesaw mechanism~\cite{Minkowski:1977sc,Yanagida:1979as,Glashow:1979nm,Gell-Mann:1979vob,Mohapatra:1979ia} and its variants), heavy neutral leptons (HNL) can be at the source of an extensive array of new phenomena, which are forbidden (or strongly suppressed) in the Standard Model (SM) of particle physics. 
These phenomena include lepton number violating (LNV) processes such as (semi-) leptonic tau and meson decays, charged lepton flavour violating (cLFV) transitions (including radiative and three-body decays, $Z$ and Higgs bosons decays, as well as rare phenomena in the presence of matter, such as neutrinoless muon-electron conversion), among others (see e.g.~\cite{Riemann:1982rq,Riemann:1999ab,Illana:1999ww,Mann:1983dv,Illana:2000ic,Alonso:2012ji,Ilakovac:1994kj,Ma:1979px,Gronau:1984ct,Deppisch:2004fa,Deppisch:2005zm,Dinh:2012bp,Hambye:2013jsa,Abada:2014kba,Abada:2015oba,Abada:2015zea,Abada:2016vzu,Calibbi:2017uvl,Abada:2018nio,Arganda:2014dta,Marcano:2019rmk,Calderon:2022alb,Abada:2018nio,Abada:2021zcm,Abada:2022asx,Crivellin:2022cve}, and references therein); several of the latter processes also reflect the possible Majorana nature of the extended neutral lepton sector.
Such heavy Majorana fermions are also expected to have a significant 
impact concerning electroweak (EW) precision observables, and also lead to striking signatures at colliders (be it at  the LHC, or at future lepton colliders). All the above new phenomena are rooted in the non-negligible mixings of the heavy Majorana states with the light active neutrinos, which in turn are responsible for modifying flavour mixings in the lepton sector. 
In addition, new leptonic CP violating phases (Dirac and Majorana) can in principle be present, %
leading to  CP violating effects
which range from an explanation of the baryon asymmetry of the Universe via leptogenesis, to contributions to an extensive array of observables.

\medskip
An accidental symmetry of the SM, lepton flavour universality (LFU) is conserved in all its gauge interactions, and  is
only broken by the charged lepton's masses (due to the non-universal interactions with the Higgs field). Once neutrino oscillation data is minimally incorporated by considering massive active neutrinos, and once left-handed leptonic mixings are encoded in the so-called Pontecorvo-Maki-Nakagawa-Sakata matrix ($U_\text{PMNS}$), LFU is naturally violated in charged current leptonic interactions; nevertheless, LFU violation (LFUV)
remains a direct consequence of having non-vanishing and non-degenerate masses for both charged and neutral leptons. 

The role of heavy neutral leptons in the violation of LFU has been explored, in particular in what concerns light meson decays, as for example upon comparison of the decay widths $P \to \ell_\alpha \nu$ 
(with $P$ denoting the decaying meson) to distinct charged lepton 
flavour final states~\cite{Shrock:1980vy,Abada:2012mc}, (semi-) leptonic tau-lepton decays and $W$-boson decays at tree-level, see for instance~\cite{Shrock:1980ct,Abada:2013aba,Abada:2017jjx}, and references therein.
Most of the above mentioned studies considered only tree-level decays in which the new heavy states played a virtual role via modifications of the leptonic mixing matrix: the usual $U_\text{PMNS}$ is no longer unitary, but rather a sub-block of an enlarged unitary mixing matrix. In addition, and if sufficiently light, the new neutral states could also be produced on-shell from the gauge boson decays, and thus induce further contributions (see, for example~\cite{Fernandez-Martinez:2016lgt,Blennow:2023mqx}).  
However, and other than tree-level effects,
the new  heavy Majorana fermions can also open the door to the mediation of higher order processes, contributing to loop-level transitions. 
In the present work we investigate to which extent these SM extensions by heavy sterile fermion states can be at the source of deviations from LFU in $W$, $Z$ and Higgs boson decays - beyond what is expected due to non-degenerate charged lepton masses-, computing all relevant one-loop contributions to the decay widths without any simplifying approximations\footnote{We have computed the higher-order contributions to the decay widths both in Feynman-’t Hooft and unitary gauges for consistency and validation of our results.}.

Addressing the violation of LFU in SM extensions via heavy neutral leptons
is also of particular interest in view of the expected experimental developments, be it at the LHC, 
or especially in view of the excellent sensitivity prospects of future lepton colliders, as the FCC-ee:
concerning the precision of electroweak measurements, the improvements might be up to 2 orders of magnitude in the reduction of the relative uncertainty~\cite{FCC:2018evy}. 
Thus, comparing (flavour-conserving) boson decays to different sets of lepton final states, as is the case of $Z\to \ell_\alpha \ell_\alpha$, or  $H\to \ell_\alpha \ell_\alpha$, will allow probing deviations from the SM expectation, and in turn hint towards the presence of new physics (NP) in the lepton sector. 
Furthermore, taking into account higher-order corrections in the computation of the theoretical predictions is crucial to match the experimental increase in precision, especially concerning new physics contributions to electroweak precision observables (as the invisible width of the $Z$ boson).

\medskip
In our work we consider beyond the SM (BSM) frameworks in which heavy sterile fermions are an intrinsic part of minimal, well-motivated mechanisms of neutrino mass generation, focusing in particular on the Inverse Seesaw (ISS) mechanism~\cite{Schechter:1980gr, Gronau:1984ct, Mohapatra:1986bd}. In its different phenomenologically viable realisations~\cite{Abada:2014vea}, the ISS offers a natural explanation to the smallness of light neutrino masses, relying on approximate lepton number conservation. 
In what follows, and for simplicity, we will always consider the so-called ISS(3,3) realisation in which two sets of three sterile fermions are added to the SM content. 

As we will argue, certain LFUV observables and $\Gamma(Z \to \text{inv.})$  can be powerful complementary probes to cLFV observables, which do still play a very important role. The presence of the new heavy sterile states can be at the origin of sizeable deviations from the SM expectation, be it in $Z$ or Higgs decays. 
As will be manifest in the numerical analysis, the inclusion of higher-order corrections to the contributions of the heavy states can induce deviations in $\Gamma(Z \to \text{inv.})$ up to $\sim5\:\mathrm{MeV}$ from the tree-level result;
in view of the current experimental precision (and its expected improvement, both at the high-luminosity phase of the LHC, and a future FCC-ee), one-loop corrections must therefore be taken into account to critically assess the viability of this class of SM extensions.
LFUV probes in $Z$ boson decays (including invisible modes) are thus poised to offer powerful complementary information to other indirect searches for NP in the lepton sector (in particular electroweak precision observables (EWPO) such as the oblique parameters), probing regimes which would otherwise lie beyond the reach of cLFV probes.

\bigskip
The manuscript is organised as follows: after discussing the underlying approach to the higher order computations in Section~\ref{sec:higherorder}, 
in Sections~\ref{sec:Wdecays}-\ref{sec:Zdecays} we present and compute the most relevant observables and quantities in what concerns LFU violation in the decays of $W$, Higgs and $Z$ bosons, respectively. Further relevant decays and EWPOs are discussed in Section~\ref{sec:EWPO:Zinv}. We then carry out a thorough numerical analysis in Section~\ref{sec:NumericalResults}, and discuss the impact of our findings as complementary to other searches for NP in Section~\ref{sec:ComplementarityProbing}. We summarise our main points in the Conclusions. The appendices include (among others) details of the renormalisation procedure, 
the expressions for the form factors which are used throughout this work, as well as a description of the model under study and the most relevant constraints on the latter.

\section{New contributions to SM boson decays from heavy neutral leptons: beyond leading order}\label{sec:higherorder}

While in the context of the SM charged current interactions are diagonal in flavour space (and so remaining up to all orders), once neutrinos acquire a mass, flavour is violated, as parametrised via the $3\times 3$ $U_\text{PMNS}$ leptonic mixing matrix. Should there be new (mass-induced) mixings involving additional neutral leptons  - as is the case of several models of neutrino mass generation - the lepton mixing matrix is enlarged: for $n_S$  additional sterile neutrino states,  
a new rectangular $3 \times (3+n_S)$  reflects the extended mixings between the active and the new heavier states, with its $3\times3$ sub-block (former $U_\text{PMNS}$) no longer being unitary.

Working in the physical (mass) basis, with $\alpha=e,\, \mu,\, \tau$ denoting the flavour of the charged lepton, and $i=1, ..., n_S$ the neutral fermion mass eigenstate (including the light, mostly active, neutrinos, and additional heavier states whose number depends on 
the SM extension being considered), the relevant terms in 
the lepton charged current Lagrangian are given by:
\begin{equation}\label{eq:lagrangian:W}
\mathcal{L}_{W^\pm}\, =\, -\frac{g_w}{\sqrt{2}} \, W^-_\mu \,
\sum_{\alpha=1}^{3} \sum_{j=1}^{3 + n_S} \mathcal{U}_{\alpha j}\, \bar \ell_\alpha\, 
\gamma^\mu \,P_L\, \nu_j \, + \, \text{H.c.}\,,
\end{equation}
with $P_{L,R} = (1 \mp \gamma_5)/2$, and $g_w$ the weak coupling constant; $\mathcal{U}$ refers to the
$(3+n_S) \times (3+n_S)$ unitary matrix which now  
parametrises mixings in the lepton sector.
Modifications are also present in 
Higgs and $Z$ leptonic interactions; the 
corresponding Lagrangian terms are now given by  
\begin{align}\label{eq:lagrangian:HZ}
& \mathcal{L}_{Z^0}^{\nu}\, = \,-\frac{g_w}{4 \cos \theta_w} \, Z_\mu \,
\sum_{i,j=1}^{3 + n_S} \bar \nu_i \,\gamma ^\mu \,\left(
P_L {C}_{ij} - P_R {C}_{ij}^* \right) \,\,\nu_j\,, \nonumber \\
& \mathcal{L}_{Z^0}^{\ell}\, = \,-\frac{g_w}{2 \cos \theta_w} \, Z_\mu \,
\sum_{\alpha=1}^{3}  \bar \ell_\alpha\, \gamma^\mu \,\left(
{\bf C}_{V} - {\bf C}_{A} \gamma_5 \right) \,\ell_\alpha\,, \nonumber \\
& \mathcal{L}_{H^0}\, = \, -\frac{g_w}{4 M_W} \, H  \,
\sum_{i\ne j= 1}^{3 + n_S}    \bar \nu_i\,\left[{C}_{ij}\,\left(
P_L m_i + P_R m_j \right) +{C}_{ij}^\ast\left(
P_R m_i + P_L m_j \right) \right] \nu_j\,,
\end{align}
with $\cos^2 \theta_w =  M_W^2 /M_Z^2$. 
The SM vector and axial-vector currents (interaction of $Z$ bosons with charged leptons) have been written in terms of the ${\bf C}_{V}$ and ${\bf C}_{A}$ coefficients,  respectively given by 
${\bf C}_{V} = -\frac{1}{2} + 2 \sin^2\theta_w$ and 
${\bf C}_{A} = -\frac{1}{2}$. Finally ${C}_{ij} $ are defined as: 
\begin{equation}\label{eq:cij}
    {C}_{ij} = \sum_{\rho = 1}^3 \,\,
  \mathcal{U}_{i\rho}^\dagger \,\,\mathcal{U}_{\rho j}^{\phantom{\dagger}}\:. 
\end{equation}
In the above, greek indices 
again denote the flavour of the charged leptons, while $i, j = 1, \dots, 3+n_S$ correspond to the physical (massive) neutrino states.
It is convenient to introduce the parameter $\eta$ (which allows to evaluate the deviations of the $U_\text{PMNS}$ from its standard form as a unitary matrix), defined as~\cite{FernandezMartinez:2007ms}
\begin{equation}
\label{eq:defPMNSeta}
U_\text{PMNS} \, \to \, \tilde U_\text{PMNS} \, = \,(\mathbb{1} - \eta)\, 
U_\text{PMNS}\,.
\end{equation}

Due to the (tree-level) non-conservation of lepton flavours, lepton flavour universality will also be violated by charged and neutral current interactions.
Moreover, the new (Majorana) states open the door to higher order corrections to the interaction vertices: $W^\pm \ell_\alpha \nu_i$,  
$Z \ell_\alpha \ell_\alpha$, and $H \ell_\alpha \ell_\alpha$, $\alpha=e,\mu,\tau$.
In our study - and as subsequently discussed in the phenomenological analysis - higher order effects in the decays of both $Z$ and Higgs bosons are of particular relevance; 
for completeness, we extend the study of the one-loop contributions to $W$ decays\footnote{We nevertheless  verified that the HNL-mediated one-loop contributions to $W$-boson decays are indeed negligible (thus confirming  the results of a previous study~\cite{Fernandez-Martinez:2015hxa}, but now including all possible contributions).}.

A thorough  evaluation of the higher order contributions requires a full renormalisation of the involved parameters (masses, mixings and couplings) and of the fields.
In our study we employ the ``on-shell'' renormalisation scheme to cancel ultraviolet (UV)-divergences via 
counterterms;
following~\cite{Denner:1991kt}, we choose the set of input parameters to be renormalised as
\begin{equation}
    M_W, \,M_Z, \,M_H, \,m_{\ell_\alpha}, \,m_{\nu_i}, \,e, \,\mathcal U_{\alpha i}\,.
\end{equation}
The renormalisation of the $C_{ij}$ matrix (defined in Eq.~(\ref{eq:cij})), 
follows directly from the renormalisation of $\mathcal U$ (and from associated unitarity relations), see Appendix~\ref{app:renormalisation}.
The counterterms for the renormalisation of the parameters  are defined as
\begin{eqnarray}
    e_0 &=& (1 + \delta Z_e)e\,,\\
    M_{W,0}^2 &=& M_W^2 + \delta M_W^2\,,\\
    M_{Z,0}^2 &=&M_Z^2 + \delta M_Z^2\,,\\
    M_{H,0}^2 &=& M_H^2 + \delta M_H^2\,,\\
    m_{\ell_\alpha, 0} &=& m_{\ell_\alpha} + \delta m_{\ell_\alpha}\,,\\
    m_{\nu_i, 0} &=& m_{\nu_i} + \delta m_{\nu_i}\,,\\
    \mathcal U_{\alpha i, 0} &=& \mathcal U_{\alpha i} + \delta \mathcal U_{\alpha i} \,,
\end{eqnarray}
in which the ``0'' subscript denotes the bare (unrenormalised) parameters. In the on-shell scheme, the weak mixing angle is a derived quantity given by
\begin{equation}
    \sin^2\theta_w \,=\, 1 - \frac{M_W^2}{M_Z^2}\,,
    \label{eqn:sw}
\end{equation}
which is computed using the renormalised gauge boson masses.
We further renormalise the involved boson and fermion fields
\begin{equation}
    W, \,Z, \,H, \,\ell_\alpha, \,\nu_i\,.
\end{equation}
The (multiplicative) boson and fermion field renormalisation constants are expanded up to one-loop order as follows (the fermion field renormalisation constants are now given by matrices, owing to the lepton mixing):
\begin{eqnarray}
    W_0^\pm &=& Z_W^{1/2} \,W^\pm = (1 + \frac{1}{2}\, \delta Z_W) \,W^\pm\,,\\
    \begin{pmatrix}Z_0\\A_0\end{pmatrix} &=& \begin{pmatrix}1 + \frac{1}{2}\,\delta Z_{ZZ} & \frac{1}{2}\, \delta Z_{ZA}\\ \frac{1}{2} \,\delta Z_{AZ} & 1 + \,\frac{1}{2} \delta Z_{AA}\end{pmatrix} \begin{pmatrix}Z\\A\end{pmatrix}\,,\\
    H_0 &=& (1 + \frac{1}{2}\,\delta Z_H)\,H\,,\\
    \ell^L_{\alpha, 0} &=& (\delta_{\alpha\beta} + \frac{1}{2} \,\delta Z_{\alpha\beta}^{\ell, L})\,\ell_\beta^L\,,\\
    \ell^R_{\alpha, 0} &=& (\delta_{\alpha\beta} + \frac{1}{2} \,\delta Z_{\alpha\beta}^{\ell, R})\,\ell_\beta^R\,,\\
    \nu_{i, 0}^L &=& (\delta_{ij} + \frac{1}{2} \,\delta Z_{ij}^{\nu, L})\,\nu_j^L\,.
\end{eqnarray}
Note in addition that the bare and renormalised Majorana neutrino fields do still fulfil the Majorana condition, $\nu_{i, 0}^L = (\nu_{i, 0}^R)^C$ and $\nu_{i}^L = (\nu_{i}^R)^C$; the renormalisation constants of the right-handed fields are thus simply given by $\delta Z_{ij}^{\nu, R} = (\delta Z_{ij}^{\nu, L})^*$.
In the presence of heavy neutral leptons, 
all of the above listed renormalisation constants (except that of the photon and the $Z$-photon mixing terms) receive new contributions, in addition to the usual SM ones. 

The renormalised Lagrangian (including the counterterms) gives rise to counter-diagrams that absorb the UV-divergences of the one-loop amplitudes. In the subsequent sections, we will provide the distinct counter-diagrams relevant to the leptonic charged and neutral current interactions.
Further details on the renormalisation procedure (constants, boson and fermion self-energies) can be found in Appendix~\ref{app:renormalisation}.

\section{LFU violation in $W^\pm$ boson decays}\label{sec:Wdecays}
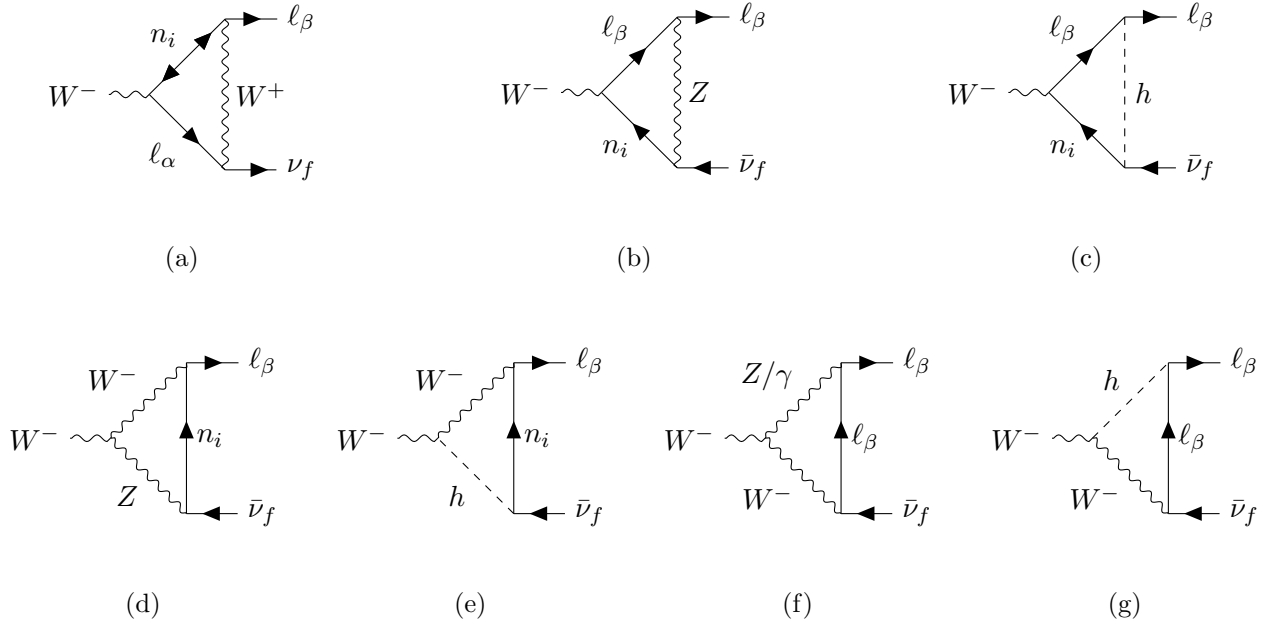
\begin{figure}[h!]
    \centering
    \begin{subfigure}[b]{0.3\textwidth}
    \centering
 \raisebox{5mm}{    \begin{tikzpicture}
    \begin{feynman}
    \vertex (a) at (0,0) {\(W^-\)};
    \vertex (b) at (1,0);
    \vertex (c) at (2,1.);
    \vertex (d) at (2,-1);
    \vertex (e) at (3,1) {\( \ell_\beta\)};
    \vertex (f) at (3,-1) {\( \nu_f\)};
    \diagram* {
    (a) -- [boson] (b),
    (b) -- [anti majorana, edge label=\(n_i\)] (c),
    (c) -- [boson, edge label=\( W^+\)] (d),
    (d) -- [anti fermion, edge label=\(\ell_\alpha\)] (b),
    (c) -- [fermion] (e),
    (f) -- [anti fermion] (d)
    };
    \end{feynman}
    \end{tikzpicture}
    }
            \caption*{(a)}
            \label{}
    \end{subfigure}
    \hfill
    \begin{subfigure}[b]{0.3\textwidth}
    \centering
 \raisebox{5mm}{    \begin{tikzpicture}
    \begin{feynman}
    \vertex (a) at (0,0) {\(W^-\)};
    \vertex (b) at (1,0);
    \vertex (c) at (2,1.);
    \vertex (d) at (2,-1);
    \vertex (e) at (3,1) {\( \ell_\beta\)};
    \vertex (f) at (3,-1) {\( \bar\nu_f\)};
    \diagram* {
    (a) -- [boson] (b),
    (b) -- [fermion, edge label=\(\ell_\beta\)] (c),
    (c) -- [boson, edge label=\( Z\)] (d),
    (d) -- [fermion, edge label=\(n_i\)] (b),
    (c) -- [fermion] (e),
    (f) -- [fermion] (d)
    };
    \end{feynman}
    \end{tikzpicture}
    }
            \caption*{(b)}
            \label{}
    \end{subfigure}
    \hfill
    \begin{subfigure}[b]{0.3\textwidth}
    \centering
\raisebox{5mm}{\begin{tikzpicture}
    \begin{feynman}
    \vertex (a) at (0,0) {\(W^-\)};
    \vertex (b) at (1,0);
    \vertex (c) at (2,1.);
    \vertex (d) at (2,-1);
    \vertex (e) at (3,1) {\( \ell_\beta\)};
    \vertex (f) at (3,-1) {\( \bar\nu_f\)};
    \diagram* {
    (a) -- [boson] (b),
    (b) -- [fermion, edge label=\(\ell_\beta\)] (c),
    (c) -- [scalar, edge label=\( h\)] (d),
    (d) -- [fermion, edge label=\(n_i\)] (b),
    (c) -- [fermion] (e),
    (f) -- [fermion] (d)
    }; 
    \end{feynman}
    \end{tikzpicture}
    }
           \caption*{(c)}
            \label{}
    \end{subfigure}
\\
\vspace{8mm}
    \begin{subfigure}[b]{0.24\textwidth}
    \centering
 \raisebox{5mm}{    \begin{tikzpicture}
    \begin{feynman}
    \vertex (a) at (0,0) {\(W^-\)};
    \vertex (b) at (1,0);
    \vertex (c) at (2,1.);
    \vertex (d) at (2,-1.);
    \vertex (e) at (3,1.) {\( \ell_\beta\)};
    \vertex (f) at (3,-1.) {\( \bar\nu_f\)};
    \diagram* {
    (a) -- [boson] (b),
    (b) -- [boson, edge label=\( W^-\)] (c),
    (c) -- [anti fermion, edge label=\(n_i\)] (d),
    (d) -- [boson, edge label=\( Z\)] (b),
    (c) -- [fermion] (e),
    (f) -- [fermion] (d)
    };
    \end{feynman}
    \end{tikzpicture}
    }
           \caption*{(d)}
            \label{}
    \end{subfigure}
    \hfill
    \begin{subfigure}[b]{0.24\textwidth}
    \centering
 \raisebox{5mm}{    \begin{tikzpicture}
    \begin{feynman}
    \vertex (a) at (0,0) {\(W^-\)};
    \vertex (b) at (1,0);
    \vertex (c) at (2,1.);
    \vertex (d) at (2,-1.);
    \vertex (e) at (3,1.) {\( \ell_\beta\)};
    \vertex (f) at (3,-1.) {\( \bar\nu_f\)};
    \diagram* {
    (a) -- [boson] (b),
    (b) -- [boson, edge label=\( W^-\)] (c),
    (c) -- [anti fermion, edge label=\(n_i\)] (d),
    (d) -- [scalar, edge label=\( h\)] (b),
    (c) -- [fermion] (e),
    (f) -- [fermion] (d)
    };
    \end{feynman}
    \end{tikzpicture}
    }
           \caption*{(e)}
            \label{}
    \end{subfigure}
    \hfill
    \begin{subfigure}[b]{0.24\textwidth}
    \centering
\raisebox{5mm}{        \begin{tikzpicture}
    \begin{feynman}
    \vertex (a) at (0,0) {\(W^-\)};
    \vertex (b) at (1,0);
    \vertex (c) at (2,1.);
    \vertex (d) at (2,-1.);
    \vertex (e) at (3,1.) {\( \ell_\beta\)};
    \vertex (f) at (3,-1.) {\( \bar\nu_f\)};
    \diagram* {
    (a) -- [boson] (b),
    (b) -- [boson, edge label=\( Z / \gamma \)] (c),
    (c) -- [anti fermion, edge label=\(\ell_\beta\)] (d),
    (d) -- [boson, edge label=\( W^-\)] (b),
    (c) -- [fermion] (e),
    (f) -- [fermion] (d)
    };
    \end{feynman}
    \end{tikzpicture}
    }
           \caption*{(f)}
            \label{}
    \end{subfigure}
        \hfill
    \begin{subfigure}[b]{0.24\textwidth}
    \centering
\raisebox{5mm}{        \begin{tikzpicture}
    \begin{feynman}
    \vertex (a) at (0,0) {\(W^-\)};
    \vertex (b) at (1,0);
    \vertex (c) at (2,1.);
    \vertex (d) at (2,-1.);
    \vertex (e) at (3,1.) {\( \ell_\beta\)};
    \vertex (f) at (3,-1.) {\( \bar\nu_f\)};
    \diagram* {
    (a) -- [boson] (b),
    (b) -- [scalar, edge label=\( h \)] (c),
    (c) -- [anti fermion, edge label=\(\ell_\beta\)] (d),
    (d) -- [boson, edge label=\( W^-\)] (b),
    (c) -- [fermion] (e),
    (f) -- [fermion] (d)
    };
    \end{feynman}
    \end{tikzpicture}
    }
           \caption*{(g)}
            \label{}
    \end{subfigure}
    \caption{Relevant one-loop diagrams contributing to $W$ decays, in unitary gauge (notice that diagram (a) violates total lepton number). 
    }
    \label{fig:Wdecays:UG}
\end{figure}

We begin by discussing leptonic charged interactions, $W \ell_\beta \nu_f$, which - and as noticed before - receive several new contributions as a consequence of the presence of the new heavy (Majorana) states: the distinct contributions are displayed in Fig.~\ref{fig:Wdecays:UG} (in unitary gauge). 
Notice that while a subset of diagrams is in essence ``SM-like" (the last two types of diagrams in the second row of Fig.~\ref{fig:Wdecays:UG}), most of the new processes reflect the presence of the new massive neutral leptons. Moreover, certain contributions exist if and only if the neutral leptons are of Majorana nature, as the corresponding interactions lead to a violation of total lepton number. This is for example the case of the first diagram on the top-most row (which can lead to $W \ell_\beta \nu_f$ and $W \ell_\beta 
\bar \nu_f$).\footnote{Notice that the $W \ell_\beta \nu_f$ decay cannot be exploited to unveil the source of lepton number violation due to the associated light neutrino mass suppression; however, it might lead to relevant vertex corrections in the case of off-shell $W$ decays into heavy neutrinos.}

The general decomposition of the $W \ell \nu$ vertex can be written as 
\begin{eqnarray}
    \bar u_\beta(p_\beta)\,  \Gamma_{W \ell_\beta \nu_f}^\mu (q) \, v_f(p_f) &= & \sum_{X = L, R} \bar u_\beta(p_\beta)\left[F_S^X \, q^\mu \,P_X + F_V^X \,\gamma^\mu \,P_X  + F_T^X \,\sigma^{\mu\nu}\, q_\nu \,P_X  \right]\,v_f(p_f)\,,
    \label{eqn:Wamplitude}
\end{eqnarray}
with $q=p_\alpha + p_f$ the momentum of the $W$ boson and $\sigma^{\mu\nu} = \frac{i}{2}[\gamma^\mu, \gamma^\nu]$; $F_{S,V,T}^{L,R}$ are scalar, vector and tensor form factors. For on-shell $W$ decays, the scalar amplitudes do not contribute (their contribution vanishes due to the Ward identity), and the decay width can be approximately cast as
\begin{eqnarray}
    \Gamma(W\to \ell_\beta \nu_f) &\simeq& \Gamma^\mathrm{tree} + \Gamma^\mathrm{tree-loop} + \Gamma^\mathrm{brems}\,,
\end{eqnarray}
in which $\Gamma^\mathrm{tree}$ is the tree-level contribution, $\Gamma^\mathrm{tree-loop}$ denotes the interference term between tree-level and 1-loop diagrams, and $\Gamma^\mathrm{brems}$ corresponds to the bremsstrahlung corrections, which are required to cancel the infrared (IR) divergences arising from the presence of virtual photons in the loop diagram (f) of Fig.~\ref{fig:Wdecays:UG}. Further details regarding the form factors for leptonic $W$ decays have been  collected in Appendix~\ref{sec:Wformfactors}.
The tree-level contribution to the $W$ decay rate is given by
\begin{eqnarray}
    \Gamma^\mathrm{tree} &=& \dfrac{\lambda^{1/2}(M_W, m_\beta, m_f)}{16 \pi^2 \,M_W^3}\dfrac{1}{3 M_W^2}\bigg\{ 2 M_W^4 - M_W^2(m_f^2+ m_\beta^2) - (m_f^2- m_\beta^2)^2\bigg\}\,|F_V^{L, \mathrm{tree}}|^2\,,
\end{eqnarray}
where $\lambda(a,b,c) = (a^2 -b^2-c^2)^2 - 4b^2c^2$ is the Käll\'en-function and the tree-level contribution is simply given by $F_V^{L, \mathrm{tree}} =-g_w/\sqrt{2} \, \mathcal{U}_{\beta f}$. The interference term can be written in terms of the above form factors as
\begin{eqnarray}
    \Gamma^\mathrm{tree-loop} &=&\dfrac{\lambda^{1/2}(M_W, m_\beta, m_f)}{16 \pi^2 \,M_W^3}\dfrac{1}{3 M_W^2} \,2 \mathrm{Re}\bigg[
    \left(2M_W^4 - M_W^2(m_f^2+ m_\beta^2) - (m_f^2- m_\beta^2)^2\right)F_V^L \,\left(F_V^{L,\mathrm{tree} }\right)^* 
    \nonumber \\
    &\phantom{=}&
    + \, 6 M_W^2 \,m_f \,m_\beta \,F_V^R \,\left(F_V^{L,\mathrm{tree} }\right)^*
    - 3i M_W^2 \,m_\beta (m_\beta^2 - m_f^2 - M_W^2) F_T^L\, \left(F_V^{L,\mathrm{tree} }\right)^*
    \nonumber \\
    &\phantom{=}&
    - \, 3i M_W^2 \,m_f (m_f^2 - m_\beta^2 - M_W^2) F_T^R\, \left(F_V^{L,\mathrm{tree} }\right)^*
    \bigg]\,.
\end{eqnarray}
(Notice that the terms involving $F_T^{L,R}$ and  $F_V^{R}$ lead to subdominant contributions.)
Finally the bremsstrahlung corrections, which originate from photonic
processes  with charged external particles (to one-loop order, only single photon radiation needs to be considered) can be found in~\cite{Denner:1991kt}, noticing that the charged current vertex has to be accordingly modified - as done throughout the present analysis. 

Upon renormalisation, and for on-shell $W$ bosons, the $W\ell \nu$ vertex exhibits the following divergence:
\begin{eqnarray}
\mathrm{div}_{W\ell_\beta \nu_f}
&=&-\frac{g^3 }{16 \pi^2 \,4 \sqrt{2}  \,\epsilon\,M_W^2 }\,\bigg\{\sum_i C_{if} \, \mathcal{U}_{\beta i} \,m_i^2 + \mathcal{U}_{\beta f} \,m_\beta^2 + \mathcal{U}_{\beta f}\,M_Z^2 + 10\, \mathcal{U}_{\beta f}\,M_W^2
\bigg\}\,,
\end{eqnarray}
which will be cancelled owing to renormalisation (which includes renormalising the matrix $\mathcal{U}$, 
$s_w$ and the electromagnetic coupling, $e$, as discussed in the previous section). For this purpose we include all the SM counterterms, and recompute the fermionic ones in the present BSM framework.
Thus, adapting the SM $W\ell\nu$ vertex renormalisation~\cite{Denner:1991kt} by taking into account renormalisation of the leptonic mixing matrix $\mathcal{U}$~\cite{Kniehl:1996bd}, the counterterm for $W$ decays is given in Eq.~(\ref{eq:WvertexRenormalisation}),
\begin{equation}\label{eq:WvertexRenormalisation}
\hspace*{-3mm} \raisebox{-12mm}
 {\begin{tikzpicture}
    \begin{feynman}
    \vertex (a) at (0,0) {\(W_\mu^-\)};
    \vertex (b) at (1,0);
    \vertex (c) at (2,1.){\( \ell_\beta^-\)};
    \vertex (d) at (2,-1.){\(\bar \nu_f\)};
    \diagram* {
    (a) -- [boson] (b),
    (c) -- [anti fermion] (b),
    (b) -- [anti fermion] (d),
    };
    \end{feynman}
    \end{tikzpicture}}
\hspace{-7mm}= -\, i\dfrac{g_w}{\sqrt{2}} \, \gamma_\mu \, \bigg\{
    \mathcal{U}_{\beta f}\, \bigg(1 - \dfrac{\delta s_w}{s_w} + \delta Z_e + \dfrac{1}{2}\delta Z_W \bigg) + \delta \mathcal{U}_{\beta f} + \dfrac{1}{2} \, \bigg( \displaystyle\sum_\alpha \,\delta Z_{\beta\alpha}^{\ell, L \dag} \, \mathcal{U}_{\alpha f}\, + \sum_i  \, \mathcal{U}_{\beta i} \,\delta Z_{if}^{\nu, L} \bigg) \bigg\} P_L\,.
\end{equation}

\bigskip
After the above (formal) considerations, we now proceed to discuss the phenomenological implications:
to address deviations from universality, we will consider 
the following ratios of decay widths
\begin{equation}
    R^W_{\alpha\beta}\, = \, \dfrac{\Gamma(W \to \ell_\alpha \nu)}{\Gamma(W \to \ell_\beta \nu)}\,, \quad \text{with } \alpha \neq \beta \, =\, e, \, \mu, \, \tau\,.
\end{equation}
In SM extensions via heavy neutral leptons, it has been shown~\cite{Fernandez-Martinez:2016lgt} that (at tree-level) $R^W_{\alpha \beta}$ displays 
the following dependency on the deviations from unitarity of the PMNS matrix, 
\begin{equation}\label{eq:RWalphabeta:analytical:eta}
 R^W_{\alpha \beta}\, \sim\, \left[1-2(\eta_{\alpha\alpha}-\eta_{\beta\beta})\right]R_{\alpha\beta}^W|_{\mathrm{SM}}\,.
 \end{equation}
The BSM predictions should thus be compared with the SM expectation, which have been  computed including corrections at one-loop level\footnote{Notice that these results were obtained for $M_H \sim 100$ GeV.} in~\cite{Kniehl:2000rb}, and with the subsequent measurements at LEP~\cite{ALEPH:2013dgf},
\begin{eqnarray}
    R_{\tau e}^W|_{\mathrm{SM}}
    &=& 0.9993\,, \quad\quad\quad 
    R_{\tau e}^W|_{\mathrm{LEP}} 
    \:=\: 1.063\pm 0.027\,, \\
    R_{\tau\mu}^W|_{\mathrm{SM}}
    &=& 0.9993\,,
    \quad\quad\quad 
    R_{\tau\mu}^W|_{\mathrm{LEP}}
    \:=\: 1.070\pm 0.026\,, 
\\
R_{\mu e}^W|_{\mathrm{SM}}
    &=& 0.9999\,, \quad\quad\quad 
    R_{\mu e}^W|_{\mathrm{LEP}}
    \:=\: 0.993\pm 0.019\,. 
\end{eqnarray}
Recently a more precise measurement has been performed by ATLAS~\cite{ATLAS:2020xea}, exhibiting a very good agreement with the SM prediction; likewise, CMS has also published new results~\cite{CMS:2022mhs}, also in good agreement with the SM (in particular concerning $W \to \tau \nu$ decays).

\section{Higgs decays to dimuons and ditaus}\label{sec:Hdecays}
Searches for lepton flavour universality violation in Higgs decays are also promising probes of new physics in the lepton sector, especially in view of the expected developments at current and future colliders:
in the coming future, the LHC Run 3 is expected to offer
a 4.5\% increase in energy and 50\% increase in the collision rate combined with luminosity levelling~\cite{Cepeda:2019klc}, and thus a potential (and more precise) measurement of the 
flavour conserving $H\to \mu\mu, \, \tau\tau$ widths.

Deviations from universality can be studied through the measurement of the ratio
\begin{equation}\label{eq:RHalphabeta:def}
    R^H_{\alpha \beta} \,= \,\frac{\Gamma(H\to \ell_\alpha \ell_\alpha)}{\Gamma(H\to \ell_\beta \ell_\beta)}\,,\quad \alpha=\mu,\quad\text{and}\quad \beta =\tau\,. 
\end{equation}
Within the SM, the $R^H_{\mu \tau}$ ratio is a theoretically very clean observable; at leading order, it reflects that in the lepton sector only the charged lepton Yukawa couplings (i.e., the charged lepton masses) are responsible for the breaking of LFU, $R^H_{\alpha \beta}|_\text{SM} \approx \mathcal{O}(m_\alpha^2/m_\beta^2)$. 
 
 Concerning the individual decay widths, and at lowest order, one finds~\cite{Dery:2013rta}
\begin{equation}
\Gamma(H\to \ell_\alpha \ell_\alpha)\,=\,
\frac{G_F\, M_H}{4 \sqrt 2 \pi}\,m_\alpha^3\,
\left( 1- \frac{4\, m_\alpha^2}{M_H^2}\right)^{3/2}\,
\left( 1 +\delta^{\ell_\alpha}_\text{QED} + \delta^{\ell_\alpha}_\text{weak} + \delta^{\ell_\alpha}_\text{HNL}\right)\,, 
\end{equation}
with the SM-like corrections (i.e. from quantum electrodynamics (QED), and weak interactions) given by
\begin{eqnarray}
    \delta_{\mathrm{QED}}^{\ell_\alpha} &=& \dfrac{3\, \alpha}{2\pi}\left(\dfrac{3}{2} - \log \dfrac{M_H^2}{m_\alpha^2}\right)\,,
    \\
    \delta_{\mathrm{weak}}^{\ell_\alpha} &=& \dfrac{G_F}{8\pi^2\sqrt{2}}\left[7m_t^2 + M_W^2\left(-5 + \dfrac{3\log c_w^2}{s_w^2}\right) - M_Z^2\dfrac{6(1-8s_w^2+16s_w^4)-1}{2}\right]\,.
\end{eqnarray}
As discussed in~\cite{Dery:2013rta}, the weak corrections do cancel out in the $R^H_{\alpha \beta}$ 
ratio; non-universal corrections, which were neglected in the above, are considerably smaller than phase-space factors and QED corrections included in $\delta^{\ell_\alpha}_\text{QED}$. Carrying out an expansion of the remaining terms up to leading order in $m_\alpha^2/M_H^2$, 
the SM prediction for $ R^H_{\alpha \beta}$ can thus be written as
\begin{equation}
    R^H_{\alpha \beta} |_\text{SM} \,\approx \, \frac{m_\alpha^2}{m_\beta^2} \,
    \left( 1+ 6\frac{m_\beta^2-m_\alpha^2}{M_H^2} - \frac{3 \alpha_e}{2 \pi} \log{\frac{m_\beta^2}{m_\alpha^2}}
    \right)\,.
\end{equation}
The presence of the additional heavy states with non-negligible mixings to the light neutrinos
can lead to deviations from the above SM prediction for $R^H_{\alpha \beta}$. A first (na\"ive) estimation of
possible LFUV NP contributions can be obtained by considering the leading order interference between the SM contribution and 
the one-loop contribution from the presence of the heavy sterile states. In particular, the latter corrections, $\delta_\mathrm{HNL}^{\ell_\alpha}$ - corresponding to the interference term between tree- and loop-levels - can be cast as
\begin{eqnarray}
    \delta_\mathrm{HNL, int}^{\ell_\alpha}&=&\bigg[\frac{G_F\, M_H}{4 \sqrt 2 \pi}\,m_\alpha^3\,
\left( 1- \frac{4\, m_\alpha^2}{M_H^2}\right)^{3/2}\bigg]^{-1}
\dfrac{1}{16\pi M_H}\sqrt{1 - \dfrac{4m_\alpha^2}{M_H^2}}
\nonumber\\
&\phantom{=}& \times 2\mathrm{Re}\bigg[-2m_\alpha^2 \left(F^\mathrm{tree }\right)^*(F_L + F_R) + (M_H^2 - 2 m_\alpha^2 )\left(F^\mathrm{tree }\right)^*(F_L + F_R)\bigg]\,,
\end{eqnarray}
where $F^\mathrm{tree} = -g_w m_\alpha/2M_W$ is 
the tree-level Higgs-lepton coupling and the 1-loop form factors $F_{L/R} = F_{L/R}^{(a)}+F_{L/R}^{(b)}$ 
(arising from the new diagrams involving the HNL, see Fig.~\ref{fig:LFCHiggsdecays:UG}) are given in 
Appendix~\ref{sec:formfactors}.
The renormalised vertex is presented in Eq.~(\ref{eq:HvertexRenormalisation}),
\begin{equation}\label{eq:HvertexRenormalisation}
 \raisebox{-12mm}{\begin{tikzpicture}
    \begin{feynman}
    \vertex (a) at (0,0) {\(H\)};
    \vertex (b) at (1,0);
    \vertex (c) at (2,1.){\( \ell_\alpha\)};
    \vertex (d) at (2,-1.){\(\ell_\beta\)};
    \diagram* {
    (a) -- [scalar] (b),
    (c) -- [anti fermion] (b),
    (b) -- [anti fermion] (d),
    };
    \end{feynman}
    \end{tikzpicture}}
= \,-\, i\dfrac{ g_w}{2 M_W} \,(\mathcal{C}_{L, \alpha \beta}^H P_L + {\mathcal{C}_{R, \alpha \beta}^H} P_R)\,.\end{equation}
In the above, we have introduced the quantities $\mathcal{C}_{L,R}^H$, which are defined as
\begin{eqnarray}
    \mathcal{C}_{L, \alpha \beta}^H &=& 
    \delta_{\alpha \beta}m_\alpha\, \bigg(1 - \dfrac{\delta s_w}{s_w} + \delta Z_e  
    +\dfrac{\delta m_\alpha}{m_\alpha} - \dfrac{\delta M_W}{M_W}
    +\dfrac{1}{2}\delta Z_{H} \bigg) 
    + \dfrac{1}{2} \, \bigg( m_\alpha \, \delta Z_{\alpha \beta}^{\ell, L \dag} \, + \delta Z_{\alpha \beta}^{\ell, R\dag} m_\beta \bigg) \,,
\nonumber \\
    \mathcal{C}_{R, \alpha \beta}^H &=& 
    \delta_{\alpha \beta}m_\alpha\, \bigg(1 - \dfrac{\delta s_w}{s_w} + \delta Z_e 
    +\dfrac{\delta m_\alpha}{m_\alpha} - \dfrac{\delta M_W}{M_W}
    +\dfrac{1}{2}\delta Z_{H} \bigg) 
    + \dfrac{1}{2} \, \bigg( m_\alpha \, \delta Z_{\alpha \beta}^{\ell, R \dag} \, + \delta Z_{\alpha \beta}^{\ell, L\dag} m_\beta \bigg) \,.
\end{eqnarray}

\section{LFU violation in $Z$ decays $R^Z_{\alpha \beta}$}\label{sec:Zdecays}
The presence of the (heavy) sterile fermion states 
can also be at the source of new contributions to the individual decay widths of $Z$ bosons, and potentially contribute to an effective violation of lepton flavour universality of $Z$-boson couplings.
As done for the Higgs decays, it is convenient to consider 
the following ratios of decay widths
\begin{equation}\label{eq:RZab}
    R^Z_{\alpha\beta}\, = \, \dfrac{\Gamma(Z \to \ell_\alpha^+\ell_\alpha^-)}{\Gamma(Z \to \ell_\beta^+\ell_\beta^-)}\,, \quad \text{with } \alpha \neq \beta \, =\, e, \, \mu, \, \tau\,,
\end{equation}
which allow cancelling QED corrections in the theoretical predictions. 
At 2-loop accuracy, one has the following SM predictions for the $R^Z_{\alpha\beta}$
 ratios~\cite{Freitas:2014hra}
\begin{eqnarray}
R_{\mu e}^Z|_{\mathrm{SM}} \, =\,  \dfrac{\Gamma(Z \to \mu^+\mu^-)^{\text{SM}}}{\Gamma(Z \to e^+e^-)^{\text{SM}}} &=& 1 \,, \nonumber \\
R_{\tau \mu }^Z|_{\mathrm{SM}} \, =\,     \dfrac{\Gamma(Z \to \tau^+\tau^-)^{\text{SM}}}{\Gamma(Z \to \mu^+\mu^-)^{\text{SM}}} &=& 0.9977\,,\nonumber\\
   R_{\tau e}^Z|_{\mathrm{SM}} \, =\,   \dfrac{\Gamma(Z \to \tau^+\tau^-)^{\text{SM}}}{\Gamma(Z \to e^+e^-)^{\text{SM}}} &=& 0.9977\,,
\end{eqnarray}
(with negligible associated uncertainties); the corresponding experimental values~\cite{ParticleDataGroup:2022pth} are
\begin{eqnarray}
    \dfrac{\Gamma(Z \to \mu^+\mu^-)^{\text{exp}}}{\Gamma(Z \to e^+e^-)^{\text{exp}}} &=& 1.0001 \pm 0.0024 \,, \nonumber \\
    \dfrac{\Gamma(Z \to \tau^+\tau^-)^{\text{exp}}}{\Gamma(Z \to \mu^+\mu^-)^{\text{exp}}} &=& 1.0010 \pm 0.0026\,,\nonumber\\
     \dfrac{\Gamma(Z \to \tau^+\tau^-)^{\text{exp}}}{\Gamma(Z \to e^+e^-)^{\text{exp}}} &=& 1.0020 \pm 0.0032\,.
\label{eq::ZdecayLFU}
\end{eqnarray} 
In what concerns the NP contributions, we estimate the modified individual partial widths as
\begin{equation}
    \Gamma (Z\to \ell^+\ell^-) \simeq \Gamma^{\text{SM}_\text{full}} + \Gamma^{\text{SM}_\text{tree}-n_S} \,,
\end{equation}
where $\Gamma^{\text{SM}_\text{full}}$ is given at 2-loop accuracy in~\cite{Freitas:2014hra}.
In our numerical evaluation we implement the parametrisation formula for $\Gamma^{\mathrm{SM}_\mathrm{full}}$ given in~\cite{Freitas:2014hra} and add the HNL contributions at 1-loop;
$\Gamma^{\text{SM}_\text{tree}-n_S}$ is the interference term between the SM tree-level contribution and the 1-loop diagrams 
reflecting the contributions from the sterile neutrinos,
\begin{eqnarray}
   \Gamma^{\text{SM}_\text{tree}-n_S} &=& \dfrac{\sqrt{M_Z^2 - (2m_\alpha)^2}}{16 \pi M_Z^2}\dfrac{1}{3} 2\mathrm{Re}\bigg[-4\left(F_L^{V}\right)^*(F_L^{\mathrm{tree}} - 2F_R^{\mathrm{tree}})m_\alpha^2 - 4\left(F_R^{V}\right)^*(-2F_L^{\mathrm{tree}} + F_R^{\mathrm{tree}})m_\alpha^2 
    \nonumber \\
    &\phantom{=}& + [2\left(F_L^{V}\right)^*F_L^{\mathrm{tree}} + 2\left(F_R^{V}\right)^*F_R^{\mathrm{tree}} + 3i\left(\left(F_L^{T}\right)^* + \left(F_R^{T}\right)^*\right)(F_L^{\mathrm{tree}} + F_R^{\mathrm{tree}})m_\alpha]q^2 
    \nonumber \\
    &\phantom{=}&
    + \dfrac{2 m_\alpha^2 q^2}{M_Z^2}\left(\left(F_L^{V}\right)^* - \left(F_R^{V}\right)^*\right)(F_L^{\mathrm{tree}} - F_R^{\mathrm{tree}})\bigg]\,.
\end{eqnarray}
The tree-level form factors are defined as $F^\text{tree}_{L/R} = -g_w/(2c_w)({\bf C}_{V} \pm {\bf C}_{A} )$.
(In the above, we have verified that terms involving $F^T_{L,R}$ lead to subdominant contributions.)
Typically, in SM extensions via heavy sterile states, the dominant contributions in general arise from the diagram with two neutral leptons in the loop (diagram (a) in Fig.~\ref{fig:LFCZdecays:UG}).

The renormalisation of the $Z \ell \ell$ vertex is given in Eq.~(\ref{eq:ZllVertexRenormalisation}):
\begin{equation}\label{eq:ZllVertexRenormalisation}
 \raisebox{-12mm}{\begin{tikzpicture}
    \begin{feynman}
    \vertex (a) at (0,0) {\(Z_\mu\)};
    \vertex (b) at (1,0);
    \vertex (c) at (2,1.){\( \ell_\alpha\)};
    \vertex (d) at (2,-1.){\( \ell_\beta\)};
    \diagram* {
    (a) -- [boson] (b),
    (c) -- [anti fermion] (b),
    (b) -- [anti fermion] (d),
    };
    \end{feynman}
    \end{tikzpicture}}
=\,-\, i\dfrac{ g_w}{ c_w} \, \gamma_\mu \,(\mathcal{C}_{L, \alpha \beta}^{Z \ell} P_L - \mathcal{C}_{R, \alpha \beta}^{Z \ell} P_R)\,,
\end{equation}
where the coefficients $\mathcal{C}_{L,R}^{Z\ell}$ are defined as
\begin{eqnarray}
    \mathcal{C}_{L, \alpha \beta}^{Z\ell} &=& g_L^Z\,\bigg[\delta_{\alpha\beta} \bigg(1 +  \dfrac{\delta g^Z_L}{g^Z_L}  + \dfrac{1}{2}\delta Z_{ZZ} \bigg)  
  + \dfrac{1}{2} \, \bigg(\delta Z_{\alpha\beta}^{\ell, L }\, + \delta Z_{\alpha\beta}^{\ell, L\dag} \bigg)\bigg] - \delta_{\alpha\beta} \dfrac{1}{2} Q_\ell\, \delta Z_{AZ} \,, 
  \nonumber \\
\mathcal{C}_{R, \alpha \beta}^{Z\ell}  &=& g_R^Z\,\bigg[\delta_{\alpha\beta} \bigg(1 +  \dfrac{\delta g^Z_R}{g^Z_R}  + \dfrac{1}{2}\delta Z_{ZZ} \bigg)  
  + \dfrac{1}{2} \, \bigg(\delta Z_{\alpha\beta}^{\ell, R }\, + \delta Z_{\alpha\beta}^{\ell, R\dag} \bigg)\bigg]- \delta_{\alpha\beta} \dfrac{1}{2} Q_\ell\, \delta Z_{AZ} \,.
\end{eqnarray}
In the above equation, the right- and left-handed $Z$ couplings to the charged leptons are
\begin{eqnarray}
    g_R^Z \:=\: - Q_\ell \,s_w^2\,,
    \quad \quad
     g_L^Z \:=\:  T^3_\ell - Q_\ell\, s_w^2\,,
\end{eqnarray}
with $T^3_\ell$ and $Q_\ell$ respectively denoting the weak isospin and electric charge. 
Details on the different couplings and wave functions can again be found in Appendix~\ref{app:renormalisation}, while the form factors are further discussed in Appendix~\ref{sec:formfactors}.

\section{Impact on EW precision observables: invisible $Z$ width and further constraints}\label{sec:EWPO:Zinv}

The presence of heavy neutral fermions  will also contribute at one-loop level to several EW (precision) observables including, among others, the oblique parameters, and the invisible $Z$ width. 
Following the PDG~\cite{ParticleDataGroup:2022pth} definition of the oblique parameters $S, T, U$, the SM contributions to the latter vanish by construction; any deviation from $0$ is thus a  clear indication of NP contributions to the electroweak gauge boson self-energies.
The oblique parameters are defined as~\cite{Peskin:1990zt,Peskin:1991sw,Fernandez-Martinez:2015hxa,ParticleDataGroup:2022pth}
\begin{eqnarray}
     \alpha_e \,S &=& \frac{4 s_w^2 \,c_w^2}{M_Z^2}\left[\hat \Sigma_{ZZ}^N(0) + \hat \Sigma_{\gamma\gamma}^N(M_Z^2) - \frac{c_w^2 - s_w^2}{c_w \,s_w}\,\hat \Sigma_{Z\gamma}^N(M_Z^2)\right]\,,\\
     \alpha_e \,T &=& \frac{\hat \Sigma_{ZZ}^N(0)}{M_Z^2} - \frac{\hat \Sigma_{WW}^N(0)}{M_W^2}\,,\\
     \alpha_e \,U &=& 4 s_w^2 \,c_w^2\left[\frac{1}{c_w^2}\,\frac{\hat \Sigma_{WW}^N(0)}{M_W^2} - \frac{\hat \Sigma_{ZZ}^N(0)}{M_Z^2} + \frac{s_w^2}{c_w^2}\,\frac{\hat \Sigma_{\gamma\gamma}^N(M_Z^2)}{M_Z^2} - 2\frac{s_w}{c_w}\,\frac{\hat \Sigma_{Z\gamma}^N(M_Z^2)}{M_Z^2}\right]\,,
\end{eqnarray}
in which the various $\hat \Sigma^N$ denote the HNL contributions to the renormalised boson self-energies.
Notice that the heavy neutral lepton contributions to the unrenormalised $\Sigma_{\gamma\gamma}$ and $\Sigma_{Z\gamma}$ are vanishing, contrary to the renormalised ones.
The definition of the renormalised self-energies in the on-shell scheme can be found in~\cite{Denner:1991kt,Fernandez-Martinez:2015hxa}, while the explicit expressions for the HNL contributions to the unrenormalised boson self-energies are given in Appendix~\ref{app:renormalisation}.
We further emphasise here that only the contributions to the $T$ parameter will be of phenomenological relevance, since contributions to $S$ and $U$ are strongly suppressed.

In the present study, we will emphasise the role of the invisible $Z$ width, in particular the impact of the new higher order contributions (for an extensive discussion on these constraints on SM extensions via sterile fermions, see~\cite{Fernandez-Martinez:2015hxa}). The relevant form factors for the invisible $Z$ decay width are also detailed in Appendix~\ref{sec:formfactors}. The renormalisation of the $Z n_a n_b$ vertex is given by
\begin{equation}\label{eq:ZnunuVertexRenormalisation}
 \raisebox{-12mm}{\begin{tikzpicture}
    \begin{feynman}
    \vertex (a) at (0,0) {\(Z_\mu\)};
    \vertex (b) at (1,0);
    \vertex (c) at (2,1.){\( n_a\)};
    \vertex (d) at (2,-1.){\(\bar n_b\)};
    \diagram* {
    (a) -- [boson] (b),
    (c) -- [anti fermion] (b),
    (b) -- [anti fermion] (d),
    };
    \end{feynman}
    \end{tikzpicture}}
=\,-\, i\dfrac{2 g_w}{4 c_w} \, \gamma_\mu \,(\mathcal{C}_{L, a b}^{Z \nu} P_L - \mathcal{C}_{R, a b}^{Z \nu} P_R)\,,
\end{equation}
where the coefficients $\mathcal{C}_{L,R}^Z$ read
\begin{align}
    \mathcal{C}_{L, a b}^{Z \nu} =  C_{a b}\, \bigg( \dfrac{s_w^2 - c_w^2}{c_w^2} \dfrac{\delta s_w}{s_w} + \delta Z_e + \dfrac{1}{2}\delta Z_{ZZ} \bigg) + \delta C_{a b} 
  + \dfrac{1}{2} \, \bigg( \displaystyle\sum_x \,\delta Z_{ax}^{\nu, L \dag} \,C_{xb}\, + \sum_y  \, C_{a y} \,\delta Z_{yb}^{\nu, L} \bigg), \nonumber \\
  \mathcal{C}_{R, a b}^{Z \nu} = C_{a b}^*\, \bigg(\dfrac{s_w^2 - c_w^2}{c_w^2} \dfrac{\delta s_w}{s_w} + \delta Z_e + \dfrac{1}{2}\delta Z_{ZZ} \bigg) + \delta C_{a b}^* + \dfrac{1}{2} \, \bigg( \displaystyle\sum_x \,\delta Z_{ax}^{\nu, R \dag} \,C_{xb}^*\, + \sum_y  \, C_{a y}^* \,\delta Z_{yb}^{\nu, R} \bigg),
\end{align}
with the $C_{ab}$ coefficients as defined in Eq.~(\ref{eq:cij}).
The 1-loop invisible $Z$ width is then given by
\begin{eqnarray}
    \Gamma(Z \to \text{inv.})&=& \sum_{a,b}^{N_\mathrm{max}}\left(1-\dfrac{1}{2}\delta_{ab}\right)\dfrac{\lambda^{1/2}(M_Z,m_a,m_b)}{48\pi M_Z^5}\,\Gamma^\mathrm{full}\,,
\end{eqnarray}
$N_\mathrm{max}$ being the heaviest neutrino kinematically allowed.  The full invisible width is the sum of the tree-level width and the interference between tree- and loop-level contributions mediated by the new states in the present model 
$\Gamma^\mathrm{full} = \Gamma^\mathrm{tree}+\Gamma^{\mathrm{tree}-n_S}$. In terms of the form factors, the widths $\Gamma^\mathrm{tree}$ and $\Gamma^{\mathrm{tree}-n_S}$  can  be written as\footnote{Notice that for identical final states, the ``cross-lined" diagrams - with $n_a \leftrightarrow n_b$ - have to be taken into account.}
\begin{eqnarray}
    \Gamma^\mathrm{tree} &=& \left(2M_Z^4 - M_Z^2(m_b^2+ m_a^2) - (m_b^2-m_a^2)^2 \right)(|F_L^{\mathrm{tree}}|^2 + |F_R^{\mathrm{tree}}|^2)
 \nonumber\\
    &\phantom{=}&
    +6M_Z^2\,m_a\,m_b\,(F_L^{\mathrm{tree}}\left(F_R^{\mathrm{tree}}\right)^*+F_R^{\mathrm{tree}}\left(F_L^{\mathrm{tree}}\right)^*)\,,
\\   
    \Gamma^{\mathrm{tree}-n_S} &=&  2\mathrm{Re}\bigg[ 
    \left(2M_Z^4 - M_Z^2(m_b^2+m_a^2) - (m_b^2-m_a^2)^2\right)(F_L^V \left(F_L^{\mathrm{tree}}\right)^* + F_R^V \left(F_R^{\mathrm{tree}}\right)^* )
    \nonumber\\
    &\phantom{=}&
    + 6M_Z^2 \,m_a \,m_b\,(F_L^V \left(F_R^{\mathrm{tree}}\right)^* + F_R^V \left(F_L^{\mathrm{tree}}\right)^*)
    \nonumber\\
    &\phantom{=}&
    - 3i M_Z^2\,m_a(m_a^2-m_b^2-M_Z^2)(F_L^T \left(F_L^{\mathrm{tree}}\right)^* + F_R^T \left(F_R^{\mathrm{tree}}\right)^* )
    \nonumber\\
    &\phantom{=}&
    - 3i M_Z^2\,m_b(m_b^2-m_a^2-M_Z^2)(F_R^T \left(F_L^{\mathrm{tree}}\right)^* + F_L^T \left(F_R^{\mathrm{tree}}\right)^* )
    \bigg]\,.
\end{eqnarray}
In the above, the tree-level couplings are given by $F_L^{\mathrm{tree}}= -2g_w/(4c_w)C_{ab}$ and $F_R^{\mathrm{tree}}= 2g_w/(4c_w)C_{ab}^*$. 
Notice that terms involving the tensor form factors $F_{L/R}^T$ are subdominant.

\bigskip
A more phenomenological approach allows revealing the important deviations from the SM expectations which are  induced from the departure from unitarity of the would-be PMNS matrix; 
this can be inferred from the tree-level expression for the invisible width (see~\cite{Fernandez-Martinez:2016lgt})
\begin{equation}\label{eq:InvZwidth:tree}
    \Gamma (Z \to \text{inv.})\, =\, \frac{G_F\, M_Z^3\, \sum_{i,j} |(\tilde U^\dagger \, \tilde U)_{ij}|^2}{12\sqrt 2 \, \pi} \, \simeq
    \frac{G_\mu\, M_Z^3}{12\sqrt 2 \, \pi}\, \big(
    3-\left(\text{Tr}(\eta) + 3 \eta_{\tau\tau}\right) 
    \big)\,,
\end{equation}
in which the parameter $\eta$ has been defined in Eq.~(\ref{eq:defPMNSeta}) and $G_{\mu}$ is the Fermi constant as measured in muon decays, introduced in Appendix~\ref{app:ISS}.

As expected, the one-loop diagrams can create further tensions with the SM, due to the sizeable contributions induced by numerous exchanges, especially those including two heavy virtual states (see diagrams (a) in Fig.~\ref{fig:Zinvdecays:UG}). 
In view of the expected improvements in the associated experimental precision, these higher order terms will play a relevant role in assessing the viability of SM extensions via HNL, as is the case of the ISS.

\section{Phenomenological analysis: the Inverse Seesaw}\label{sec:NumericalResults}
After having discussed the formal aspects of the  contributions of the (heavy) sterile states regarding the 
leptonic interaction vertices, we now proceed to illustrate  the effects of  the HNL in a well-motivated UV-complete NP model, which naturally incorporates them: the Inverse Seesaw mechanism of neutrino mass generation, which can be realised at comparatively low scales.
In this section we thus present the results of our phenomenological analysis of the ISS(3,3), a realisation in which 3 right-handed neutrinos and 3 other sterile states are added to the SM content  (see Appendix~\ref{app:ISS} for a detailed presentation of the model).  We focus on several LFUV and EW precision observables, being particularly interested in regimes for which the (in general very constraining) bounds from cLFV observables are superseded by the flavour conserving probes.

\subsection{Exploring the ISS(3,3) parameter space}
Despite the different parametrisations of the ISS(3,3) which are 
available in the literature, and which have been extensively used to study its parameter space, a first exploration showed that the latter were not well-suited for our purposes.
On the one hand, the ``standard'' Casas-Ibarra parametrisation~\cite{Casas:2001sr} can scale exponentially if complex angles in the arbitrary orthogonal matrix are considered, thus quickly leading to regimes of non-perturbative Yukawa couplings.
Moreover, one does not have control over flavour-violating couplings, such that regimes in which one has large flavour non-universality but small flavour-violation correspond to an extreme fine-tuning of the Casas-Ibarra parameters.
On the other hand, while the 
``$\mu_X$''-parametrisation~\cite{Arganda:2014dta} solves some of these problems, 
only certain benchmark points were considered, as to maximise flavour violation in a certain ``direction''.
Therefore, we have considered a more general formulation of the ``$\mu_X$''-parametrisation of~\cite{Arganda:2014dta} and~\cite{Garnica:2023ccx}, which despite its simplicity, efficiently allows to fully explore flavour non-universal regimes, while retaining control over flavour-violating configurations. We thus cast
\begin{equation}\label{eq:parametrisation:muX1}
\mu_X\, = \,M_R^T \,m_D^{(-1)} \,U_\text{PMNS}^\ast\, \mathrm{diag}(m_{\nu_1}, m_{\nu_2}, m_{\nu_3}) \,U_\text{PMNS}^\dagger \, (m_D^T)^{(-1)} \,M_R\,,
    \end{equation}
with     
\begin{equation}\label{eq:parametrisation:muX1:mD}
    m_D \,= \, v \,Y_D  \,=\, v \,\mathrm{diag}(y_1, y_2, y_3)\mathcal V\, M_R^\dagger\,,
\end{equation}
and in which $\mathcal V$ is an arbitrary (special-) unitary matrix, $M_R$ is an arbitrary complex matrix (which for simplicity, and without loss of generality, is assumed to be diagonal and real), $U_\text{PMNS}$ is the $(3\times3)$-PMNS matrix and $m_{\nu_i}$ are the light (active) neutrino masses.
From the above it is clear that $Y_D$ must be invertible, 
and so this parametrisation is {\it only} valid for $(3,n)$ realisations of the ISS; thus, at least 3 heavy states must be present\footnote{The parametrisation of Eqs.~(\ref{eq:parametrisation:muX1}, \ref{eq:parametrisation:muX1:mD}) can be likely generalised to more general ISS configurations, however we did not explore further possibilities, as our analysis was focused on the (3,3) realisation.}. 
Recalling the definition of the parameter $\eta$ (see Eq.~(\ref{eq:defPMNSeta})), 
one can readily verify that the above parametrisation allows to access regimes for which $\eta$ will be approximately diagonal:
\begin{equation}
    \eta \,\simeq\, \frac{1}{2}m_D^* \left(M_R^{-1}\right)^{\dagger}\left(M_R^{-1}\right)m_D^T\, =\, \frac{1}{2}\mathrm{diag}(y_1^2, y_2^2, y_3^2)\,;
\end{equation}
in particular, for mass-degenerate heavy states, flavour violation is absent by construction.
Furthermore, the unitary matrix $\mathcal V$ in Eq.~(\ref{eq:parametrisation:muX1:mD}) controls the off-diagonal terms in the ``active-sterile'' mixing matrix.
At leading order in the perturbative diagonalisation of the full mass-matrix, and under the assumption of a symmetric block-matrix $M_R$, one has
\begin{equation}
    \mathcal U_{3\times 6} \,\simeq\, \frac{1}{\sqrt{2}}\left(i m_D^\ast \, (M_R^\dagger)^{-1}\, , \, m_D^\ast \, (M_R^\dagger)^{-1}\right)\,.
\end{equation}
Again, for a real symmetric (or even diagonal) $M_R$,
and upon inserting the parametrisation of $m_D$ given in Eq.~\eqref{eq:parametrisation:muX1:mD}, it is evident that the only off-diagonal structure appearing in the ``active-sterile'' mixing matrix is entirely controlled by $\mathcal V$.
In particular,  one then finds
\begin{equation}
    \mathcal U_{3\times 6} \,\simeq \,\frac{1}{\sqrt{2}}\left(i\, \mathrm{diag}(y_1, y_2, y_3) \,\mathcal V^\ast, \mathrm{diag}(y_1, y_2, y_3) \,\mathcal V^\ast \right)\,.
\end{equation}
If one now chooses an Euler-parametrisation of $\mathcal V$ (analogously to the PMNS-matrix), as
\begin{equation}
    \mathcal V\, = \,R_{23}\,R_{13}\,R_{12}\,,
\end{equation}
the ``directions'' of charged lepton flavour violating transitions are controlled by the size of the mixing angles appearing in the rotations $R_{ij}$, such that for instance the angle in $R_{12}$ controls the size of $\mu-e$ cLFV transitions. The approach offered by the parametrisation of Eqs.~(\ref{eq:parametrisation:muX1},  \ref{eq:parametrisation:muX1:mD}) does allow to efficiently access and explore regimes with very distinctive features, both cLFV- and LFUV-wise.

\bigskip
Before presenting the results of the numerical study, we describe the scanning procedure, as well as the different steps of the analysis.

\noindent As seen from Eqs.~(\ref{eq:parametrisation:muX1}, \ref{eq:parametrisation:muX1:mD}), the input parameters are the unitary matrix $\mathcal V$, the heavy sterile masses $M_R$, the diagonal Yukawa couplings, and finally, the active neutrino data (i.e., the light state masses, their ordering, as well as the PMNS matrix angles).  
We thus vary the input parameters as follows:
\begin{itemize}
    \item matrix $\mathcal V$ - 3 angles and 9 phases in the range $(-\pi, \pi)$;
    \item diagonal entries of $M_R$ - taken in the range $(0.5, 20)$~TeV (logarithmically varied), no hierarchy is assumed;
    \item lightest neutrino mass - $m_0$ in the range $10^{-10}, 10^{-3}$eV (logarithmically varied);
    \item PMNS angles and active mass splittings fixed to best fit values (always for a normal ordering of the light neutrino spectrum), with PMNS phases sampled in the range $(-\pi, \pi)$~\cite{Esteban:2020cvm};
    \item diagonal Yukawa couplings ($y_1, y_2, y_3$) taken in the range $(10^{-3}, 100)$ - as a logarithmic prior -, no hierarchy assumed. 
\end{itemize}

The analysis then proceeds as follows. All input parameters are randomly generated in the chosen ranges, ensuring perturbativity of all the entries of the Yukawa couplings ($ |Y^D_{ij}|\leq \sqrt{4\pi}$). As detailed in Appendix~\ref{app:ISS}, numerous theoretical and phenomenological constraints are then enforced.
Firstly, perturbative unitarity constraints are applied to the heavy neutrino spectrum: as a first estimate, the width of each $N_i$ (i.e. $\Gamma_{N_i}$) is computed at tree-level from the leading decay channels, $N_i\to W\ell$, $N_i\to Z N_j$, $N_i\to H N_j$; regimes violating $\Gamma_{N_i} \leq \frac{1}{2} m_{N_i}$ are thus excluded.
Compatibility with numerous flavour observables is subsequently imposed, in particular concerning the cLFV leptonic decays $\mu\to e\gamma$, $\mu\to eee$, $\mu-e$ conversion, $\tau\to \mu\gamma$, $\tau\to e\gamma$, $\tau\to\mu\mu\mu$, and $\tau\to eee$. 
Likewise, the bounds from numerous flavour-universality observables are imposed, including  $\Delta r_K$, $\Delta r_\pi$ (defined in  Appendix~\ref{app:ISS}), and universality in tau-lepton decays. A first estimate (at tree-level) of the invisible $Z$ width and 
of lepton flavour universality violation in $W$ decays are also carried out, and regimes leading to violation of the latter bounds at the $3\sigma$ level are excluded. 

Once this first set of constraints is applied, we compute the loop-corrections to $\Gamma(Z\to \mathrm{inv.})$ and the LFU ratios of the $W$-decay modes (which are affected by the presence of the HNL already at tree-level), as well as the loop-corrections to the neutral boson decays $H\to \ell\ell$ and $Z\to \ell\ell$.

A detailed description of (current) bounds relevant for the discussion can be found in Table~\ref{tab:obs}.
\renewcommand{\arraystretch}{1.3}
\begin{table}[h!]
    \centering
    \hspace*{-2mm}{\small\begin{tabular}{|c|c|c|}
    \hline
    Observable & Exp. Measurement & SM prediction  \\
    \hline\hline
    $R_{\mu e}(W\to\ell\nu)$ & $0.993\pm 0.019$ (LEP~\cite{ALEPH:2013dgf}) & $0.9999$ ~\cite{Kniehl:2000rb}\\
    & $1.002\pm0.006$ (PDG~\cite{ParticleDataGroup:2022pth}) &\\
    $R_{\tau e}(W\to\ell\nu)$ & $1.062\pm0.027$ (LEP~\cite{ALEPH:2013dgf}) & $0.9993$ ~\cite{Kniehl:2000rb}\\
    & $1.015\pm0.020$ (PDG~\cite{ParticleDataGroup:2022pth}) &\\
    $R_{\tau \mu}(W\to\ell\nu)$ & $1.070\pm0.026$ (LEP~\cite{ALEPH:2013dgf}) & $0.9993$ ~\cite{Kniehl:2000rb}\\
    & $1.002\pm0.020$ (PDG~\cite{ParticleDataGroup:2022pth}) &\\
    \hline
    $R_{\mu e}(Z\to\ell\ell)$ & $1.0001\pm 0.0024$ (LEP~\cite{ALEPH:2005ab}) & $1.0$~\cite{Freitas:2014hra}\\
    $R_{\tau e}(Z\to\ell\ell)$ & $1.0020\pm0.0032$ (LEP~\cite{ALEPH:2005ab}) & $0.9977$~\cite{Freitas:2014hra}\\
    $R_{\tau \mu}(Z\to\ell\ell)$ & $1.0010\pm 0.0026$ (LEP~\cite{ALEPH:2005ab}) & $0.9977$~\cite{Freitas:2014hra}\\
    \hline
    $\Gamma(Z\to e^+e^-)$ & $83.91\pm0.12\:\mathrm{MeV}$ (LEP~\cite{ALEPH:2005ab}) & $83.965\pm0.016\:\mathrm{MeV}$~\cite{Freitas:2014hra}\\
    $\Gamma(Z\to \mu^+\mu^-)$ & $83.99\pm0.18\:\mathrm{MeV}$ (LEP~\cite{ALEPH:2005ab}) & $83.965\pm0.016\:\mathrm{MeV}$~\cite{Freitas:2014hra}\\
    $\Gamma(Z\to \tau^+\tau^-)$ & $84.08\pm0.22\:\mathrm{MeV}$ (LEP~\cite{ALEPH:2005ab}) & $83.775\pm0.016\:\mathrm{MeV}$~\cite{Freitas:2014hra}\\
    \hline
    $\Gamma(Z\to\mathrm{inv.})$ & $499.0 \pm 1.5\:\mathrm{MeV}$ (PDG~\cite{ParticleDataGroup:2022pth})& $501.45\pm 0.05\:\mathrm{MeV}$~\cite{Freitas:2014hra}\\
    \hline
    $\mathrm{BR}(H\to\tau^+\tau^-)$ & $0.06_{-0.007}^{+0.008}$ (PDG~\cite{ParticleDataGroup:2022pth}) & $0.0624\pm0.0035$~\cite{Denner:2011mq}\\
    % \hline
    $\mathrm{BR}(H\to\mu^+\mu^-)$ & $(2.6 \pm 1.3)\times 10^{-4}$ (PDG~\cite{ParticleDataGroup:2022pth}) & $(2.17 \pm0.13)\times 10^{-4}$~\cite{Denner:2011mq}\\
    \hline
    $R_{\mu e}(\tau\to\ell\nu\nu)$ & $0.9762\pm0.0028$ (PDG~\cite{ParticleDataGroup:2022pth}) & $0.972559\pm0.00005$~\cite{Pich:2013lsa}\\
    \hline
    \end{tabular}}
    \caption{Experimental  values and SM predictions  for several LFUV and EW observables discussed in the phenomenological analysis. All uncertainties are given at 68\% C.L., while for the SM predictions of the universality ratios, the parametric uncertainties are negligible.}
    \label{tab:obs}
\end{table}
\renewcommand{\arraystretch}{1.}

The FCC collaboration estimates that, due to an increase in statistics by five orders of magnitude with respect to LEP, it is likely that uncertainties be reduced by at least one order (in some cases even more than two orders) of magnitude~\cite{FCC:2018evy,FCC:2018byv}.
In what regards these future sensitivities for the observables under consideration, in this work we assume a more modest improvement, and conservatively take a reduction of the uncertainties by a factor 4. In the following discussion, the intervals corresponding to the future sensitivities are established 
by fixing the central value\footnote{Whenever the future sensitivity of FCC-ee is displayed, we consider the intervals centred on the current experimental central value (and thus in tension with the SM). Should future measurements exhibit better agreement with the SM expectation, the impact on our results can be evaluated by re-centring the FCC-ee lines around the SM value.  } of the future measurements to the current averages, and scaling their uncertainties accordingly (i.e., dividing the $1\sigma$ uncertainty by $4$).
This leads us to 
\begin{eqnarray}
    R_{\tau e}(Z\to \ell\ell)|_\mathrm{FCC} &\simeq& 1.0020\pm 0.0008 \\
    R_{\tau\mu}(Z\to \ell\ell)|_\mathrm{FCC} &\simeq& 1.0010\pm 0.0007 \\
    \Gamma(Z\to\tau^+\tau^-)|_\mathrm{FCC} &\simeq& 84.08\pm0.06\:\mathrm{MeV}\\
    \Gamma(Z\to\mathrm{inv.})|_\mathrm{FCC} &\simeq& 499.0\pm0.38\:\mathrm{MeV}\,,
\end{eqnarray}
in which we have also assumed that the uncertainty of the indirect determination of the invisible $Z$ width will also improve by a factor 4.

\bigskip
Before we proceed with the discussion of the observables, let us briefly notice that throughout the investigated parameter space (as indirectly accessed via the parametrisation of Eqs.~(\ref{eq:parametrisation:muX1}, \ref{eq:parametrisation:muX1:mD})), the generated Yukawa couplings typically range from tiny values to the imposed perturbativity cutoff.

Likewise, and concerning the (diagonal) entries of the matrix $\eta$ (cf. Eq.~(\ref{eq:defPMNSeta})), one finds that the maximal obtained values are $\eta_{ee}^\text{max}\approx 0.004$, 
$\eta_{\mu\mu}^\text{max} \approx 0.003$,
$\eta_{\tau\tau}^\text{max} \approx 0.01$
while in agreement with all the imposed constraints\footnote{Notice that we are not doing a full fit including hadronic observables (other than those specifically mentioned), nor do we include bounds on the entries of the CKM quark mixing matrix. Should that be the case, $\eta_{\tau\tau}$ would be more constrained than the regimes here allowed (see, for example,~\cite{1605.08774,Blennow:2023mqx}). In the present study, our goal is not to infer bounds on $\eta_{\alpha\beta}$, but rather to compare the constraining power of distinct observables.}
 (following the above discussion, all off-diagonal entries 
$\eta_{\alpha \beta}$ are compatible with zero, 
$\eta^\text{max}_{\alpha \beta} \lesssim 10^{-19}$, for $\alpha \neq \beta$).

\subsection{Numerical results}
We begin the discussion of our results by considering several observables directly related to the violation of lepton flavour universality in $Z$ boson decays. We then briefly comment upon the prospects of the ISS(3,3) in what concerns LFUV in Higgs and $W$ decays. Upon presentation of the results, we will often rely on the (pseudo-) parameters $M_\mathrm{max}$ and $Y_\mathrm{max}$, which are respectively defined as the heaviest mass eigenstate and the largest entry in $Y_D$ (defined in Eq.~(\ref{eq:parametrisation:muX1:mD})).

\bigskip

\begin{figure}[h!]
    \centering
    \includegraphics[width=0.48\textwidth]{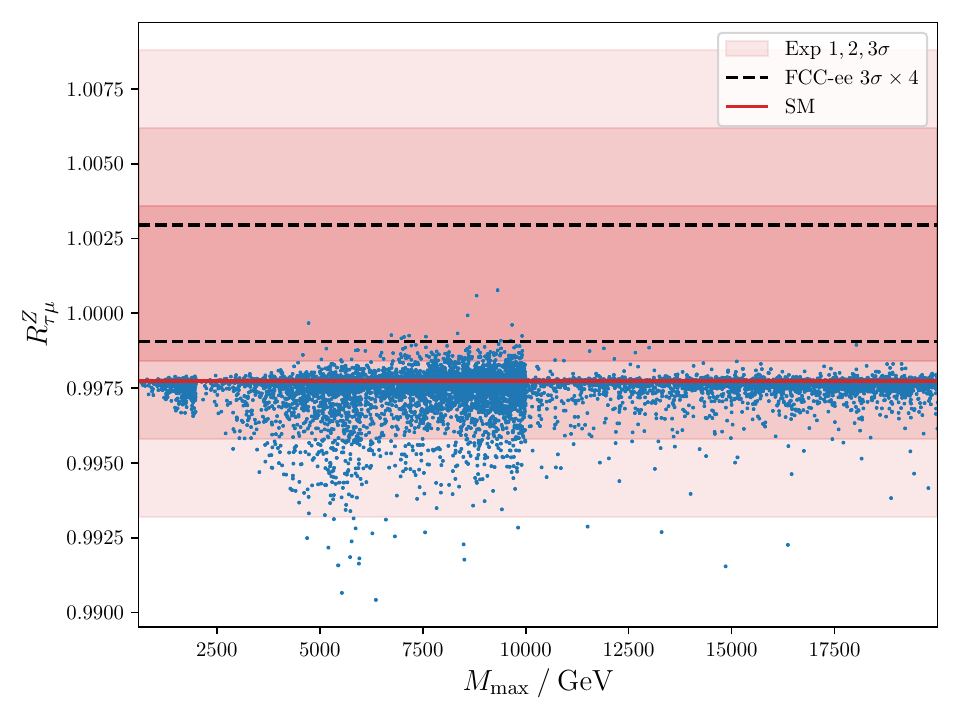}\includegraphics[width=0.48\textwidth]{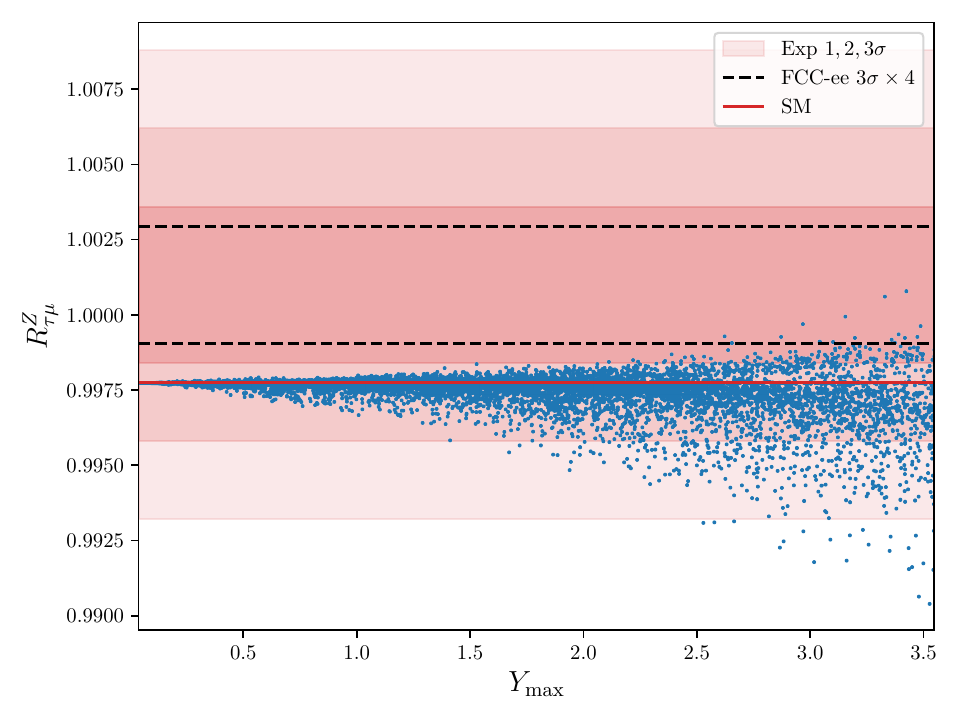}
    \caption{Contributions to $R^Z_{\tau \mu}$ in the ISS(3,3): on the left panel, as a function of the mass of the heaviest sterile state, $M_\text{max}$; on the right, dependency on the largest entry of the Yukawa couplings, $Y_\text{max}$. For both panels the horizontal coloured bands correspond to the experimental $1\sigma$-$3\sigma$ intervals (from darker to lighter), while
    the dashed horizontal lines denote the FCC-ee future sensitivity. The red solid line denotes the SM expectation.}
    \label{fig:RZ_tau_mu}
\end{figure}

\paragraph{LFUV in $Z$ decays}
Defined in Eq.~(\ref{eq:RZab}), the observable $R^Z_{\tau \mu}$ compares the decay widths of the $Z$ into pairs of tau-leptons and muons. In Fig.~\ref{fig:RZ_tau_mu} (left panel), we display the ISS(3,3) contributions to $R^Z_{\tau \mu}$ as a function of the mass of the heaviest sterile state, $M_\text{max}$. 
All points displayed comply with the distinct constraints referred to above. The coloured bands correspond to the experimental $1\sigma$, 
$2\sigma$ and $3\sigma$ intervals (from darker to lighter), and
the dashed horizontal lines denote the expected FCC-ee future sensitivity (assuming the current central experimental value, and a 4-fold increase in precision). 
Throughout the investigated parameter space, and for all masses considered, most of the contributions saturate around the SM expectation ($R^Z_{\tau \mu}|_\text{SM} \sim 0.9975$), itself already revealing a slight tension with the experimentally determined ratio. 
However, one also encounters values of $R^Z_{\tau \mu}$ considerably smaller: this occurs typically for a heavy spectrum around 6~TeV, and such a behaviour is associated with regimes for which the entries (at least one) of the Yukawa couplings are very large, close to its maximal allowed value. This can be confirmed from inspection of the right panel of Fig.~\ref{fig:RZ_tau_mu}, in which we present $R^Z_{\tau \mu}$ vs. $Y_\text{max}$. As visible, the most significant deviations with respect to the SM-like value occur for $Y_\text{max}$ close to the perturbativity bound.

For a lighter HNL spectrum (i.e. lower values of $M_\text{max}$), 
regimes leading to a strong violation of universality in $Z$ decays - in association with large values of the Yukawa couplings - would be already in conflict with other LFUV observables, such as 
$R^\ell_K$ or $R^\ell_\pi$. Moreover, such regimes would also typically lead to contributions to the invisible $Z$ width excluded by data (already at tree level). 
Larger values of $M_\text{max}$ are associated with SM-like predictions, as the mixings become smaller (a consequence of the intrinsic structure of the ISS). 
In association with the expected improvement in experimental sensitivity, FCC-ee is expected
to probe most of the ISS(3,3) contributions to the LFUV-sensitive observable $R^Z_{\tau \mu}$.

\begin{figure}[h!]
    \centering
    \includegraphics[width=0.52\textwidth]{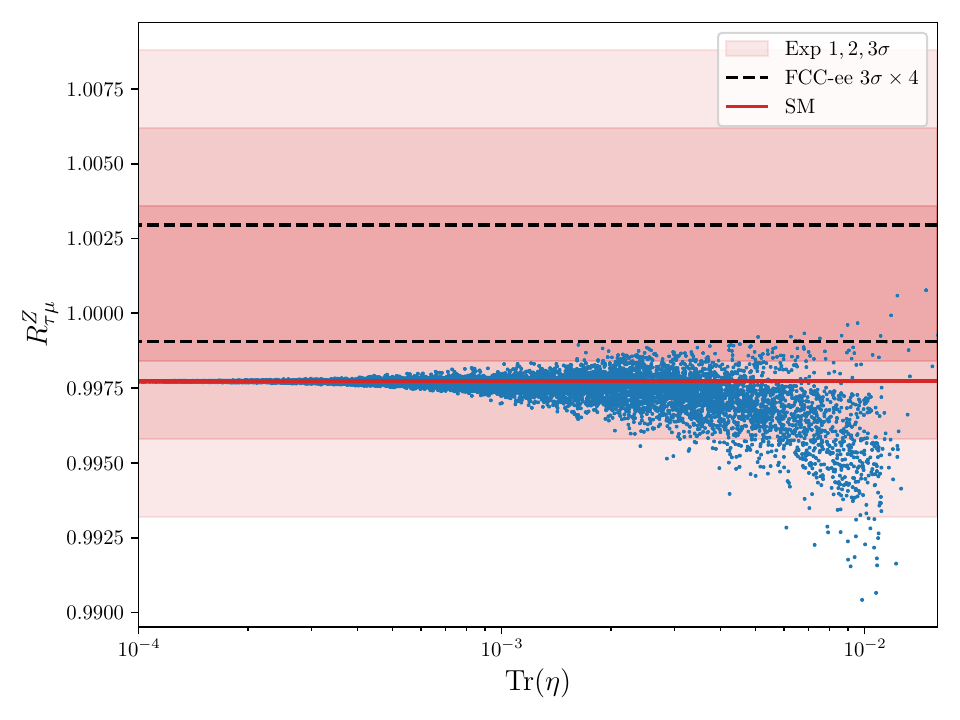}
    \caption{Contributions to $R^Z_{\tau \mu}$ in the ISS(3,3), as a function of Tr($\eta$).
    Line and colour code as in Fig.~\ref{fig:RZ_tau_mu}.}
    \label{fig:RZ_tau_mu:Treta}
\end{figure}

A complementary way to understand the possible deviations of the ISS(3,3) contributions to $R^Z_{\tau \mu}$ with respect to the SM is to consider the dependency of this observable  on $\eta$.
This is displayed in Fig.~\ref{fig:RZ_tau_mu:Treta}, in which we present $R^Z_{\tau \mu}$ as a function of the trace of the matrix $\eta$. 
Clearly, the larger the values of Tr($\eta$), the stronger the deviations, with the  entry $\eta_{\tau \tau}$ 
driving the deviations from zero. As can be also inferred from the comparison of Figs.~\ref{fig:RZ_tau_mu} and~\ref{fig:RZ_tau_mu:Treta}, the maximal values of $\eta$ occur for  regimes of heavy sterile masses between 4~TeV and 10~TeV, in association with large values of the Yukawa couplings. 

\begin{figure}[h!]
    \centering
    \includegraphics[width=0.48\textwidth]{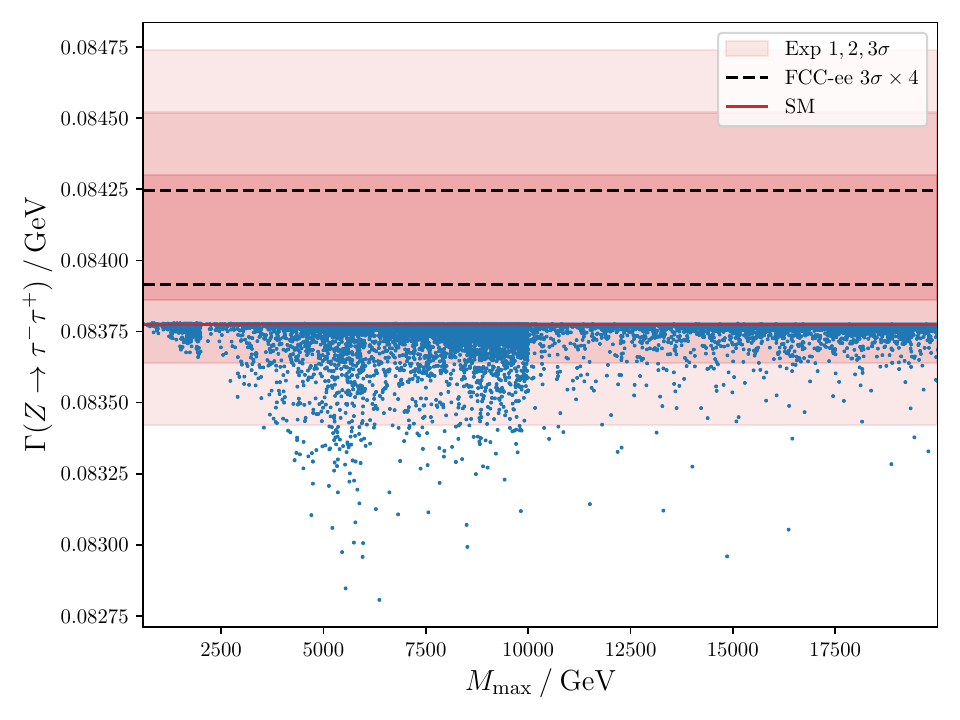}
    \includegraphics[width=0.48\textwidth]{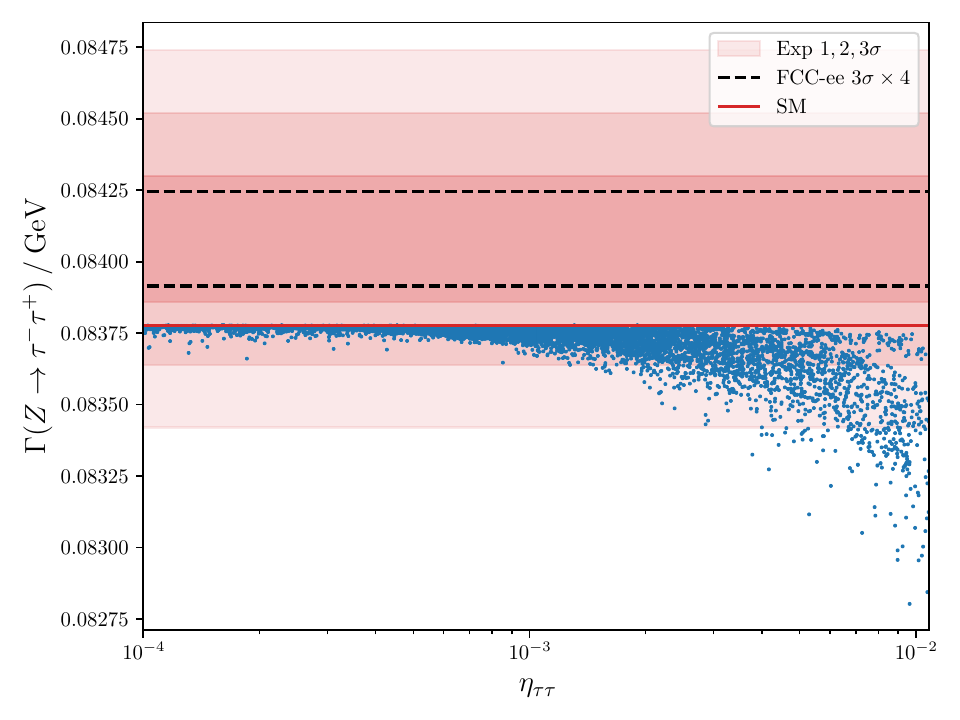}
    \caption{Contributions to the width of the $Z$ decay into a pair of taus, $\Gamma (Z \to \tau \tau)$ in the ISS(3,3): on the left panel, as a function of the mass of the heaviest sterile state, $M_\text{max}$; on the right, dependency on $\eta_{\tau\tau}$. 
Line and colour code as in Fig.~\ref{fig:RZ_tau_mu}.}
    \label{fig:Gamma_Z_tautau}
\end{figure}

\medskip
The above situation is again manifest  if one individually considers  the decays of  the $Z$ into a pair of taus, $\Gamma (Z \to \tau \tau)$.
The results are depicted in Fig.~\ref{fig:Gamma_Z_tautau}, 
in which the ISS contributions are displayed as a function of $M_\text{max}$ on the left panel, and for completeness versus the diagonal entry  $\eta_{\tau\tau}$ on the right one. 
In agreement with the results encountered for $R^Z_{\tau \mu}$, 
the right panel confirms that a strong departure from lepton flavour universality is indeed associated with a large $\eta_{\tau\tau}$ entry, as observed on the left panel, for masses between 4~TeV and 10~TeV.

\begin{figure}[h!]
    \centering
    \includegraphics[width=0.48\textwidth]{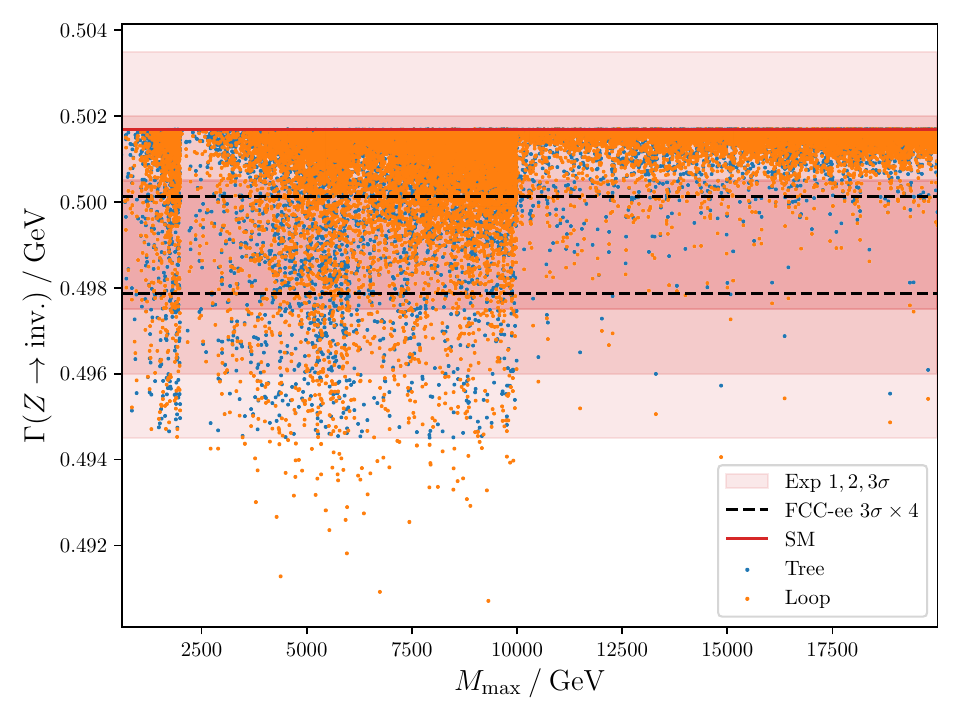}\includegraphics[width=0.48\textwidth]{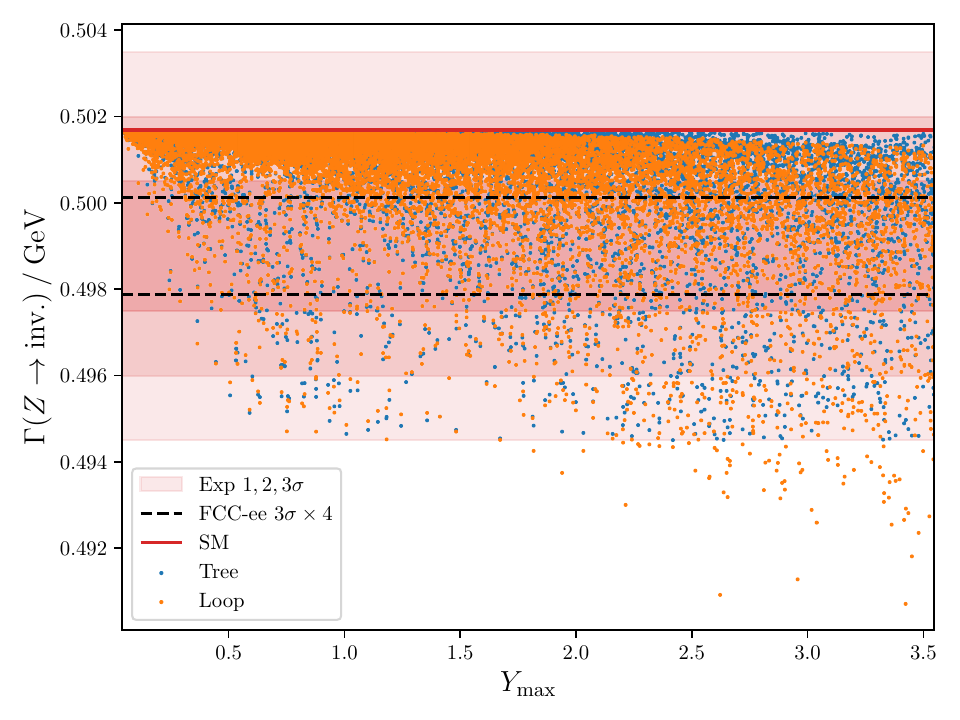}
    \caption{Contributions to the invisible $Z$ width, $\Gamma (Z \to \text{inv.})$ in the ISS(3,3): on the left panel, as a function of the mass of the heaviest sterile state, $M_\text{max}$; on the right, dependency on the largest entry of the Yukawa couplings, $Y_\text{max}$. For both panels, the tree-level (one-loop) contributions correspond to the blue (orange) points, with the SM prediction being denoted by a horizontal red line. As before, the horizontal coloured bands correspond to the experimental $1\sigma$-$3\sigma$ intervals (from darker to lighter), while
    the dashed horizontal lines denote the FCC-ee future sensitivity.}
    \label{fig:Gamma_Z_inv}
\end{figure}

\bigskip
We now turn our attention to the contributions of the ISS(3,3) to the invisible $Z$ width. On the left (right) panel of Fig.~\ref{fig:Gamma_Z_inv}, the new contributions are displayed as a function of $M_\text{max}$ ($Y_\text{max}$). We separately present the ISS(3,3) contributions obtained from the na\"ive tree-level estimation (blue points), and from the full one-loop computation (orange). 
As can be seen from the left panel of Fig.~\ref{fig:Gamma_Z_inv}, a first striking result concerns the 
relevance of considering the full one-loop computation upon evaluation of the new physics contributions to the invisible $Z$ width. The comparison of ``stacked" blue and orange points clearly reveals that the tree-level estimate is at the origin of much smaller deviations of  $\Gamma (Z \to \text{inv.})$ from the SM expectation. Notice that the difference between the tree-level estimate and the full one-loop computation of the width  can  easily lie around the few~MeV level, which is comparable - or even larger - than the current experimental precision.
In particular, one can observe that while relying on a tree-level calculation all regimes are within the $3\sigma$ interval, this is no longer the case when carrying out the full one-loop evaluation. 
As can also be seen, the expected future sensitivity of the FCC-ee 
will be instrumental in probing (and in potentially ruling out) important 
regions of the ISS(3,3) parameter space in association with the $\Gamma (Z \to \text{inv.})$ observable.

\begin{figure}[h!]
    \centering
    \includegraphics[width=0.48\textwidth]{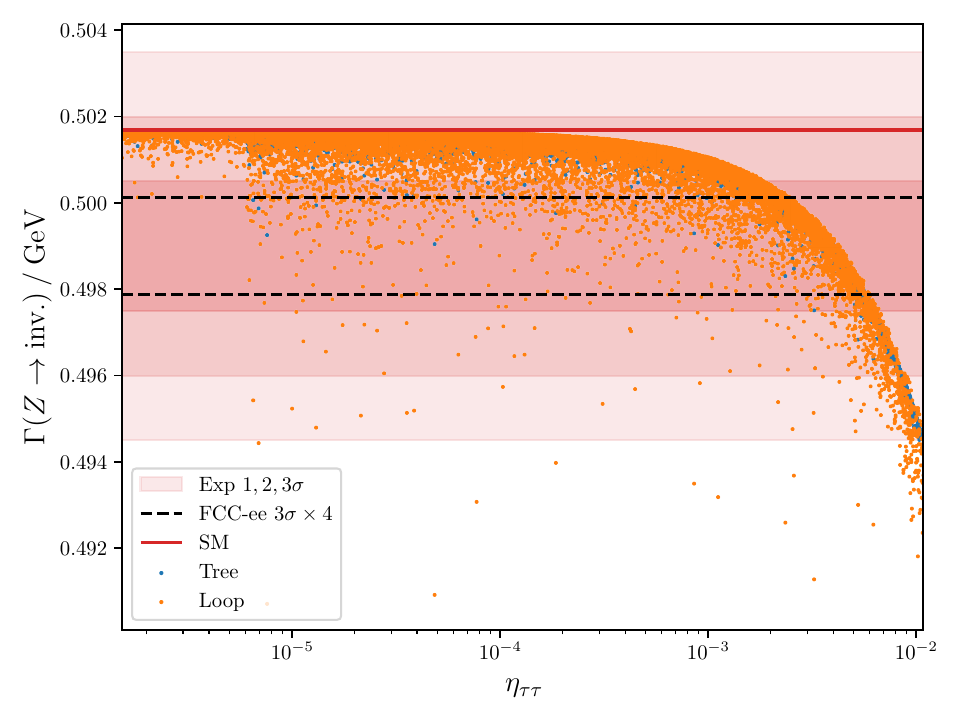}
    \includegraphics[width=0.48\textwidth]{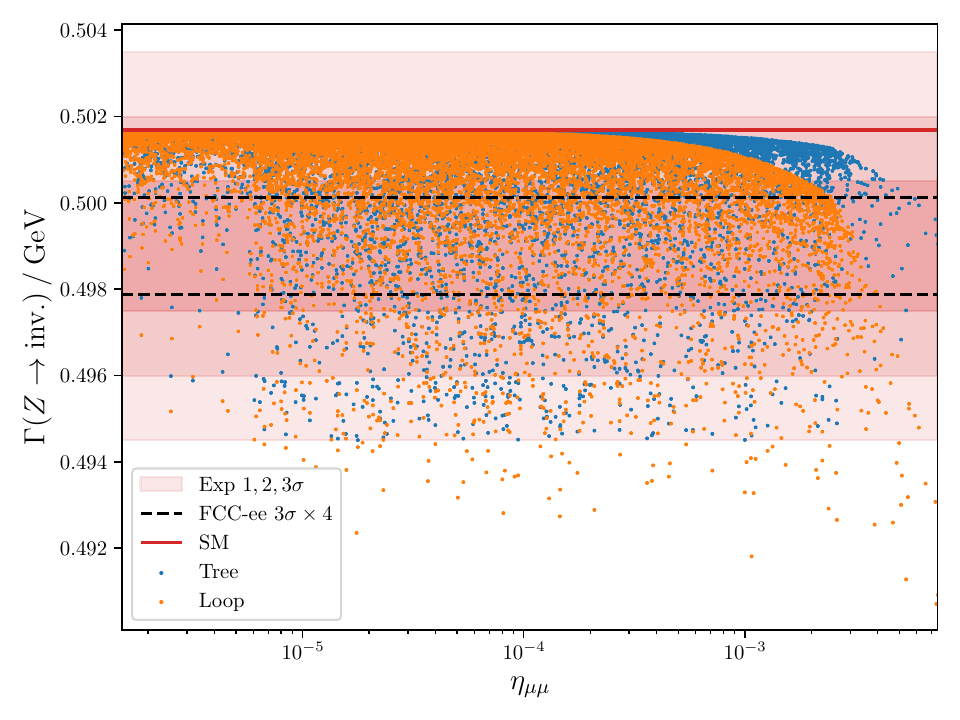}
    \caption{Contributions to the invisible $Z$ width, $\Gamma (Z \to \text{inv.})$ in the ISS(3,3), as a function of $\eta_{\tau\tau}$ ($\eta_{\mu\mu}$) on the left (right) panel. Line and colour code as in Fig.~\ref{fig:Gamma_Z_inv}.}
    \label{fig:Gamma_Z_inv:eta}
\end{figure}

Figure~\ref{fig:Gamma_Z_inv:eta} highlights the impact of the departure from unitarity of the PMNS in the new contributions to $\Gamma (Z \to \text{inv.})$. The left panel displays the very clear correlation between $\Gamma (Z \to \text{inv.})$ and $\eta_{\tau\tau}$, as expected from the tree-level expression of Eq.~(\ref{eq:InvZwidth:tree}), from which one had $\Gamma (Z \to \text{inv.}) \propto\left(
    3-(\text{Tr}(\eta) + 3 \eta_{\tau\tau}) 
    \right) $~\cite{Fernandez-Martinez:2016lgt}.
 The other diagonal entries - i.e. $\eta_{\mu\mu}$ and $\eta_{ee}$ both have a non-negligible impact regarding the observed differences between tree-level vs. one-loop contributions to $\Gamma (Z \to \text{inv.})$. We illustrate the role of $\eta_{\mu\mu}$ on the right panel
of  Fig.~\ref{fig:Gamma_Z_inv:eta}.
We emphasise that, contrary to the other observables here considered, for the decay $\Gamma(Z\to\mathrm{inv.})$ the HNL contributions at one-loop have a significant impact, even for much smaller entries in $\eta$.

\paragraph{LFUV in $H$ decays}
In view of the experimental prospects to observe and measure leptonic Higgs decays, here we only consider LFUV probes of the ISS(3,3) in $H \to \tau\tau $ and $H \to \mu\mu $ decays, studying the new contributions which we always normalise with respect to the SM width. 

\begin{figure}[h!]
    \centering
    \includegraphics[width=0.48\textwidth]{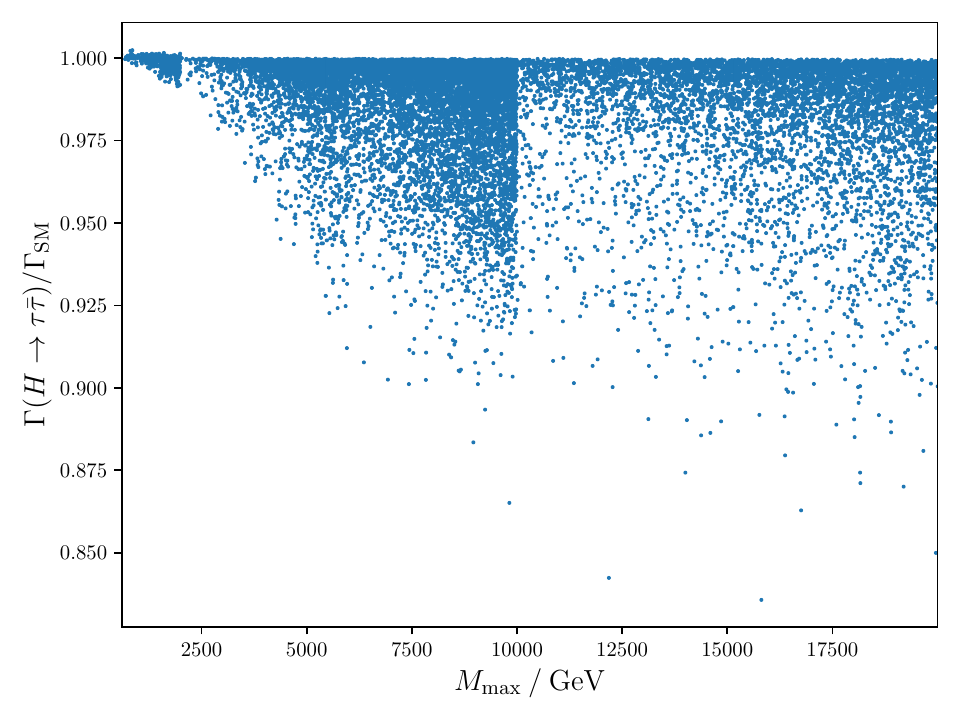}
    \caption{Prospects for the leptonic Higgs decays in the ISS(3,3), normalised to the SM prediction, $\Gamma (H \to \tau \tau)/\Gamma (H \to \tau\tau)|_\text{SM}$. }
    \label{fig:Gamma_H_tau_mu_divSM}
\end{figure}

\begin{figure}[h!]
    \centering
    \includegraphics[width=0.48\textwidth]{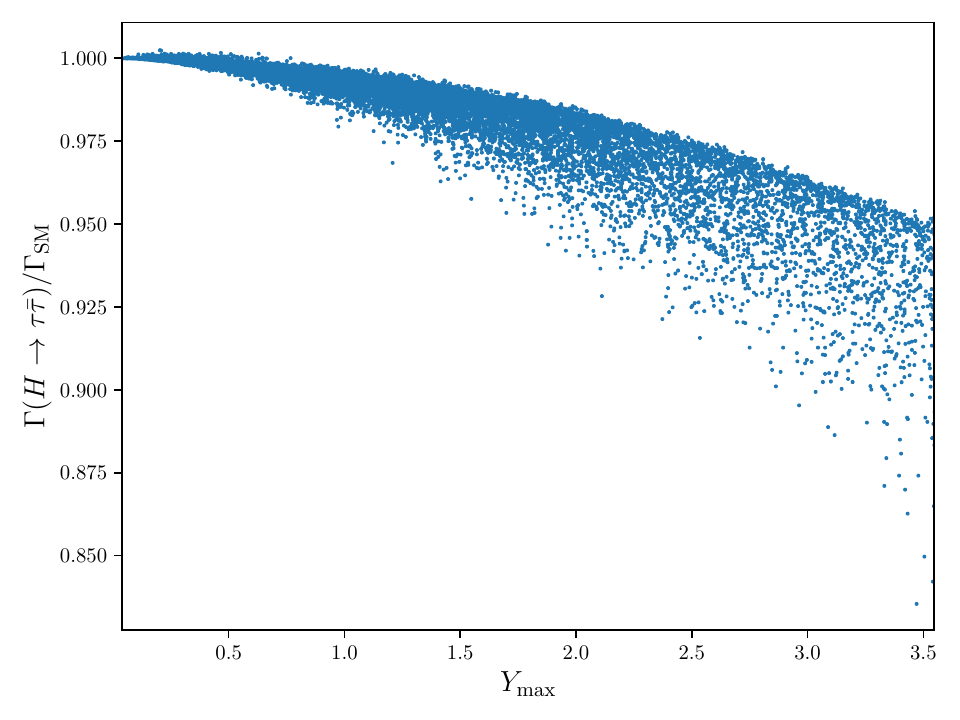}\includegraphics[width=0.48\textwidth]{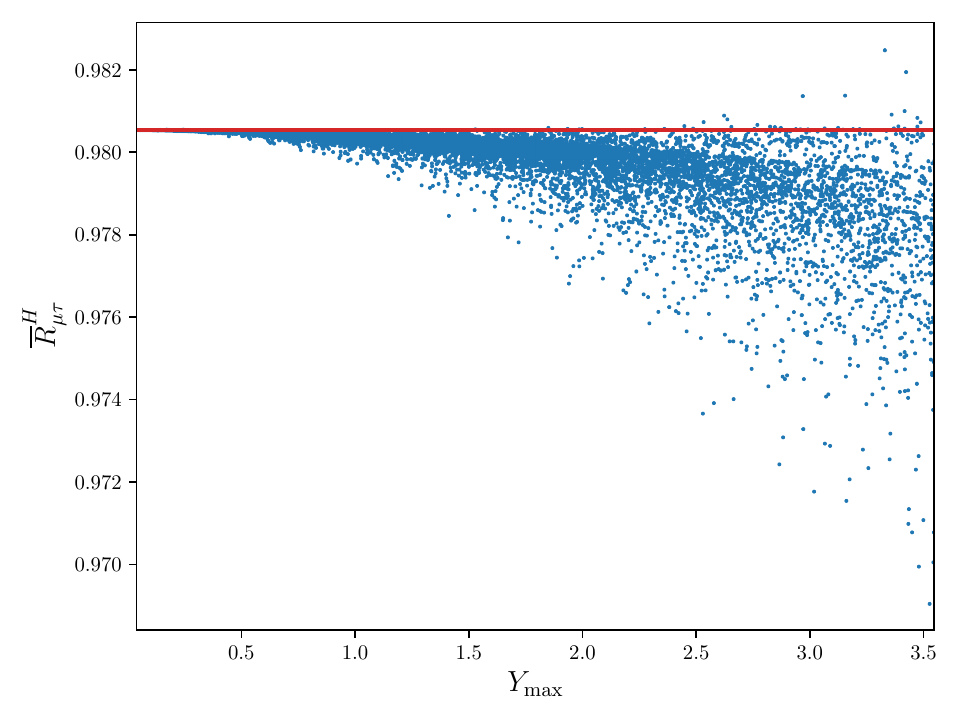}
    \caption{Prospects for the leptonic Higgs decays in the ISS(3,3): on the left, dependency of $\Gamma (H \to \tau \tau)/\Gamma (H \to \tau \tau)|_\text{SM}$ on $Y_\text{max}$; on the right, comparison of the rates to ditaus and dimuons (each normalised to the SM-expected value), also as a function of $Y_\text{max}$.
    On the right panel, the horizontal red line denotes the SM expectation for $\bar R^{H}_{\mu \tau}$.}
    \label{fig:Gamma_H_tau_mu_Ymax}
\end{figure}

The observable $\Gamma (H \to \tau \tau)/\Gamma (H \to \tau \tau)|_\text{SM}$, is displayed in Fig.~\ref{fig:Gamma_H_tau_mu_divSM}, as a function of the mass of the heaviest sterile state, $M_\text{max}$. 
Although not displayed here, we notice that $H \to \mu \mu $ exhibits the same behaviour as  $H\to \tau\tau$.
Due to the (effective) Higgs interactions with Majorana sterile states, the impact on the decay rates becomes more pronounced as the mass of the new mediators increases; this is visible in Fig.~\ref{fig:Gamma_H_tau_mu_divSM}, in which deviations $\sim 5\%$ in the (normalised) width are manifest for states with $M_\text{max} \simeq 5$~TeV. The (expected) driving role of the Yukawa couplings is also visible in Fig.~\ref{fig:Gamma_H_tau_mu_Ymax}: on its left panel one confirms that the largest deviations from the SM expectation are indeed associated with the regimes of large Yukawas. 
In the right panel of   
Fig.~\ref{fig:Gamma_H_tau_mu_Ymax} we consider the universality ratios, for which the contributions of the heavy ISS states are also clearly relevant. 
For convenience, here we considered the ``normalised'' universality ratio, defined as
\begin{equation}
    \bar R_{\mu\tau}^H \,= \,\frac{m_\tau^2}{m_\mu^2}\,R_{\mu\tau}^H\,,
\end{equation}
in which $R_{\mu\tau}^H$ had been defined in Eq.~\eqref{eq:RHalphabeta:def}.

\begin{figure}[h!]
    \centering
    \includegraphics[width=0.48\textwidth]{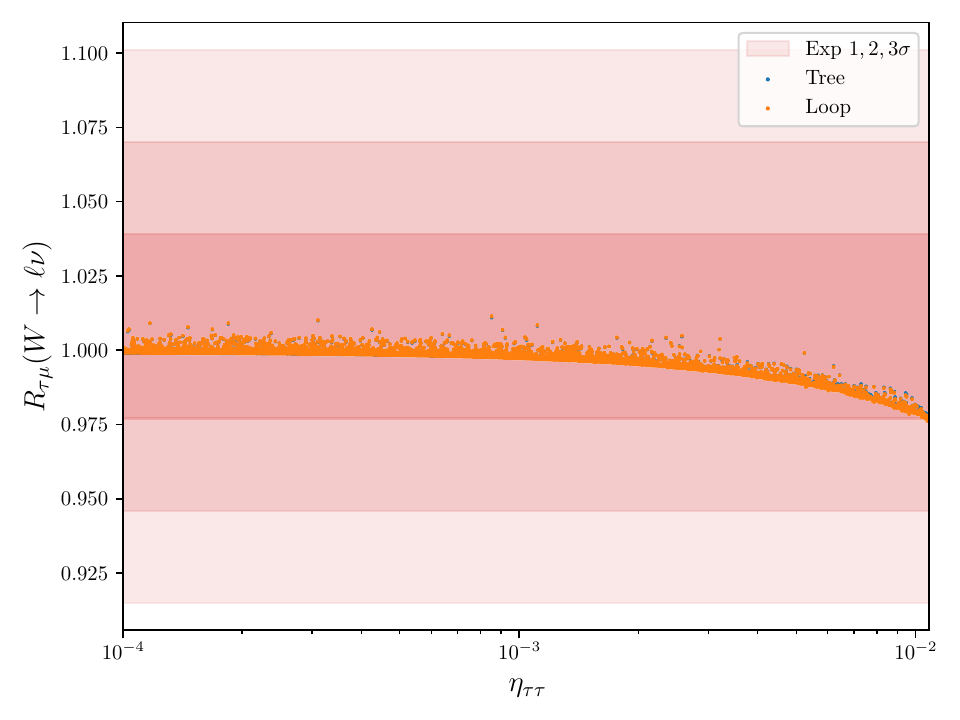}
    \caption{Prospects for LFUV in $W$  decays in the ISS(3,3): $R^W_{\tau\mu}$ as a function of $\eta_{\tau\tau}$. 
   As before, the tree-level (one-loop) contributions correspond to the blue (orange) points, and the horizontal coloured bands denote to the experimental $1\sigma$-$3\sigma$ intervals (from darker to lighter).}
    \label{fig:Wellnu}
\end{figure}

\paragraph{LFUV in $W \to \ell \nu$ decays}
In order to conclude the discussion, we briefly discuss the contributions of the sterile states concerning  charged current decays. In Fig.~\ref{fig:Wellnu}, we illustrate the ISS(3,3) impact on LFUV in leptonic $W$ decays, presenting $R^W_{\tau\mu}$ as a function of $\eta_{\tau\tau}$. As can be seen, there is a good agreement with the SM expectation (and with current experimental measurements), which in turn renders this LFUV-sensitive observable comparatively less powerful than others previously discussed here. 
While there is indeed a difference between the tree-level and one-loop predictions for the individual decay widths (typically around 0.7\%), the impact of taking into account the higher order corrections never exceeds 0.2\% when ratios are considered. The observed behaviour is also in good agreement with the (tree-level) expectation, as given by Eq.~(\ref{eq:RWalphabeta:analytical:eta}) - notice that the dependency on $\eta$ only becomes apparent for $\eta_{\tau\tau}\gtrsim 10^{-3}$.

\section{Probing power: LFUV vs. cLFV}\label{sec:ComplementarityProbing}
As mentioned throughout this work, in many models of neutrino mass generation with a strong phenomenological impact (as is the case of several low-scale realisations of the type I seesaw, and its variants), the most stringent constraints on the parameter space usually arise from cLFV observables, together with certain EWPO, as is the case of the invisible $Z$ width. 
In this final section, we discuss the interplay of the former with LFU violation probes in constraining the ISS(3,3) parameter space. Moreover, we highlight regimes for which (flavour conserving) LFUV and EWPO turn out to play the leading constraining roles.

\begin{figure}[h!]
    \centering
    \includegraphics[width=0.48\textwidth]{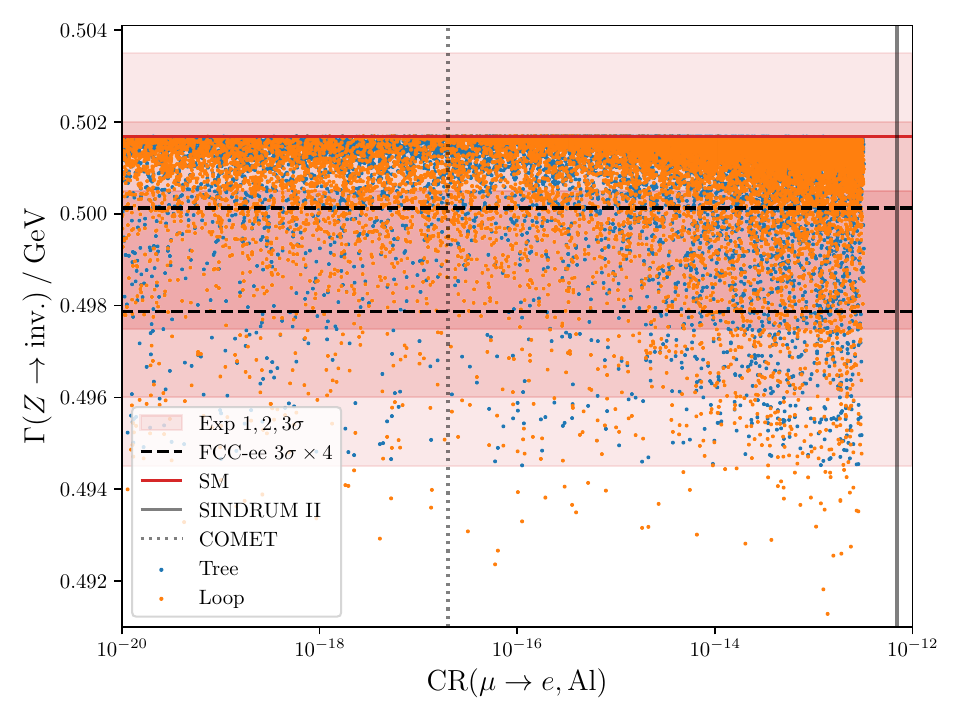}\includegraphics[width=0.48\textwidth]{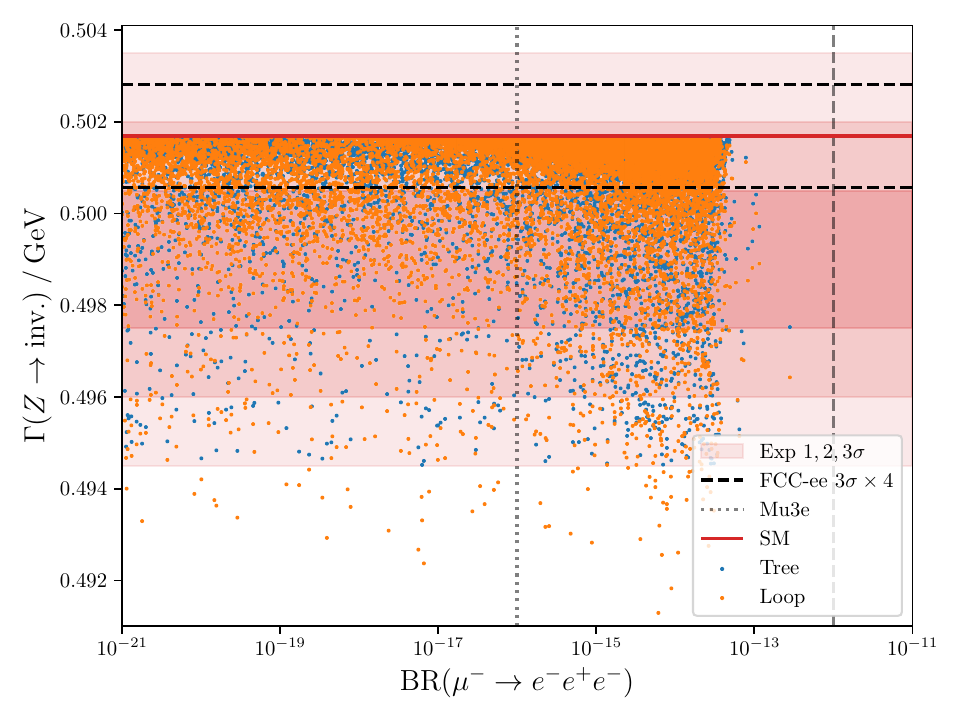}
    \caption{Complementarity of distinct observables in constraining the ISS(3,3) parameter space: $\Gamma (Z \to \text{inv.})$ vs. CR($\mu-e$, Al) (left panel) and vs. BR($\mu \to 3e$) (right panel). Colour code as in Fig.~\ref{fig:Gamma_Z_inv}, with the additional vertical lines now referring to current cLFV bounds and future sensitivities. }
    \label{fig:GammaZinv_cLFV}
\end{figure}

\bigskip
In Fig.~\ref{fig:GammaZinv_cLFV}, we compare the probing power of cLFV observables in the $\mu-e$ sector (usually responsible for some of the most stringent constraints in SM extensions via HNL) with that of invisible $Z$ decays. In particular, we consider the impact of the invisible $Z$ decays and that of 
CR($\mu-e$, Al)\footnote{Notice that the apparently artificial vertical cut in the right-hand side of the left panel of Fig.~\ref{fig:GammaZinv_cLFV} 
is due to imposing that all points be in agreement with the current best bound on atomic $\mu-e$ conversion, which has been obtained for Gold nuclei: 
CR($\mu-e$, Au)$\lesssim 7 
\times 10^{-13}$~\cite{Bertl:2006up}.} and BR($\mu \to 3e$). In both cases, it is clear that the future sensitivity to the invisible $Z$ decays offers the possibility to probe and explore regimes that are beyond the reach of COMET/Mu2e, and especially of Mu3e (see Table~\ref{tab:cLFV_lep}). Although - and as expected -  the ISS(3,3) gives important contributions to muon cLFV observables, an EW observable as $\Gamma (Z \to \text{inv.})$ allows probing regimes both associated with sizeable cLFV contributions or truly negligible ones - in fact as small as $10^{-20}$ (or below). It is also interesting to emphasise that the relevance of taking into account the loop corrections for $\Gamma (Z \to \text{inv.})$ holds for all explored regimes - be it those leading to approximate charged lepton flavour conservation, or those leading to cLFV signals within future reach.
(Although not shown here, the new contributions to $\mu-\tau$ cLFV observables are comparatively small, lying only marginally within future Belle II reach; for the latter case LFUV observables can be of relevance.)

\begin{figure}[h!]
    \centering
    \includegraphics[width=0.48\textwidth]{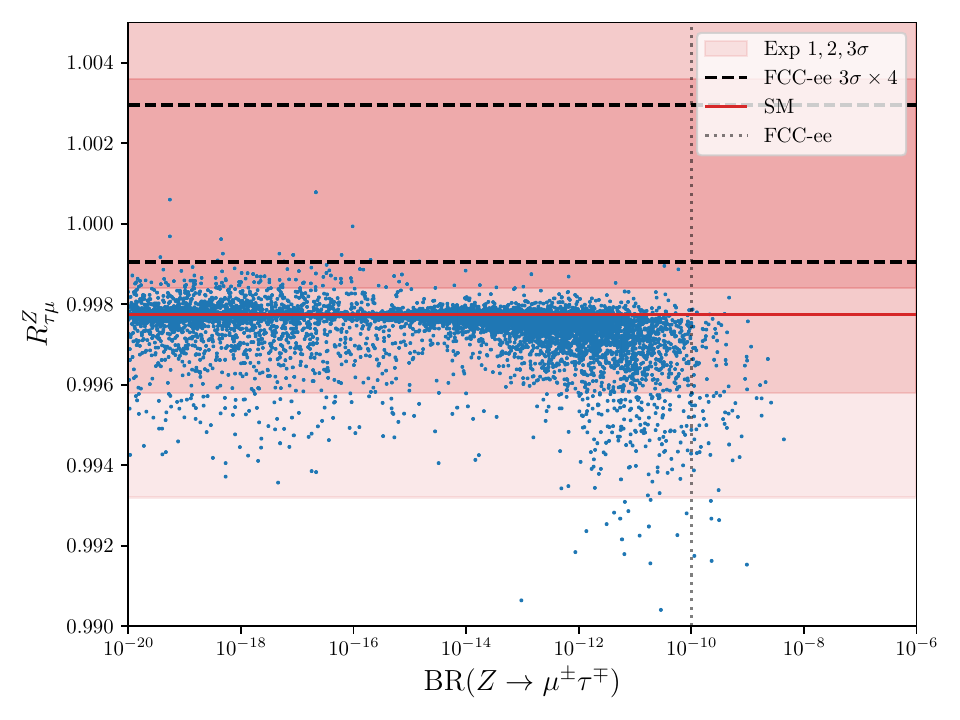}
    \caption{Complementarity of distinct observables in constraining the ISS(3,3) parameter space: LFUV $R^Z_{\tau\mu}$ vs. BR($Z \to \mu \tau$). 
Line and colour code as in Fig.~\ref{fig:RZ_tau_mu}, with the additional vertical line corresponding to the cLFV  $Z$ decay future sensitivity.}
    \label{fig:RZmutau_cLFVZmutau}
\end{figure}

It is also interesting to compare LFUV and cLFV observables directly associated with (visible) leptonic $Z$ decays. This is done in Fig.~\ref{fig:RZmutau_cLFVZmutau}, in which we present 
the ISS(3,3) contributions in the plane spanned by the LFUV probe $R^Z_{\tau\mu}$ and the cLFV  decay rate $Z \to \mu \tau$ (see Table~\ref{tab:cLFV_lep} for current bounds and future sensitivities for the cLFV decays). Interestingly, LFUV in $Z$ decays allows  probing important regions of the parameter space, especially in comparison to $Z \to \mu \tau$ (both at a future FCC-ee).

\begin{figure}[h!]
    \centering
    \includegraphics[width=0.48\textwidth]{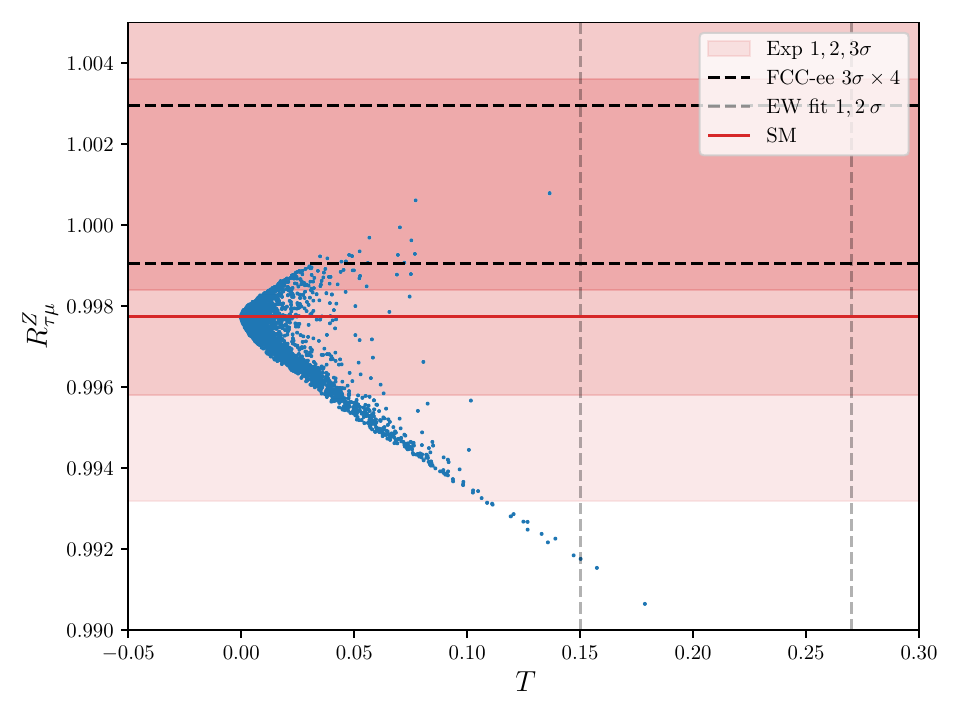}\includegraphics[width=0.48\textwidth]{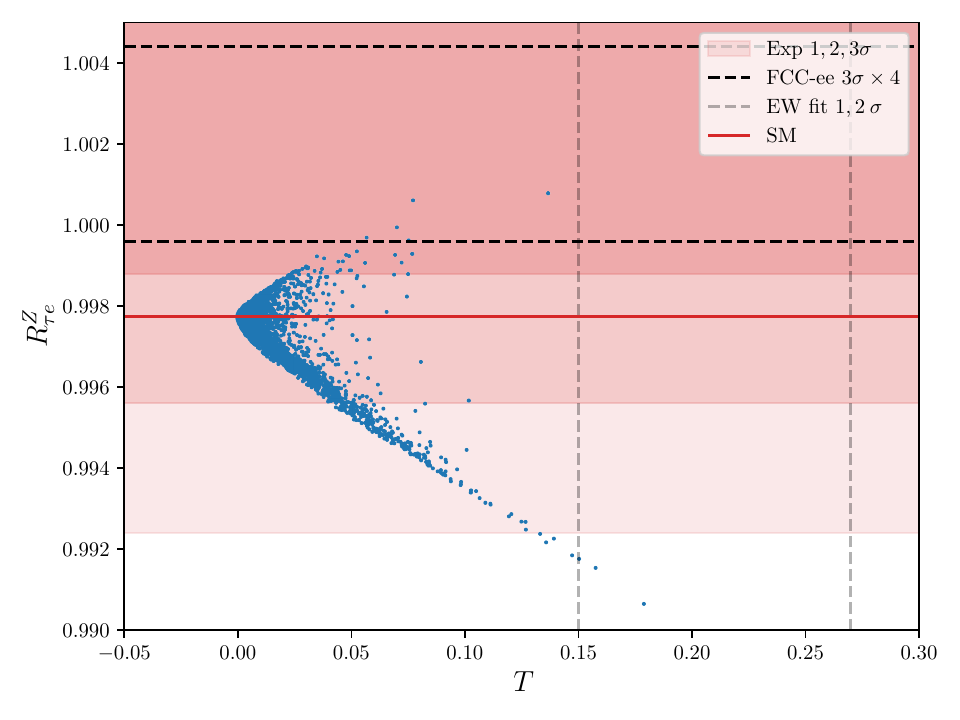}
  \includegraphics[width=0.48\textwidth]{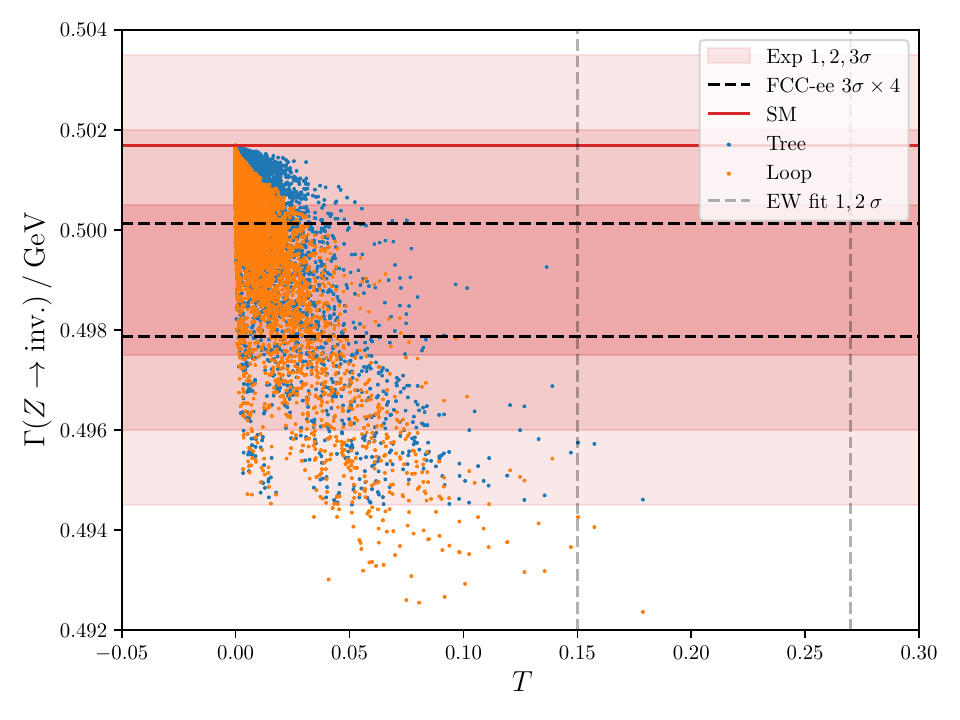}
    \caption{Complementarity of distinct observables in constraining the ISS(3,3) parameter space: on the upper row, LFUV $R^Z_{\tau\mu}$ (left) and $R^Z_{\tau e}$ (right) vs. the oblique $T$ parameter. Line and colour code as in Fig.~\ref{fig:RZ_tau_mu} (the additional vertical dashed lines denote the current $1\sigma$ and $2\sigma$ bounds from the EW fit~\cite{ParticleDataGroup:2022pth} for $T$).
    On the lower row, $\Gamma(Z \to \text{inv.})$, also as a function of $T$. Line and colour code as in Fig.~\ref{fig:Gamma_Z_inv}.    }
    \label{fig:RZ_Zinv_T}
\end{figure}

Finally, in Fig.~\ref{fig:RZ_Zinv_T}, we present a combination of several LFUV $Z$-observables, all depicted versus the expected impact for the oblique $T$ parameter. 
By themselves, and as of today, the LFUV bounds are more constraining than the (indirect) bounds on the oblique parameters arising from the EW fit~\cite{ParticleDataGroup:2022pth}.
Although we do not include such limits in Fig.~\ref{fig:RZ_Zinv_T}, 
the future combined sensitivity of HL-LHC together with FCC-ee is expected to constrain the $T$ parameter to the permille level~\cite{deBlas:2019rxi}.

In summary, the results here collected offer numerous examples of the powerful role of LFUV observables (together with the invisible $Z$-decay width) in constraining important regimes of the ISS(3,3), either for mostly (charged lepton) flavour conserving regimes, or for those associated with cLFV within future reach.

\section{Conclusions}
In this work we have considered the impact of heavy neutral leptons on $Z$, $W$ and Higgs bosons decays, highlighting their role in what concerns lepton flavour universality violation, both at tree-level and at higher order. We have focused on UV-complete extensions of the SM model including new, heavy sterile states, and we have also investigated the prospects for EW precision observables such as the invisible width of the $Z$ boson.  

We have provided detailed expressions for the different decay widths. These are valid for - and can easily be 
adapted and applied to - any phenomenological study of UV-complete SM extensions by HNL (in association with an enlarged leptonic mixing matrix), in particular to low-scale type I seesaw models, and variations thereof, as is the case of the linear seesaw and inverse seesaw. For a numerical illustration, we have focused on a realisation of the ISS via three right-handed neutrinos and three additional sterile states, the so-called ISS(3,3).

As extensively studied in the literature, cLFV observables (be it current bounds or future projected sensitivities) are usually at the origin of the most stringent constraints for such new physics models. However, and as substantiated by our study, this need not always be the case. 
Not only in regimes associated with negligible contributions to flavour violation, but also for cLFV within future experimental reach, LFU violation in vector and scalar boson decays, as well as the invisible width of the $Z$ can play an important - if not the leading - 
role.

In our study, we have shown that in this class of SM extension via HNL, sizeable deviations from universality expectations can be found, especially in association with a significant departure from unitarity of the PMNS matrix (driven by regimes of large Yukawa couplings) in association with a moderately heavy sterile spectrum, between 4~TeV and 10~TeV. The tensions are well evidenced via several LFUV probes, in particular $R^Z_{\tau \mu}$, as well as in Higgs decays, with departures from the SM as large as a few percent in the latter (possibly reaching, in certain cases, more than~10\%).
The invisible $Z$ width is one of the so-called EWPO that is particularly sensitive to the presence of extensions of the SM neutral lepton sector, and thus to states potentially playing a role in the mechanism of neutrino mass generation. As discussed here, the contributions of the heavy sterile states can induce deviations from the SM of around a few~MeV, even for a small departure from unitarity of the would-be PMNS matrix.

As aforementioned, cLFV observables remain privileged probes for low-scale seesaw realisations relying on additional heavy sterile fermions, in particular concerning flavour violation in the $\mu-e$ sector, as a consequence of the  excellent future experimental prospects. Nevertheless, and as confirmed here, LFUV observables and $\Gamma(Z \to \text{inv.})$  can be powerful complementary probes to $\mu-e$ cLFV: for  example,  $\Gamma(Z \to \text{inv.})$ can complement searches for regimes lying beyond Mu3e sensitivity.
The role of LFUV probes is all the more relevant for $\mu-\tau$ sector observables, whose improvements in what regards future experimental sensitivity are expected to be somewhat more modest (typically one to two orders of magnitude): as shown by our study, the future FCC-ee sensitivity to $R^Z_{\tau \mu}$ allows probing regimes, in which BR($Z\to\mu\tau$) is beyond the FCC-ee reach.
  
In comparison with other EWPO, both $R^Z_{\tau \mu (e)}$ and 
$\Gamma(Z \to \text{inv.})$ clearly offer an important complementarity concerning   (indirect) searches
for new physics in the lepton sector, as they would allow for a clearer interpretation of  a possible future signal in the oblique parameters.
In the context of low-scale seesaw models, these LFUV observables can  become even more constraining than the oblique parameters and cLFV bounds alone, notably in the absence of a signal in the latter two.
All the above strongly suggests that regimes associated with large Yukawas, even if not necessarily leading to sizeable cLFV signals (or excessive deviations concerning EWPO) can be further constrained by LFUV-probing observables, as well as invisible $Z$ decays. 

Another crucial point that emerges from this study is that taking into account higher order contributions is also of paramount importance, as the one-loop corrections can lead to a shift in the predicted widths that can be comparable to - or even larger than - the (current) experimental uncertainty. This is all the more important in view of the expected breakthroughs in precision with the onset of FCC-ee.

\bigskip
As visible in the plots illustrating our most important results, and as argued throughout the discussion, the impact of the LFUV and EWPO observables is captive to the evolution of the SM predictions vs. experimental measurement. The future precision (especially with the potential advent of the FCC-ee) will certainly improve dramatically; however, the issue lies in whether or not the SM (theory) prediction and central experimental value will converge, or whether the tensions between them do remain. In the former case, several of the here investigated LFUV observables, together with the 
invisible $Z$ width, will efficiently probe extensive regions of the ISS 
parameter space (especially in association with regimes at the source of little to no cLFV contributions); in the latter situation, low-scale models of neutrino mass generation, as the one here considered, are excellent candidates to account for the (persistent) tensions. 

Despite its intrinsic flavour structure (associated with providing a natural, and minimal set-up for neutrino mass generation at comparatively low scales), the conclusions here drawn in association with the ISS(3,3) realisation can be potentially generalised to other mechanisms of neutrino mass generation, in which the SM is extended via HNLs. Lepton flavour universality probes, as well as the invisible $Z$ width (taking into account higher order contributions) are of paramount importance in constraining this class of SM extensions.

\section*{Acknowledgements}
This project has received support from the European Union's Horizon 2020 research and innovation programme under the Marie Sk\l{}odowska-Curie grant agreement No.~860881 (HIDDeN network) and from the IN2P3 (CNRS) Master Project, ``Flavour probes: lepton sector and beyond'' (16-PH-169).
JK is supported by the Slovenian Research Agency under the research grants N1-0253 and in part by J1-4389. JK is grateful to Miha Nemevšek and Svjetlana Fajfer for many useful discussions.

\appendix

\section{Details on the renormalisation procedure}
\label{app:renormalisation}
The renormalisation constants necessary for the counterterm Lagrangian are fixed by the renormalisation conditions; in the on-shell scheme here followed, the latter are formulated for on-mass-shell external fields.
Following~\cite{Denner:1991kt}, these are fixed using the one-particle irreducible two-point functions, which we summarise in the following.

Before we present the structure of the renormalisation constants and the explicit results of the two-point functions, we briefly describe our input parameter scheme.
We choose the following independent set of parameters (in addition to the fermion masses) fixed by experimental measurements~\cite{ParticleDataGroup:2022pth}:
\begin{eqnarray}
    G_\mu &=& (1.1663787 \pm 0.0000006)\times 10^{-5}\,,\\
    M_W &=& (80.379 \pm 0.012)\:\mathrm{GeV}\,,\\
    M_Z &=& (91.1876 \pm 0.0021)\:\mathrm{GeV}\,,\\
    M_H &=& (125.1 \pm 0.14)\:\mathrm{GeV}\,.
\end{eqnarray}
This choice of input parameters follows the recommendation of~\cite{Brivio:2021yjb}, and offers the advantage of being independent of light fermion contributions in the photon self-energy function at zero momentum, thus also omitting contributions proportional to the hadronic vacuum polarisation (which we will discuss in the following)\footnote{For a recent discussion concerning the impact of different input schemes for NLO corrections to electroweak observables (in the context of SMEFT predictions) for the FCC-ee precision era, see e.g.~\cite{2305.03763}.}.

\subsection{Renormalisation constants}
Before we discuss the explicit expressions of the renormalisation constants necessary for the renormalisation of masses, fermion mixings and fields, we discuss the ``derived'' renormalisation constants for the weak mixing angle and the electric charge.
Due to the on-shell definition of the weak mixing angle (see Eq.~\eqref{eqn:sw}), which holds to all orders in perturbation theory, a counterterm derived from the mass renormalisation constants of the $W$- and $Z$-boson masses must be introduced, and is given by
\begin{eqnarray}
    \dfrac{\delta s_w}{s_w}&=& -\dfrac{c_w^2}{s_w^2}\dfrac{\delta c_w}{c_w} = -\dfrac{1}{2}\dfrac{c_w^2}{s_w^2}\,\mathrm{Re}\left(\dfrac{\delta M_W^2}{M_W^2} - \dfrac{\delta M_Z^2}{M_Z^2} \right) \nonumber \\
    &=& -\dfrac{1}{2}\dfrac{c_w^2}{s_w^2}\,\mathrm{Re}\left(\dfrac{\Sigma_T^W(M_W^2)}{M_W^2} - \dfrac{\Sigma_T^{ZZ}(M_Z^2)}{M_Z^2}\right)\,,
\end{eqnarray}
where in the second line we have already inserted the expressions for the mass renormalisation constants which will be subsequently discussed and presented.

The renormalisation constant for the electric charge can be derived from the three-point function correcting the photon vertex.
However, by virtue of the Ward-identity, it can also be expressed via self-energies of the photon and of the photon-$Z$-mixing term~\cite{Denner:1991kt}, resulting in
\begin{eqnarray}
    \delta Z_e &=& \dfrac{1}{2}\dfrac{\partial \Sigma^{AA}_T(k^2)}{\partial k^2}\bigg|_{k^2=0} - \dfrac{s_w}{c_w} \dfrac{\Sigma_T^{AZ}(0)}{M_Z^2}\,.
\end{eqnarray}
In the first term of $\delta Z_e$ one has,  in addition to the bosonic contributions to the photon self-energy, also the contribution of the fermion loops.
Special care must be devoted to the light quark contributions at vanishing $k^2$ (the so-called hadronic vacuum polarisation), which can only be calculated either via lattice field theory methods or via data-driven approaches  from $e^+e^-$ scattering data (due to the optical theorem).
However, in our input scheme, $\alpha_e$ and therefore $e$ is a derived quantity from $G_\mu$ and the $W$- and $Z$-boson masses.
At the tree-level the relation between the latter  quantities is given by
\begin{equation}
    \alpha_e^{G_\mu}\,= \,\dfrac{\sqrt{2}\,G_\mu\, M_W^2}{\pi}\left(1 - \dfrac{M_W^2}{M_Z^2}\right)\,.
\end{equation}
The determination of $G_\mu$ from the muon lifetime must be also ``loop-corrected'' (see e.g.~\cite{1912.06823} and references therein).
The EW corrections at next-to-leading order (NLO) to the Michel decay of the muon are quantified in $\Delta r^{(1)}$
\begin{eqnarray}
    \Delta r^{(1)} = \dfrac{\partial \Sigma^{AA}_T(k^2)}{\partial k^2}\bigg|_{k^2=0} + 2 \dfrac{\delta s_w}{s_w}  + \dfrac{\Sigma^W_T(0) - \Sigma^W_T(M_W^2)}{M_W^2} + 2\dfrac{c_w}{s_w}\dfrac{\Sigma^{AZ}_T(0)}{M_Z^2} + \dfrac{\alpha_e(0)}{4\pi s_w^2}\left(6 + \dfrac{7 - 4 s_w^2}{2 s_w^2}\log c_W^2\right),
\end{eqnarray}
in which $\alpha_e(0)^{-1} = 137.035999180(10)$~\cite{ParticleDataGroup:2022pth} is the average of low-energy determinations at $k^2 = 0$, and the photon-$Z$-mixing two-point function is taken from the SM calculation (see e.g.~\cite{Denner:1991kt} for the relevant  expressions).
The modified charge renormalisation constant is then given by~\cite{1912.06823}
\begin{equation}
    \delta Z_e = \dfrac{1}{2}\dfrac{\partial \Sigma^{AA}_T(k^2)}{\partial k^2}\bigg|_{k^2=0} - \dfrac{s_w}{c_w} \dfrac{\Sigma_T^{AZ}(0)}{M_Z^2} - \dfrac{1}{2} \Delta r^{(1)}\,.
\end{equation}
As clear from the above, there is an exact cancellation of  the contributions of the photon self-energy at vanishing momentum transfer.
Furthermore, the potentially sizeable finite contributions from $\delta s_w/s_w$ in weak NLO corrections also cancel out at the amplitude level.

\subsection{Boson self energies}
\begin{figure}[h!]
    \centering
        \begin{subfigure}[b]{0.32\textwidth}
    \centering
   \raisebox{0.7mm}{\begin{tikzpicture}
    \begin{feynman}
    \vertex (a) at (0,0) {\(W^-\)};
    \vertex (b) at (1.2,0);
    \vertex (c) at (2.2,0);
    \vertex (d) at (3.4,0) {\(W^-\)};
    \diagram* {
    (a) -- [boson] (b),
    (b) -- [fermion, half right, edge label'=\(\vphantom{\frac{1}{2}}\ell_\alpha\)] (c),
    (b) -- [anti fermion, half left, edge label=\(n_i\)] (c),
    (c) -- [boson] (d),
    };
    \end{feynman}
    \end{tikzpicture}}
            \label{}
    \end{subfigure}
    \hfill
        \begin{subfigure}[b]{0.32\textwidth}
    \centering
   \raisebox{0.7mm}{\begin{tikzpicture}
    \begin{feynman}
    \vertex (a) at (0,0) {\(Z\)};
    \vertex (b) at (1.2,0);
    \vertex (c) at (2.2,0);
    \vertex (d) at (3.4,0) {\(Z\)};
    \diagram* {
    (a) -- [boson] (b),
    (b) -- [fermion, half right, edge label'=\(\vphantom{\frac{1}{2}}n_i\)] (c),
    (b) -- [anti fermion, half left, edge label=\(n_j\)] (c),
    (c) -- [boson] (d),
    };
    \end{feynman}
    \end{tikzpicture}}
            \label{}
    \end{subfigure}
    \hfill
        \begin{subfigure}[b]{0.32\textwidth}
    \centering
   \raisebox{0.7mm}{\begin{tikzpicture}
    \begin{feynman}
    \vertex (a) at (0,0) {\(H\)};
    \vertex (b) at (1.2,0);
    \vertex (c) at (2.2,0);
    \vertex (d) at (3.4,0) {\(H\)};
    \diagram* {
    (a) -- [scalar] (b),
    (b) -- [fermion, half right, edge label'=\(\vphantom{\frac{1}{2}}n_i\)] (c),
    (b) -- [anti fermion, half left, edge label=\(n_j\)] (c),
    (c) -- [scalar] (d),
    };
    \end{feynman}
    \end{tikzpicture}}
            \label{}
    \end{subfigure}
    \caption{Feynman diagrams of the new contributions to the $W$, $Z$ and Higgs boson self-energies, in unitary gauge.}
    \label{fig:BosonSelfenergies:UG}
\end{figure}
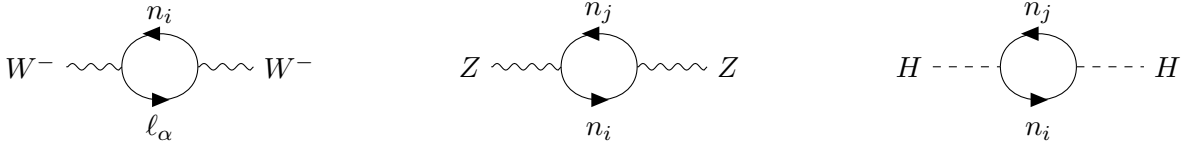

In Fig.~\ref{fig:BosonSelfenergies:UG} we schematically 
present the 
neutral lepton contributions to the boson self-energies.

The unrenormalised boson self-energies are defined by
\begin{eqnarray}
 \raisebox{-6mm}{
 \begin{tikzpicture}
    \begin{feynman}
    \vertex[blob] (a) at (1.5 ,0) {\contour{black}{ }};
    \vertex (f1) at (0,0) {\(a,\,\mu\)};
    \vertex (f2) at (3,0){\(b,\,\nu\)};
    \diagram* {
    (f1) -- [boson, edge label'=\(\vphantom{\frac{1}{2}}k\)] (a) --  [boson] (f2),
};
    \end{feynman}
\end{tikzpicture}} 
\,=\, i \Sigma_{\mu\nu}^{ab}(k) \,=\, -i \left[ g_{\mu\nu} - \dfrac{k_\mu k_\nu}{k^2} \right] \Sigma_T^{ab}(k^2) - i \dfrac{k_\mu k_\nu}{k^2} \Sigma_L^{ab}(k^2)\,,
\end{eqnarray}
where $a,b = A,Z, W$.
The NP contributions to the transverse part of the boson self-energies  are given by (we integrate in $D=4-2\varepsilon$ dimensions)
\begin{eqnarray}
    \Sigma^W_T (k^2) = - \dfrac{g_w^2}{16 \pi^2} \sum_{\ell, n} \, \mathcal{U}_{\ell n}\, \mathcal{U}_{\ell n}^* \,\left\{k^2 (B_1(k^2,m_n^2,m_\ell^2)+B_{11}(k^2,m_n^2,m_\ell^2)) + (D-2)B_{00}(k^2,m_n^2,m_\ell^2) \right\},
\end{eqnarray}
\begin{eqnarray}
    \Sigma^{ZZ}_T (k^2) &=& -\dfrac{g_w^2}{64 \pi^2 c_w^2} \sum_{i,j} \, 
    \big\{ m_i m_j (C_{ij}^2 + C_{ij}^{* 2})B_0(k^2,m_j^2,m_i^2) \nonumber 
    \\ 
    &\phantom{=}& + 
    2 C_{ij} C_{ij}^* [(D-2)B_{00}(k^2,m_j^2,m_i^2) 
    + k^2 B_1(k^2,m_j^2,m_i^2) + k^2 B_{11}(k^2,m_j^2,m_i^2) ]
    \big\}\,,
\end{eqnarray}
in which we notice that the SM parts of the two point functions, with the exception of  light (active) neutrino terms, have been  taken from~\cite{Denner:1991kt}.

The scalar Higgs unrenormalised self energy is simply given by
\begin{eqnarray}
 \raisebox{-6mm}{
 \begin{tikzpicture}
    \begin{feynman}
    \vertex[blob] (a) at (1.5 ,0) {\contour{black}{ }};
    \vertex (f1) at (0,0) {\(H\)};
    \vertex (f2) at (3,0){\(H\)};
    \diagram* {
    (f1) -- [scalar, edge label'=\(\vphantom{\frac{1}{2}}k\)] (a) --  [scalar] (f2),
};
    \end{feynman}
\end{tikzpicture}} 
\,=\, i \Sigma^{H}(k^2)\,,
\end{eqnarray}
where the neutral lepton contribution is 
\begin{eqnarray}
    \Sigma^H(k^2) &=& -\frac{g_w^2 }{16 \pi^2 m_j^2}
    \bigg\{2 \left(C_{ij}^2 m_i m_j+C_{ij} C_{ij}^* \left(m_i^2+m_j^2\right)+C_{ij}^{*2} m_i m_j\right) (D B_{00}+k^2 B_1+k^2 B_{11})
    \nonumber \\
    &\phantom{=}&
    +m_i m_j \left(C_{ij}^2 \left(m_i^2+m_j^2\right)+4 C_{ij} C_{ij}^* m_i m_j+C_{ij}^{*2} \left(m_i^2+m_j^2\right)\right) B_0\bigg\}\,,
\end{eqnarray}
with the Passarino Veltman functions $B_{rs}=B_{rs}(k^2,m_j^2,m_i^2)$.

The wave function and mass counterterms are then given by 
\begin{align}
    \delta M_W^2 &= \mathrm{Re}\,\Sigma_T^W(M_W^2)\,,   &\delta Z_W &= - \mathrm{Re}\dfrac{\partial\Sigma_T^W(k^2)}{\partial k^2}\bigg|_{k^2=M_W^2}\,,\\
    \delta M_Z^2 &= \mathrm{Re}\,\Sigma_T^{ZZ}(M_Z^2)\,,   &\delta Z_{ZZ} &= - \mathrm{Re}\dfrac{\partial\Sigma_T^{ZZ}(k^2)}{\partial k^2}\bigg|_{k^2=M_Z^2}\,,\\
   \delta M_H^2 &= \mathrm{Re}\,\Sigma^{H}(M_H^2)\,,   &\delta Z_{H} &= - \mathrm{Re}\dfrac{\partial\Sigma^{H}(k^2)}{\partial k^2}\bigg|_{k^2=M_H^2}\,.
\end{align}

\subsection{Fermion self-energies}
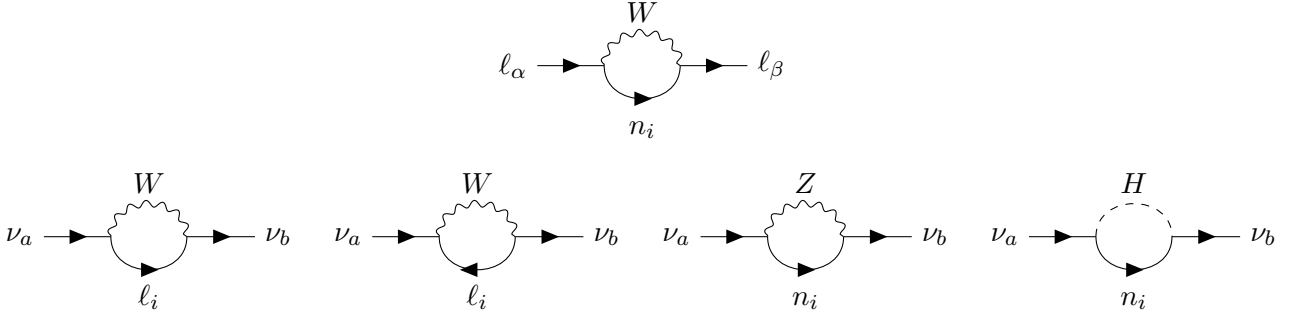
\begin{figure}[h!]
    \centering
    \begin{subfigure}[b]{\textwidth}
    \centering
  \raisebox{1mm}{ \begin{tikzpicture}
    \begin{feynman}
    \vertex (a) at (0,0) {\(\ell_\alpha\)};
    \vertex (b) at (1.2,0);
    \vertex (c) at (2.2,0);
    \vertex (d) at (3.4,0) {\(\ell_\beta\)};
    \diagram* {
    (a) -- [fermion] (b),
    (b) -- [fermion, half right, edge label'=\(\vphantom{\frac{1}{2}}n_i\)] (c),
    (b) -- [boson, half left, edge label=\(W\)] (c),
    (c) -- [fermion] (d),
    };
    \end{feynman}
    \end{tikzpicture}}
            \label{}
    \end{subfigure}
    \\
    \begin{subfigure}[b]{0.23\textwidth}
    \centering
   \begin{tikzpicture}
    \begin{feynman}
    \vertex (a) at (0,0) {\(\nu_a\)};
    \vertex (b) at (1.2,0);
    \vertex (c) at (2.2,0);
    \vertex (d) at (3.4,0) {\(\nu_b\)};
    \diagram* {
    (a) -- [fermion] (b),
    (b) -- [fermion, half right, edge label'=\(\vphantom{\frac{1}{2}}\ell_i\)] (c),
    (b) -- [boson, half left, edge label=\(W\)] (c),
    (c) -- [fermion] (d),
    };
    \end{feynman}
    \end{tikzpicture}
            \label{}
    \end{subfigure}
    \hfill
        \begin{subfigure}[b]{0.23\textwidth}
    \centering
   \begin{tikzpicture}
    \begin{feynman}
    \vertex (a) at (0,0) {\(\nu_a\)};
    \vertex (b) at (1.2,0);
    \vertex (c) at (2.2,0);
    \vertex (d) at (3.4,0) {\(\nu_b\)};
    \diagram* {
    (a) -- [fermion] (b),
    (b) -- [anti fermion, half right, edge label'=\(\vphantom{\frac{1}{2}}\ell_i\)] (c),
    (b) -- [boson, half left, edge label=\(W\)] (c),
    (c) -- [fermion] (d),
    };
    \end{feynman}
    \end{tikzpicture}
            \label{}
    \end{subfigure}
    \hfill
    \begin{subfigure}[b]{0.23\textwidth}
    \centering
  \begin{tikzpicture}
    \begin{feynman}
    \vertex (a) at (0,0) {\(\nu_a\)};
    \vertex (b) at (1.2,0);
    \vertex (c) at (2.2,0);
    \vertex (d) at (3.4,0) {\(\nu_b\)};
    \diagram* {
    (a) -- [fermion] (b),
    (b) -- [fermion, half right, edge label'=\(\vphantom{\frac{1}{2}}n_i\)] (c),
    (b) -- [boson, half left, edge label=\(Z\)] (c),
    (c) -- [fermion] (d),
    };
    \end{feynman}
    \end{tikzpicture}
            \label{}
    \end{subfigure}
    \hfill
    \begin{subfigure}[b]{0.23\textwidth}
    \centering
  \begin{tikzpicture}
    \begin{feynman}
    \vertex (a) at (0,0) {\(\nu_a\)};
    \vertex (b) at (1.2,0);
    \vertex (c) at (2.2,0);
    \vertex (d) at (3.4,0) {\(\nu_b\)};
    \diagram* {
    (a) -- [fermion] (b),
    (b) -- [fermion, half right, edge label'=\(\vphantom{\frac{1}{2}}n_i\)] (c),
    (b) -- [scalar, half left, edge label=\(H\)] (c),
    (c) -- [fermion] (d),
    };
    \end{feynman}
    \end{tikzpicture}
            \label{}
    \end{subfigure}
    \hfill
    \caption{Feynman diagrams of the NP contributions to the charged lepton (upper line) and neutrino (second line) off-diagonal self energies, in unitary gauge.}
    \label{fig:LeptonSelfenergies:UG}
\end{figure}

The two-point functions of charged and neutral leptons  (which depend on the NP contributions) are presented in Fig.~\ref{fig:LeptonSelfenergies:UG}:  two diagrams are mediated by a $W$-boson, with an opposite charged fermion flow, due to the Majorana nature of the neutrinos. 
The unrenormalised fermion self-energy~\cite{Denner:1991kt} is defined as 
\begin{eqnarray}
 \raisebox{-6mm}{
 \begin{tikzpicture}
    \begin{feynman}
    \vertex[blob] (a) at (1.5 ,0) {\contour{black}{ }};
    \vertex (f1) at (0,0) {\(f_i\)};
    \vertex (f2) at (3,0){\(f_f\)};
    \diagram* {
    (f1) -- [fermion, edge label'=\(\vphantom{\frac{1}{2}}p\)] (a) --  [fermion] (f2),
};
    \end{feynman}
\end{tikzpicture}} 
\,=\, i \Sigma_{fi}(p) \,=\, i\slashed{p} \left[ \Sigma_{fi}^{ R}(p^2) \,P_R +  \Sigma_{fi}^{ L}(p^2) \,P_L \right] + i \Sigma_{fi}^{SR}(p^2) \,P_R + i \Sigma_{fi}^{SL}(p^2) \,P_L \,,
\end{eqnarray}
with the corresponding charged lepton off-diagonal ($\ell_\alpha \to \ell_\beta$ transitions) coefficients 
\begin{eqnarray}
    \Sigma_{\beta \alpha}^{\ell L}(p^2) &=& -\dfrac{g_w^2}{32 \pi^2 M_W^2} \sum_n \mathcal{U}_{\beta n} \mathcal{U}_{\alpha n}^* (m_n^2 + (D-2)M_W^2)B_1(p^2, m_n^2,M_W^2)\,,
\end{eqnarray}
\begin{eqnarray}
    \Sigma_{\beta \alpha}^{\ell R}(p^2) &=& -\dfrac{g_w^2}{32 \pi^2 M_W^2} m_\alpha m_\beta \sum_n \mathcal{U}_{\beta n} \mathcal{U}_{\alpha n}^*  B_1(p^2, m_n^2,M_W^2)\,,
\end{eqnarray}
\begin{eqnarray}
    \Sigma_{\beta \alpha}^{\ell SL}(p^2) &=& -\dfrac{g_w^2}{32 \pi^2 M_W^2} m_\beta m_n^2  \sum_n \mathcal{U}_{\beta n} \mathcal{U}_{\alpha n}^*  B_0(p^2, m_n^2,M_W^2)\,,
\end{eqnarray}
\begin{eqnarray}
    \Sigma_{\beta \alpha}^{\ell SR}(p^2) &=& -\dfrac{g_w^2}{32 \pi^2 M_W^2} m_\alpha m_n^2  \sum_n \mathcal{U}_{\beta n} \mathcal{U}_{\alpha n}^*  B_0(p^2, m_n^2,M_W^2)\,.
\end{eqnarray}
The associated diagonal contributions can be computed by 
setting $\beta =\alpha$.
The neutrino off-diagonal self energy contributions (for $n_a \to n_b$ transitions) are 
\begin{eqnarray}
    \Sigma_{ba}^{\nu, L}(p^2) &=& 
    -\frac{g_w^2}{4 c_w^2 M_W^2} \bigg\{\sum_n \bigg[c_w^2 (C_{bn} m_n+C_{bn}^* m_b) (C_{na} m_n+C_{na}^* m_a) B_1(p^2,m_n^2,M_H^2)
    \nonumber \\
    &\phantom{=}&
    + \big[C_{bn} \left(C_{na} c_w^2 m_n^2+C_{na} (D-2) M_W^2+C_{na}^* c_w^2 m_a m_n\right)
    \nonumber \\
    &\phantom{=}&
    + C_{bn}^* c_w^2 m_b (C_{na} m_n+C_{na}^* m_a)\big]B_1(p^2,m_n^2,M_Z^2)\bigg]
    \nonumber \\
    &\phantom{=}& +2 c_w^2 \sum_\ell \left(m_a m_b \mathcal{U}_{\ell a}^* \mathcal{U}_{\ell b}+\mathcal{U}_{\ell a} \mathcal{U}_{\ell b}^* \left(m_\ell^2+(D-2) M_W^2\right)\right) B_1(p^2,m_\ell^2,M_W^2) \bigg\}
    \,,
\end{eqnarray}
\begin{eqnarray}
    \Sigma_{ba}^{\nu, R}(p^2) &=& 
   -\frac{g_w^2}{4 c_w^2 M_W^2} \bigg\{\sum_n \bigg[c_w^2  (C_{bn} m_b+C_{bn}^* m_n) (C_{na} m_a+C_{na}^* m_n) B_1(p^2,m_n^2,M_H^2)
   \nonumber \\
    &\phantom{=}& 
   +\big[C_{bn} c_w^2 m_b (C_{na} m_a+C_{na}^* m_n)
    \nonumber \\
    &\phantom{=}& 
   +C_{bn}^* \left(C_{na} c_w^2 m_a m_n+C_{na}^* c_w^2 m_n^2+C_{na}^* (D-2) M_W^2\right)\big] B_1(p^2,m_n^2,M_Z^2) \bigg]
   \nonumber \\
    &\phantom{=}& 
   +2 c_w^2 \sum_\ell  \left(m_a m_b \mathcal{U}_{\ell a} \mathcal{U}_{\ell b}^*+m_\ell^2 \mathcal{U}_{\ell a}^* \mathcal{U}_{\ell b}+(D-2) M_W^2 \mathcal{U}_{\ell a}^* \mathcal{U}_{\ell b}\right) B_1(p^2,m_\ell^2,M_W^2)\bigg\}
    \,,
\end{eqnarray}
\begin{eqnarray}
    \Sigma_{ba}^{\nu,, SL}(p^2) &=& 
   \frac{g_w^2 }{4 c_w^2 M_W^2} \bigg\{m_n \sum_n \bigg[c_w^2 (C_{bn} m_b+C_{bn}^* m_n) (C_{na} m_n+C_{na}^* m_a) B_0(p^2,m_n^2,M_H^2)
   \nonumber \\
    &\phantom{=}& 
    - \left(C_{bn} c_w^2 m_b (C_{na} m_n+C_{na}^* m_a)+C_{bn}^* \left(C_{na} c_w^2 m_n^2-C_{na} D M_W^2+C_{na}^* c_w^2 m_a m_n\right)\right)B_0(p^2,m_n^2,M_Z^2)\bigg]
   \nonumber \\
    &\phantom{=}& 
   -2 c_w^2 m_\ell^2 \sum_\ell  (m_a \mathcal{U}_{\ell a}^* \mathcal{U}_{\ell b}+m_b \mathcal{U}_{\ell a} \mathcal{U}_{\ell b}^*)B_0(p^2,m_\ell^2,M_W^2)\bigg\}
    \,,
\end{eqnarray}
\begin{eqnarray}
    \Sigma_{ba}^{\nu, SR}(p^2) &=& 
   \frac{g_w^2}{4 c_w^2 M_W^2} \bigg\{m_n \sum_n \bigg[c_w^2 (C_{bn} m_n+C_{bn}^* m_b) (C_{na} m_a+C_{na}^* m_n) B_0(p^2,m_n^2,M_H^2)
   \nonumber \\
    &\phantom{=}& 
    - \left(C_{bn} \left(C_{na} c_w^2 m_a m_n+C_{na}^* c_w^2 m_n^2-C_{na}^* D M_W^2\right)+C_{bn}^* c_w^2 m_b (C_{na} m_a+C_{na}^* m_n)\right)B_0(p^2,m_n^2,M_Z^2)\bigg]
   \nonumber \\
    &\phantom{=}& 
   -2 c_w^2 m_\ell^2 \sum_\ell  (m_a \mathcal{U}_{\ell a} \mathcal{U}_{\ell b}^*+m_b \mathcal{U}_{\ell a}^* \mathcal{U}_{\ell b})B_0(p^2,m_\ell^2,M_W^2) \bigg\}
    \,.
\end{eqnarray}
The diagonal terms can easily be obtained by the appropriate replacements.
The fermion wave function renormalisation is then finally  given by
\begin{eqnarray}
    \delta Z_{fi}^L = \dfrac{2}{m_f^2 - m_i^2} \left[ 
    m_i^2\Sigma_{fi}^{ L}(m_i^2)  +
    m_i m_f\Sigma_{fi}^{ R}(m_i^2)  +
   m_f \Sigma_{fi}^{S L}(m_i^2) + 
    m_i \Sigma_{fi}^{S R}(m_i^2)\right]\,,
\end{eqnarray}
\begin{eqnarray}
    \delta Z_{fi}^R = \dfrac{2}{m_f^2 - m_i^2} \left[ 
    m_i^2\Sigma_{fi}^{R}(m_i^2)  +
    m_i m_f\Sigma_{fi}^{L}(m_i^2)  +
   m_f \Sigma_{fi}^{S R}(m_i^2) + 
    m_i \Sigma_{fi}^{S L}(m_i^2)\right]\,,
\end{eqnarray}
for $f_i \to f_f$ transitions, both for Dirac and Majorana fermions. The diagonal parts read
\begin{eqnarray}
\delta Z_{ff}^{L} &=& 
-\Sigma_{ff}^{  L}(m_f^2) 
+ \dfrac{1}{2 m_f}\left[\Sigma_{ff}^{SL}(m_f^2) - \Sigma_{ff}^{SR}(m_f^2) \right] 
\nonumber\\
&\phantom{=}&
-  m_f^2\, \left[\partial \Sigma_{ff}^{  L}(m_f^2)+\partial \Sigma_{ff}^{ R}(m_f^2) \right]
- m_f\, \left[\partial \Sigma_{ff}^{SL}(m_f^2)+\partial \Sigma_{ff}^{SR}(m_f^2)\right]\,,
\end{eqnarray}
\begin{eqnarray}
\delta Z_{ff}^{R} &=& 
-\Sigma_{ff}^{R}(m_f^2) 
- \dfrac{1}{2 m_f}\left[\Sigma_{ff}^{SL}(m_f^2) - \Sigma_{ff}^{SR}(m_f^2) \right] 
\nonumber\\
&\phantom{=}&
-  m_f^2\, \left[\partial \Sigma_{ff}^{  L}(m_f^2)+\partial \Sigma_{ff}^{ R}(m_f^2) \right]
- m_f\, \left[\partial \Sigma_{ff}^{SL}(m_f^2)+\partial \Sigma_{ff}^{SR}(m_f^2)\right]\,.
\end{eqnarray}

The new (enlarged) mixing matrix $\mathcal{U}$ also needs to be renormalised, as well as the $C_{ij}$ (in an analogous way to what is usually done for the quark sector).
One thus has
\begin{eqnarray}
\delta \mathcal{U}_{\beta f} &=& \dfrac{1}{4}\, \sum_\alpha^{e, \mu,\tau}\, \big[ \delta Z_{\beta \alpha}^{\ell, L} - \delta Z_{\beta \alpha}^{\ell, L \dag} \big]\, \mathcal{U}_{\alpha f}
- \dfrac{1}{4} \,\sum_{i=1}^{3+n_s} \, \mathcal{U}_{\beta i} \, \big[ \delta Z_{i f}^{\nu, L} - \delta Z_{i f}^{\nu, L \dag} \big]\,,
\end{eqnarray}
\begin{eqnarray}
\delta \mathcal{U}_{\beta f}^* &=& \dfrac{1}{4}\, \sum_\alpha^{e, \mu,\tau}\, \big[ \delta Z_{\alpha \beta }^{\ell, L \dag} - \delta Z_{\alpha \beta }^{\ell, L } \big]\, \mathcal{U}_{\alpha f}^*
- \dfrac{1}{4} \,\sum_{i=1}^{3+n_s} \, \mathcal{U}_{\beta i} \, \big[ \delta Z_{f i }^{\nu, L \dag } - \delta Z_{ f i }^{\nu, L} \big]\,,
\end{eqnarray}
\begin{eqnarray}\label{eq:cij_renormalisation}
\delta C_{ab} &=& \dfrac{1}{4}\, \sum_{x=1}^{3+n_s}\, \big[ \delta Z_{ax}^{\nu, L} - \delta Z_{ax}^{\nu, L \dag} \big]\, C_{xb}
- \dfrac{1}{4} \,\sum_{y=1}^{3+n_s} \, C_{a y} \, \big[ \delta Z_{yb}^{\nu, L} - \delta Z_{yb}^{\nu, L \dag} \big]\,.
\end{eqnarray}
The renormalisation of the left and right $Z$-charged lepton couplings are given by
\begin{eqnarray}
\delta g_R^Z \:=\: - Q_\ell \,s_w^2\bigg[\delta Z_e + \dfrac{1}{c_w^2} \dfrac{\delta s_w}{s_w}\bigg]\,, \quad\quad
    \delta g_L^Z \:=\: T^3_\ell \bigg[\delta Z_e + \dfrac{s_w^2-c_w^2}{c_w^2} \dfrac{\delta s_w}{s_w}\bigg] + \delta g_R^Z\,.
\end{eqnarray}

\mathversion{bold}
\section{Form factors for $W$ decays }\label{sec:Wformfactors}
\mathversion{normal}
The vector and tensor form factors of the $W\to \ell_\beta \nu_f$ decay are given below 
\begin{eqnarray}
F_L^{V (a)} &=& \dfrac{g_w^3}{32 \pi^2 \sqrt{2} M_W^2}\sum_{\alpha}\sum_{i}
\mathcal{U}_{\alpha i}\,\mathcal{U}_{\alpha f}^*\,\mathcal{U}_{\beta i} \,m_f\, m_i
\bigg\{ 
-(D-3) M_W^2 C_{0}-(D-2) M_W^2 C_{2}-(D-2) M_W^2 C_{1}
\nonumber\\
&\phantom{=}&
- m_\beta^2 C_{11} -m_f^2 C_{22}+\left(-m_\beta^2-m_f^2\right) C_{12}-2 C_{00}+B_0(q^2,m_i^2,m_\alpha^2)
\bigg\}\,,
\end{eqnarray}
\begin{eqnarray}
F_R^{V (a)} &=& \dfrac{g_w^3}{32 \pi^2 \sqrt{2} M_W^2}\sum_{\alpha}\sum_{i}
\mathcal{U}_{\alpha i}\,\mathcal{U}_{\alpha f}^*\,\mathcal{U}_{\beta i} \,m_i \,m_\beta
\bigg\{ 
\left(m_\alpha^2-m_f^2-2 M_W^2\right) C_{2}+\left(m_\alpha^2-m_f^2-2 M_W^2\right) C_{1}
\nonumber\\
&\phantom{=}&
-2 M_W^2 C_{0} -m_f^2 C_{22}-2 m_f^2 C_{12}-m_f^2 C_{11}
\bigg\}\,,
\end{eqnarray}
\begin{eqnarray}
F_L^{T (a)} &=& i\dfrac{g_w^3}{32 \pi^2 \sqrt{2} M_W^2}\sum_{\alpha}\sum_{i}
\mathcal{U}_{\alpha i}\,\mathcal{U}_{\alpha f}^*\,\mathcal{U}_{\beta i} \,m_i\, m_\beta\, m_f
\bigg\{ 
C_{12}+C_{11}
\bigg\}\,,
\end{eqnarray}
\begin{eqnarray}
F_R^{T (a)} &=& i\dfrac{g_w^3}{32 \pi^2 \sqrt{2} M_W^2}\sum_{\alpha}\sum_{i}
\mathcal{U}_{\alpha i}\,\mathcal{U}_{\alpha f}^*\,\mathcal{U}_{\beta i} \,m_i 
\bigg\{ 
\left(-m_\alpha^2+m_f^2+2 M_W^2\right) C_{2}+(D-2) M_W^2 C_{1}
\nonumber\\
&\phantom{=}&
+2 M_W^2 C_{0}+m_f^2 C_{22}+m_f^2 C_{12}
\bigg\}\,,
\end{eqnarray}
with the Passarino-Veltman functions defined as $C_{r,s} = C_{r,s}(m_\beta^2,q^2,m_f^2,M_W^2,m_i^2,m_\alpha^2)$.

\begin{eqnarray}
F_L^{V (b)} &=& \dfrac{g_w^3}{64 \pi^2 \sqrt{2} M_W^2}\sum_{i}
\mathcal{U}_{\beta i} 
\bigg\{ 
C_{if} \bigg[
\left(m_\beta^2 \left(-m_f^2+m_i^2-2 M_Z^2\right)+6 m_f^2 \left(M_Z^2-2 M_W^2\right)\right) C_{2}
\nonumber\\
&\phantom{=}&
+\left(m_\beta^2 \left(-m_f^2+m_i^2-8 M_W^2+2 M_Z^2\right)+2 m_f^2 \left(M_Z^2-2 M_W^2\right)\right) C_{1}
\nonumber\\
&\phantom{=}&
+C_{0} \left(4 m_\beta^2 \left(M_W^2-M_Z^2\right)+\left(2 M_W^2-M_Z^2\right) \left(-2 m_f^2+2 m_i^2+(D-6) M_Z^2-2 q^2\right)\right)
\nonumber\\
&\phantom{=}&
+\left(2 M_W^2-M_Z^2\right) \left(2 B_0\left(m_\beta^2,M_Z^2,m_\beta^2\right)+(D-6) B_0(q^2,m_\beta^2,m_i^2)+2 B_0\left(m_f^2,M_Z^2,m_i^2\right)\right)
\nonumber\\
&\phantom{=}&
+m_\beta^2 \left(-\left(m_f^2+(D-2) \left(2 M_W^2-M_Z^2\right)\right)\right) C_{11}-m_f^2 \left(m_\beta^2+(D-2) \left(2 M_W^2-M_Z^2\right)\right) C_{22}
\nonumber\\
&\phantom{=}&
+\left(m_\beta^2 \left(-2 m_f^2-(D-2) \left(2 M_W^2-M_Z^2\right)\right)-(D-2) m_f^2 \left(2 M_W^2-M_Z^2\right)\right) C_{12}
\nonumber\\
&\phantom{=}&
-2 (D-2) \left(2 M_W^2-M_Z^2\right) C_{00}\bigg]
\nonumber\\
&\phantom{=}&
+C_{if}^* \bigg[2 m_f m_i \left(M_Z^2-2 M_W^2\right) C_{0}+2 m_f m_i \left(M_Z^2-2 M_W^2\right) C_{2}+2 m_f m_i \left(M_Z^2-2 M_W^2\right) C_{1}
\nonumber\\
&\phantom{=}&
- m_\beta^2 m_f m_i C_{22}-2 m_\beta^2 m_f m_i C_{12}-m_\beta^2 m_f m_i C_{11}\bigg]
\bigg\}\,,
\end{eqnarray}
\begin{eqnarray}
F_R^{V (b)} &=& \dfrac{g_w^3}{64\pi^2 \sqrt{2} M_W^2}\sum_{i}
\mathcal{U}_{\beta i} m_\beta 
\bigg\{ 
C_{if} \bigg[-m_f \left(m_f^2+(D-2) \left(2 M_W^2-M_Z^2\right)\right) C_{22}
\nonumber\\
&\phantom{=}&
-m_f \left(m_\beta^2+m_f^2+2 (D-2) \left(2 M_W^2-M_Z^2\right)\right) C_{12}-m_f \left(m_\beta^2+(D-2) \left(2 M_W^2-M_Z^2\right)\right) C_{11}
\nonumber\\
&\phantom{=}&
-2 (D-2) m_f M_W^2 C_{2}-2 (D-2) m_f M_W^2 C_{1}-(D-3) m_f M_Z^2 C_{0}-2 m_f C_{00}+m_f B_0(q^2,m_\beta^2,m_i^2)\bigg]
\nonumber\\
&\phantom{=}&
+C_{if}^* \bigg[-(D-2) m_i \left(2 M_W^2-M_Z^2\right) C_{2}-(D-2) m_i \left(2 M_W^2-M_Z^2\right) C_{1}-(D-3) m_i M_Z^2 C_{0}
\nonumber\\
&\phantom{=}&
-m_\beta^2 m_i C_{11}-m_f^2 m_i C_{22}-m_i \left(m_\beta^2+m_f^2\right) C_{12}-2 m_i C_{00}+m_i B_0(q^2,m_\beta^2,m_i^2)\bigg]
\bigg\}\,,
\end{eqnarray}
\begin{eqnarray}
F_L^{T (b)} &=& i\dfrac{g_w^3}{64 \pi^2 \sqrt{2} M_W^2}\sum_{i}
\mathcal{U}_{\beta i} m_\beta 
\bigg\{ 
C_{if} \bigg[\left(m_f^2-m_i^2+2 M_Z^2\right) C_{2}+\left(4 M_W^2+(D-4) M_Z^2\right) C_{1}
\nonumber\\
&\phantom{=}&
+\left(m_f^2+(D-2) \left(2 M_W^2-M_Z^2\right)\right) C_{12}+(D-2) \left(2 M_W^2-M_Z^2\right) C_{11}+2 M_Z^2 C_{0}+m_f^2 C_{22}\bigg]
\nonumber\\
&\phantom{=}&
+C_{if}^* m_f m_i \left[C_{22} + C_{12}\right]
\bigg\}\,,
\end{eqnarray}
\begin{eqnarray}
F_R^{T (b)} &=& i\dfrac{g_w^3}{64 \pi^2 \sqrt{2} M_W^2}\sum_{i}
\mathcal{U}_{\beta i} 
\bigg\{ 
C_{if} \bigg[D m_f \left(2 M_W^2-M_Z^2\right) C_{2}+(D-2) m_f \left(2 M_W^2-M_Z^2\right) C_{22}+m_\beta^2 m_f C_{11}
\nonumber\\
&\phantom{=}&
+m_f \left(m_\beta^2+(D-2) \left(2 M_W^2-M_Z^2\right)\right) C_{12}+m_f \left(4 M_W^2-2 M_Z^2\right) C_{0}+m_f \left(4 M_W^2-2 M_Z^2\right) C_{1}\bigg]
\nonumber\\
&\phantom{=}&
+C_{if}^* \bigg[(D-2) m_i \left(2 M_W^2-M_Z^2\right) C_{2}+m_i \left(4 M_W^2-2 M_Z^2\right) C_{0}+m_i \left(4 M_W^2-2 M_Z^2\right) C_{1}
\nonumber\\
&\phantom{=}&
+m_\beta^2 m_i C_{12}+m_\beta^2 m_i C_{11}\bigg]
\bigg\}\,,
\end{eqnarray}
with the Passarino-Veltman functions $C_{r,s} = C_{r,s}(m_\beta^2,q^2,m_f^2,M_Z,m_\beta,m_i)$.

\begin{eqnarray}
F_L^{V (c)} &=& \dfrac{g_w^3}{64 \pi^2 \sqrt{2} M_W^2}\sum_{i}
\mathcal{U}_{\beta i} m_\beta^2
\bigg\{ 
C_{if} \bigg[2 \left(m_f^2+m_i^2\right) C_{0}+\left(3 m_f^2+m_i^2\right) C_{2}+\left(3 m_f^2+m_i^2\right) C_{1}+m_f^2 C_{22}
\nonumber\\
&\phantom{=}&
+2 m_f^2 C_{12}+m_f^2 C_{11}\bigg]
+C_{if}^* \,m_f\,m_i\bigg[4  C_{0}+4  C_{2}+ C_{22}+4  C_{1}+2  C_{12}+ C_{11}\bigg]
\bigg\}\,,
\end{eqnarray}
\begin{eqnarray}
F_R^{V (c)} &=& \dfrac{g_w^3}{64 \pi^2 \sqrt{2} M_W^2}\sum_{i}
\mathcal{U}_{\beta i} m_\beta
\bigg\{ 
C_{if} m_f\bigg[- M_H^2 C_{0}+ \left(m_\beta^2+m_f^2\right) C_{12}+m_\beta^2  C_{11}+2  C_{00}+m_f^2 C_{22}
\nonumber\\
&\phantom{=}&
- B_0(q^2,m_\beta^2,m_i^2)\bigg]
\nonumber\\
&\phantom{=}&
+C_{if}^* m_i\bigg[-M_H^2  C_{0}+m_\beta^2  C_{11}+m_f^2  C_{22}+ \left(m_\beta^2+m_f^2\right) C_{12}+2 C_{00}- B_0(q^2,m_\beta^2,m_i^2)\bigg]
\bigg\}\,,
\end{eqnarray}
\begin{eqnarray}
F_L^{T (c)} &=& -i\dfrac{g_w^3}{64 \pi^2 \sqrt{2} M_W^2}\sum_{i}
\mathcal{U}_{\beta i} m_\beta
\bigg\{ 
C_{if} \bigg[\left(m_f^2+m_i^2\right) C_{2}+m_f^2 C_{22}+m_f^2 C_{12}\bigg]
\nonumber\\
&\phantom{=}& + C_{if}^* m_f\,m_i\bigg[2 C_{2}+ C_{22}+ C_{12}\bigg]
\bigg\}\,,
\end{eqnarray}
\begin{eqnarray}
F_R^{T (c)} &=& -i\dfrac{g_w^3}{64 \pi^2 \sqrt{2} M_W^2}\sum_{i}
\mathcal{U}_{\beta i} m_\beta^2
\bigg\{ 
C_{if} m_f \bigg[2  C_{1}+ C_{12}+ C_{11}\bigg]
+C_{if}^* m_i \bigg[2  C_{1}+ C_{12}+ C_{11}\bigg]
\bigg\}\,, \nonumber \\
\end{eqnarray}
where the Passarino-Veltman functions are $C_{r,s}=C_{r,s}(m_\beta^2,q^2,m_f^2,M_H^2,m_\beta^2,m_i^2)$.

\begin{eqnarray}
F_L^{V (d)} &=& \dfrac{g_w^3}{64 \pi^2 \sqrt{2} M_W^2}\sum_{i}
\mathcal{U}_{\beta i} 
\nonumber\\
&\phantom{=}&
\times\bigg\{ 
C_{if} \bigg[\left(m_\beta^2 \left(-\left(2 \left(c_w^2-1\right) M_Z^2+m_f^2+m_i^2-2 M_W^2\right)\right)-2 m_f^2 \left(m_i^2+(D-7) M_W^2\right)\right) C_{2}
\nonumber\\
&\phantom{=}&
+\left(m_\beta^2 \left(-\left(2 \left(\left(c_w^2-1\right) M_Z^2+m_i^2+(D-7) M_W^2\right)+m_f^2\right)\right)-m_f^2 \left(m_i^2-2 M_W^2\right)\right) C_{1}
\nonumber\\
&\phantom{=}&
+\left(-2 c_w^2 m_i^2 M_Z^2-m_\beta^2 \left(m_i^2-4 M_W^2\right)-m_f^2 \left(m_i^2-4 M_W^2\right)+4 m_i^2 M_W^2+2 m_i^2 M_Z^2-4 M_W^4-4 M_W^2 M_Z^2\right) C_{0}
\nonumber\\
&\phantom{=}&
+m_\beta^2 \left(-\left(m_f^2+m_i^2+2 (D-2) M_W^2\right)\right) C_{11}-m_f^2 \left(m_\beta^2+m_i^2+2 (D-2) M_W^2\right) C_{22}
\nonumber\\
&\phantom{=}&
+\left(m_\beta^2 \left(-\left(2 m_f^2+m_i^2+2 (D-2) M_W^2\right)\right)-m_f^2 \left(m_i^2+2 (D-2) M_W^2\right)\right) C_{12}
\nonumber\\
&\phantom{=}&
-2 \left(m_i^2+2 (D-2) M_W^2\right) C_{00}-4 M_W^2 \left(B_0(m_\beta^2,m_i^2,M_W^2)+B_0(m_f^2,m_i^2,M_Z^2)-B_0(q^2,M_W^2,M_Z^2)\right)\bigg]
\nonumber\\
&\phantom{=}&
+C_{if}^* m_f \,m_i \bigg[- \left(m_\beta^2+m_i^2-2 (D-1) M_W^2\right) C_{0}- \left(2 m_\beta^2+m_f^2+m_i^2-2 D M_W^2+2 M_W^2\right) C_{2}
\nonumber\\
&\phantom{=}&
- \left(3 m_\beta^2+m_i^2-2 (D-1) M_W^2\right) C_{1}-2 m_\beta^2  C_{11}- \left(m_\beta^2+m_f^2\right) C_{22}
\nonumber\\
&\phantom{=}&
- \left(3 m_\beta^2+m_f^2\right) C_{12}-2  C_{00}\bigg]
\bigg\}\,,
\end{eqnarray}
\begin{eqnarray}
F_R^{V (d)} &=& \dfrac{g_w^3}{64 \pi^2 \sqrt{2} M_W^2}\sum_{i}
\mathcal{U}_{\beta i} m_\beta
\bigg\{ 
C_{if} m_f \bigg[- \left(m_f^2+3 m_i^2+2 (D-2) M_W^2\right) C_{2}
\nonumber\\
&\phantom{=}&
-\left(m_f^2+m_i^2+2 (D-2) M_W^2\right) C_{22}-\left(m_\beta^2+3 m_i^2+2 (D-2) M_W^2\right) C_{1}
\nonumber\\
&\phantom{=}&
- \left(m_\beta^2+m_f^2+2 m_i^2+4 (D-2) M_W^2\right) C_{12}- \left(m_\beta^2+m_i^2+2 (D-2) M_W^2\right) C_{11}
-2  m_i^2 C_{0}-2  C_{00}\bigg]
\nonumber\\
&\phantom{=}&
+C_{if}^* m_i \bigg[- \left(2 \left(c_w^2 M_Z^2-(D-1) M_W^2 -M_Z^2\right)+m_f^2+m_i^2\right) C_{0}
\nonumber\\
&\phantom{=}&
- \left(2 \left(c_w^2 M_Z^2-(D-1) M_W^2 -M_Z^2\right)+3 m_f^2+m_i^2\right) C_{2}
\nonumber\\
&\phantom{=}&
- \left(2 c_w^2 M_Z^2+m_\beta^2+2 m_f^2+m_i^2-2 (D -1)M_W^2 -2 M_Z^2\right) C_{1}-2 m_f^2  C_{22}- \left(m_\beta^2+3 m_f^2\right) C_{12}
\nonumber\\
&\phantom{=}&
- \left(m_\beta^2+m_f^2\right) C_{11}-2  C_{00}\bigg]
\bigg\}\,,
\end{eqnarray}
\begin{eqnarray}
F_L^{T (d)} &=& i \dfrac{g_w^3}{64 \pi^2 \sqrt{2} M_W^2}\sum_{i}
\mathcal{U}_{\beta i} m_\beta
\bigg\{ 
C_{if} \bigg[\left(2 \left(c_w^2-1\right) M_Z^2+m_f^2+m_i^2-2 M_W^2\right) C_{2}
\nonumber\\
&\phantom{=}&
+2 \left(m_i^2+(D-4) M_W^2\right) C_{1}+\left(m_f^2+m_i^2+2 (D-2) M_W^2\right) C_{12}+\left(m_i^2+2 (D-2) M_W^2\right) C_{11}
\nonumber\\
&\phantom{=}&
+m_i^2 C_{0}+m_f^2 C_{22}\bigg]
+C_{if}^* m_f\,m_i\bigg[ C_{0}+2  C_{2}+ C_{22}+2  C_{1}+2  C_{12}+ C_{11}\bigg]
\bigg\}\,,
\end{eqnarray}
\begin{eqnarray}
F_R^{T (d)} &=& i \dfrac{g_w^3}{64 \pi^2 \sqrt{2} M_W^2}\sum_{i}
\mathcal{U}_{\beta i} 
\bigg\{ 
C_{if} m_f \bigg[2  \left(m_i^2+(D-4) M_W^2\right) C_{2}+ \left(m_i^2+2 (D-2) M_W^2\right) C_{22}
\nonumber\\
&\phantom{=}&
+ \left(m_\beta^2+m_i^2+2 (D-2) M_W^2\right) C_{12}+ \left(m_\beta^2+m_i^2-2 M_W^2\right) C_{1}+ m_i^2 C_{0}+m_\beta^2 C_{11}\bigg]
\nonumber\\
&\phantom{=}&
+C_{if}^* m_i\bigg[ \left(2 \left(c_w^2 M_Z^2-D M_W^2+M_W^2-M_Z^2\right)+m_f^2+m_i^2\right) C_{2}+\left(m_i^2-2 (D-4) M_W^2\right) C_{0}
\nonumber\\
&\phantom{=}&
+ \left(m_\beta^2+m_i^2-2 (D-1) M_W^2\right) C_{1}+m_\beta^2  C_{11}+m_f^2 C_{22}+ \left(m_\beta^2+m_f^2\right) C_{12}\bigg]
\bigg\}\,,
\end{eqnarray}
with the following Passarino-Veltman functions $C_{r,s}=C_{r,s}(m_\beta^2,q^2,m_f^2,m_i^2,M_W^2,M_Z^2)$.

\begin{eqnarray}
F_L^{V (e)} &=& \dfrac{g_w^3}{64 \pi^2 \sqrt{2} M_W^2}\sum_{i}
\mathcal{U}_{\beta i} 
\bigg\{ 
C_{if} \bigg[m_i^2 \left(-m_\beta^2+m_f^2+2 M_W^2\right) C_{0}+\left(m_\beta^2 \left(m_f^2-m_i^2\right)-2 m_f^2 M_W^2\right) C_{2}
\nonumber\\
&\phantom{=}&
+\left(m_\beta^2 \left(m_f^2-2 m_i^2\right)+m_f^2 \left(m_i^2-2 M_W^2\right)\right) C_{1}+m_\beta^2 \left(m_f^2-m_i^2\right) C_{11}+m_f^2 \left(m_\beta^2-m_i^2\right) C_{22}
\nonumber\\
&\phantom{=}&
+\left(m_\beta^2 \left(2 m_f^2-m_i^2\right)-m_f^2 m_i^2\right) C_{12}-2 m_i^2 C_{00}\bigg]
\nonumber\\
&\phantom{=}&
+C_{if}^* m_f\,m_i \bigg[ \left(-m_\beta^2+m_i^2+2 M_W^2\right) C_{0}- \left(m_f^2-m_i^2+2 M_W^2\right) C_{2}
\nonumber\\
&\phantom{=}&
+ \left(-m_\beta^2+m_i^2-2 M_W^2\right) C_{1}+ \left(m_\beta^2-m_f^2\right) C_{22}+ \left(m_\beta^2-m_f^2\right) C_{12}-2  C_{00}\bigg]
\bigg\}\,,
\end{eqnarray}
\begin{eqnarray}
F_R^{V (e)} &=& \dfrac{g_w^3}{64 \pi^2 \sqrt{2} M_W^2}\sum_{i}
\mathcal{U}_{\beta i} m_\beta
\bigg\{ 
C_{if} \bigg[m_f \left(m_\beta^2-m_i^2\right) C_{1}+m_f \left(m_\beta^2+m_f^2-2 m_i^2\right) C_{12}
\nonumber\\
&\phantom{=}&
+m_f \left(m_\beta^2-m_i^2\right) C_{11}+2 m_f C_{00}+\left(m_f^3-m_f m_i^2\right) C_{2}+\left(m_f^3-m_f m_i^2\right) C_{22}\bigg]
\nonumber\\
&\phantom{=}&
+C_{if}^* m_i\bigg[\left(m_i^2-m_f^2\right) C_{0}+\left(m_i^2-m_f^2 \right) C_{2}+ \left(m_\beta^2-2 m_f^2+m_i^2\right) C_{1}+ \left(m_\beta^2-m_f^2\right) C_{12}
\nonumber\\
&\phantom{=}&
+ \left(m_\beta^2-m_f^2\right) C_{11}+2  C_{00}\bigg]
\bigg\}\,,
\end{eqnarray}
\begin{eqnarray}
F_L^{T (e)} &=& i \dfrac{g_w^3}{64 \pi^2 \sqrt{2} M_W^2}\sum_{i}
\mathcal{U}_{\beta i} m_\beta
\bigg\{ 
C_{if} \bigg[m_i^2 C_{0}+\left(m_i^2-m_f^2\right) C_{2}+2 m_i^2 C_{1}+\left(m_i^2-m_f^2\right) C_{12}
\nonumber\\
&\phantom{=}&
+m_i^2 C_{11}-m_f^2 C_{22}\bigg]
+C_{if}^* m_f\,m_i\bigg[ C_{0}- C_{22}+2  C_{1}+ C_{11}\bigg]
\bigg\}\,,
\end{eqnarray}
\begin{eqnarray}
F_R^{T (e)} &=& i \dfrac{g_w^3}{64 \pi^2 \sqrt{2} M_W^2}\sum_{i}
\mathcal{U}_{\beta i} 
\bigg\{ 
C_{if} m_f\bigg[- \left(m_\beta^2+m_i^2-2 M_W^2\right) C_{1}- m_i^2 C_{0}+ m_i^2 C_{22}
\nonumber\\
&\phantom{=}&
+ \left(m_i^2-m_\beta^2\right) C_{12} - m_\beta^2  C_{11}\bigg]
\nonumber\\
&\phantom{=}&
+C_{if}^* m_i \bigg[-m_i^2 C_{0}- \left(m_\beta^2+m_i^2-2 M_W^2\right) C_{1}+ \left(m_f^2-m_i^2\right) C_{2}+m_f^2  C_{22}
\nonumber\\
&\phantom{=}&
+ \left(m_f^2-m_\beta^2\right) C_{12}-m_\beta^2  C_{11}\bigg]
\bigg\}\,,
\end{eqnarray}
with the following Passarino-Veltman functions $C_{r,s}=C_{r,s}(m_\beta^2,q^2,m_f^2,m_i^2,M_W^2,M_H^2)$.

\begin{eqnarray}
F_L^{V (f),Z} &=&  \dfrac{g_w^3}{32 \pi^2 \sqrt{2} M_W^2 M_Z^2}
\mathcal{U}_{\beta f} 
\nonumber\\
&\phantom{=}&
\times \bigg\{ 
-4 M_W^2 \left(2 M_W^2-M_Z^2\right) \left[B_0\left(m_\beta^2,m_\beta^2,M_Z^2\right)+B_0\left(m_f^2,m_\beta^2,M_W^2\right)-B_0(q^2,M_Z^2,M_W^2)\right]
\nonumber\\
&\phantom{=}&
-m_\beta^2 \left[M_Z^2 \left(m_\beta^2+m_f^2\right)+4 (D-2) M_W^4-2 (D-2) M_W^2 M_Z^2\right] C_{11}
\nonumber\\
&\phantom{=}&
-2 m_f^2 \left[m_\beta^2 M_Z^2+2 (D-2) M_W^4-(D-2) M_W^2 M_Z^2\right] C_{22}
\nonumber\\
&\phantom{=}&
+\left[-2 m_\beta^2 M_Z^2-8 (D-2) M_W^4+4 (D-2) M_W^2 M_Z^2\right] C_{00}
\nonumber\\
&\phantom{=}&
+\big[-m_\beta^4 M_Z^2+m_\beta^2 \left(-\left(M_Z^2 \left(m_f^2+2 M_Z^2\right)+4 (D-4) M_W^4+2 (3-2 D) M_W^2 M_Z^2\right)\right)
\nonumber\\
&\phantom{=}&
-4 M_W^2 \left(2 M_W^2-M_Z^2\right) \left(-m_f^2+M_W^2+M_Z^2\right)\big] C_{0}
\nonumber\\
&\phantom{=}&
+\big[-m_\beta^4 M_Z^2+m_\beta^2 \left(-3 m_f^2 M_Z^2-4 (D-2) M_W^4+2 (2 D-3) M_W^2 M_Z^2\right)
\nonumber\\
&\phantom{=}&
-2 m_f^2 \left(2 M_W^2-M_Z^2\right) \left((D-6) M_W^2-M_Z^2\right)\big] C_{2}
\nonumber\\
&\phantom{=}&
-2 \left[m_\beta^4 M_Z^2+m_\beta^2 \left(m_f^2 M_Z^2+4 (D-4) M_W^4-3 (D-3) M_W^2 M_Z^2\right)+m_f^2 M_Z^2 \left(M_Z^2-2 M_W^2\right)\right] C_{1}
\nonumber\\
&\phantom{=}&
+\big[-m_\beta^4 M_Z^2+m_\beta^2 \left(-3 m_f^2 M_Z^2-4 (D-2) M_W^4+2 (D-2) M_W^2 M_Z^2\right)
\nonumber\\
&\phantom{=}&
-2 (D-2) m_f^2 M_W^2 \left(2 M_W^2-M_Z^2\right)\big] C_{12}
\bigg\}\,,
\end{eqnarray}
\begin{eqnarray}
F_R^{V (f),Z} &=&  \dfrac{g_w^3}{32 \pi^2 \sqrt{2} M_W^2 M_Z^2}
\mathcal{U}_{\beta f} m_\beta \, m_f
\bigg\{ 
\left[-M_Z^2 \left(3 m_\beta^2+m_f^2-4 M_Z^2\right)-8 (D-2) M_W^4+2 (3 D-8) M_W^2 M_Z^2\right] C_{2}
\nonumber\\
&\phantom{=}&
+\left[-M_Z^2 \left(m_\beta^2+m_f^2\right)-4 (D-2) M_W^4+2 (D-2) M_W^2 M_Z^2\right] C_{22}
\nonumber\\
&\phantom{=}&
+\left[-M_Z^2 \left(3 m_\beta^2+m_f^2\right)-8 (D-2) M_W^4+4 (D-2) M_W^2 M_Z^2\right] C_{12}
\nonumber\\
&\phantom{=}&
+\left[-2 m_\beta^2 M_Z^2-4 (D-2) M_W^4+2 (D-2) M_W^2 M_Z^2\right] C_{11}
\nonumber\\
&\phantom{=}&
+\left[-2 m_\beta^2 M_Z^2-4 (D-2) M_W^4+4 (D-3) M_W^2 M_Z^2+4 M_Z^4\right] C_{0}
\nonumber\\
&\phantom{=}&
+\left[-4 m_\beta^2 M_Z^2-8 (D-2) M_W^4+2 (3 D-8) M_W^2 M_Z^2+4 M_Z^4\right] C_{1}-2 M_Z^2 C_{00}
\bigg\}\,,
\end{eqnarray}
\begin{eqnarray}
F_L^{T (f),Z} &=&  i \dfrac{g_w^3}{32 \pi^2 \sqrt{2} M_W^2 M_Z^2}
\mathcal{U}_{\beta f} m_\beta
\bigg\{ 
\left[m_\beta^2 M_Z^2+4 (D-4) M_W^4-4 (D-4) M_W^2 M_Z^2\right] C_{0}
\nonumber\\
&\phantom{=}&
+\left[M_Z^2 \left(m_\beta^2+m_f^2\right)+4 (D-2) M_W^4+2 (3-2 D) M_W^2 M_Z^2\right] C_{2}
\nonumber\\
&\phantom{=}&
+\left[2 M_Z^2 \left(m_\beta^2-2 M_Z^2\right)+8 (D-3) M_W^4+2 (10-3 D) M_W^2 M_Z^2\right] C_{1}
\nonumber\\
&\phantom{=}&
+\left[M_Z^2 \left(m_\beta^2+m_f^2\right)+4 (D-2) M_W^4-2 (D-2) M_W^2 M_Z^2\right] C_{12}
\nonumber\\
&\phantom{=}&
+\left[m_\beta^2 M_Z^2+4 (D-2) M_W^4-2 (D-2) M_W^2 M_Z^2\right] C_{11}+m_f^2 M_Z^2 C_{22}
\bigg\}\,,
\end{eqnarray}
\begin{eqnarray}
F_R^{T (f),Z} &=&  i \dfrac{g_w^3}{32 \pi^2 \sqrt{2} M_W^2 M_Z^2}
\mathcal{U}_{\beta f} \, m_f
\bigg\{ 
2 M_Z^2 \left[m_\beta^2-2 M_W^2+M_Z^2\right] C_{1}
\nonumber\\
&\phantom{=}&
+\left[2 m_\beta^2 M_Z^2+4 (D-4) M_W^4-2 (D-4) M_W^2 M_Z^2\right] C_{2}
\nonumber\\
&\phantom{=}&
+\left[m_\beta^2 M_Z^2+4 (D-2) M_W^4-2 (D-2) M_W^2 M_Z^2\right] C_{22}
\nonumber\\
&\phantom{=}&
+\left[2 m_\beta^2 M_Z^2+4 (D-2) M_W^4-2 (D-2) M_W^2 M_Z^2\right] C_{12}+m_\beta^2 M_Z^2 C_{0}+m_\beta^2 M_Z^2 C_{11}
\bigg\}\,, \nonumber \\
\end{eqnarray}
with the Passarino-Veltman functions defined as $C_{r,s}=C_{r,s}(m_\beta^2,q^2,m_f^2,m_\beta^2,M_Z^2,M_W^2)$.

\begin{eqnarray}
F_L^{V (f),\gamma} &=&  \dfrac{g_w^3 (M_W^2-M_Z^2)}{16 \pi^2 \sqrt{2}  M_Z^2}
\mathcal{U}_{\beta f} 
\bigg\{ 
\left[(D-4) m_\beta^2+2 \left(-m_f^2+M_\gamma^2+M_W^2\right)\right] C_{0}
\nonumber\\
&\phantom{=}&
+2 \left[B_0\left(m_\beta^2,m_\beta^2,M_\gamma^2\right)+B_0\left(m_f^2,m_\beta^2,M_W^2\right)-B_0(q^2,M_\gamma^2,M_W^2)\right]
\nonumber\\
&\phantom{=}&
+2 (D-4) m_\beta^2 C_{1}+(D-2) m_\beta^2 C_{11}+\left[(D-2) m_\beta^2+(D-6) m_f^2\right] C_{2}
\nonumber\\
&\phantom{=}&
+(D-2) m_f^2 C_{22}+(D-2) \left(m_\beta^2+m_f^2\right) C_{12}+2 (D-2) C_{00}
\bigg\}\,,
\end{eqnarray}
\begin{eqnarray}
F_R^{V (f),\gamma} &=&  \dfrac{g_w^3 (M_W^2-M_Z^2)}{16 \pi^2 \sqrt{2}  M_Z^2}
\mathcal{U}_{\beta f} m_\beta \, m_f \, (D-2)
\bigg\{ 
C_{0}+2 C_{2}+C_{22}+2 C_{1}+2 C_{12}+C_{11}
\bigg\}\,,
\end{eqnarray}
\begin{eqnarray}
F_L^{T (f),\gamma} &=&  -i \dfrac{g_w^3 (M_W^2-M_Z^2)}{16 \pi^2 \sqrt{2}  M_Z^2}
\mathcal{U}_{\beta f} m_\beta
\bigg\{ 
(D-4) C_{0}+2 (D-3) C_{1}+(D-2) \big[C_{2} + C_{12}+ C_{11}\big]
\bigg\}\,,\nonumber \\
\end{eqnarray}
\begin{eqnarray}
F_R^{T (f),\gamma} &=&  -i\dfrac{g_w^3 (M_W^2-M_Z^2)}{16 \pi^2 \sqrt{2}  M_Z^2}
\mathcal{U}_{\beta f}  m_f
\bigg\{ 
(D-4) C_{2}+(D-2) C_{22}+(D-2) C_{12}
\bigg\}\,,
\end{eqnarray}
with the Passarino-Veltman functions $C_{r,s}=C_{r,s}(m_\beta^2,q^2,m_f^2,m_\beta^2,M_\gamma^2,M_W^2)$.

\begin{eqnarray}
F_L^{V (g)} &=&  \dfrac{g_w^3 }{32 \pi^2 \sqrt{2}  M_W^2}
\mathcal{U}_{\beta f} m_\beta^2 
\bigg\{ 
-2 M_W^2 C_{1}+\left(m_\beta^2-m_f^2+2 M_W^2\right) C_{0}+\left(m_\beta^2-m_f^2-2 M_W^2\right) C_{2}
\nonumber\\
&\phantom{=}&
+\left(m_f^2-m_\beta^2\right) C_{12}+\left(m_f^2-m_\beta^2\right) C_{11}-2 C_{00}
\bigg\}\,,
\end{eqnarray}
\begin{eqnarray}
F_R^{V (g)} &=&  \dfrac{g_w^3 }{32 \pi^2 \sqrt{2}  M_W^2}
\mathcal{U}_{\beta f} m_\beta\, m_f 
\bigg\{ 
\left(m_f^2-m_\beta^2\right) \big[C_{2}+C_{22}+C_{12}\big]+2 C_{00}
\bigg\}\,,
\end{eqnarray}
\begin{eqnarray}
F_L^{T (g)} &=&  i \dfrac{g_w^3 }{32 \pi^2 \sqrt{2}  M_W^2}
\mathcal{U}_{\beta f} m_\beta
\bigg\{ 
\left(-m_\beta^2-m_f^2+2 M_W^2\right) C_{2}+m_\beta^2 C_{11}-m_\beta^2 C_{0}-m_f^2 C_{22}
\nonumber\\
&\phantom{=}&
+\left(m_\beta^2-m_f^2\right) C_{12}
\bigg\}\,,
\end{eqnarray}
\begin{eqnarray}
F_R^{T (g)} &=&  i\dfrac{g_w^3 }{32 \pi^2 \sqrt{2}  M_W^2}
\mathcal{U}_{\beta f} m_\beta^2 \, m_f
\bigg\{ 
C_{0}+2 C_{2}+C_{22}-C_{11}
\bigg\}\,,
\end{eqnarray}
where the Passarino-Veltman functions are $C_{r,s}=C_{r,s}(m_\beta^2,q^2,m_f^2,m_\beta^2,M_H^2,M_W^2)$.

\mathversion{bold}
\section{Form factors for  $Z$ and Higgs leptonic decays }\label{sec:formfactors}
\mathversion{normal}
In this section we collect several expressions which are relevant for the computation of the NP contributions to key observables; in particular, we detail the form factors for the invisible $Z$ decay width, as well as those concerning the flavour conserving decays of the $Z$ and Higgs bosons
into charged lepton pairs, $Z, H \to \ell_\alpha \ell_\alpha$.
All are computed in Feynman gauge, for a UV-complete SM extension via HNL (as in the case of the ISS used in the numerical analysis in the main body of this study).

\mathversion{bold}
\subsection{LFC $H\to \ell_\alpha \ell_\alpha$ form factors}\label{app:HiggsFF}
\mathversion{normal}

\begin{figure}[h!]
    \centering
    \begin{subfigure}[b]{0.38\textwidth}
    \centering
 \raisebox{5mm}{    \begin{tikzpicture}
    \begin{feynman}
    \vertex (a) at (0,0) {\(H\)};
    \vertex (b) at (1,0);
    \vertex (c) at (2,1.);
    \vertex (d) at (2,-1);
    \vertex (e) at (3,1) {\( \ell_\alpha\)};
    \vertex (f) at (3,-1) {\( \bar\ell_\alpha\)};
    \diagram* {
    (a) -- [scalar] (b),
    (b) -- [fermion, edge label=\(n_i\)] (c),
    (c) -- [boson, edge label=\( W^\pm\)] (d),
    (d) -- [fermion, edge label=\(n_j\)] (b),
    (c) -- [fermion] (e),
    (f) -- [fermion] (d)
    };
    \end{feynman}
    \end{tikzpicture}
    }
            \caption*{(a)}
            \label{}
    \end{subfigure}
    \begin{subfigure}[b]{0.38\textwidth}
    \centering
\raisebox{5mm}{        \begin{tikzpicture}
    \begin{feynman}
    \vertex (a) at (0,0) {\(H\)};
    \vertex (b) at (1,0);
    \vertex (c) at (2,1.);
    \vertex (d) at (2,-1.);
    \vertex (e) at (3,1.) {\( \ell_\alpha\)};
    \vertex (f) at (3,-1.) {\( \bar\ell_\alpha\)};
    \diagram* {
    (a) -- [scalar] (b),
    (b) -- [boson, edge label=\( W^\pm\)] (c),
    (c) -- [anti fermion, edge label=\(n_i\)] (d),
    (d) -- [boson, edge label=\( W^\pm\)] (b),
    (c) -- [fermion] (e),
    (f) -- [fermion] (d)
    };
    \end{feynman}
    \end{tikzpicture}
    }
            \caption*{(b)}
            \label{}
    \end{subfigure}
    \caption{Feynman diagrams contributing to LFC Higgs decays (displayed for simplicity in unitary gauge).}
    \label{fig:LFCHiggsdecays:UG}
\end{figure}
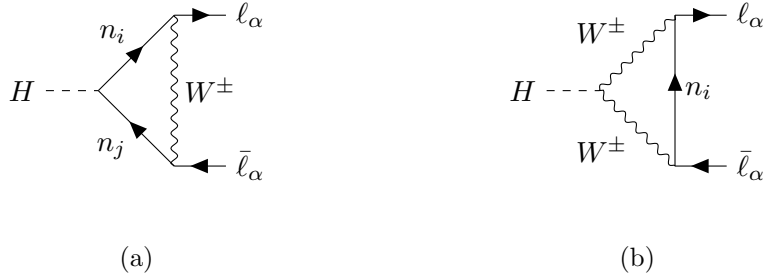
Below we provide the form factors at the origin of the one-loop corrections to the di-lepton Higgs decay, as shown in Fig.~\ref{fig:LFCHiggsdecays:UG}. 

\begin{eqnarray}
 F^{(a)}_L &=& 
\dfrac{g^3 m_\alpha}{16\pi^2 4  M_W^3}\,\sum_{i,j} \mathcal{U}_{\alpha i}\, \mathcal{U}_{\alpha j}^* \bigg\{ 
C_{ij} \big[\left(m_\alpha^2 \left(m_i^2+m_j^2\right)+m_i^2 \left((D-2) M_W^2-2 m_j^2\right)+(D-2) m_j^2 M_W^2\right) C_{1} 
\nonumber\\ &\phantom{=}&
+(D-1) m_j^2 M_W^2 C_{0} +m_j^2 B_0 \big]
\nonumber\\ &\phantom{=}&
+\,C_{ij}^* m_i m_j \big[-\left(-2 m_\alpha^2+m_i^2+m_j^2-2 (D-2) M_W^2 \right) C_{1} +(D-1) M_W^2 C_{0} +B_0 \big]
 \bigg\}\,,
\end{eqnarray}

\begin{eqnarray}
 F^{(a)}_R &=& 
\dfrac{g^3 m_\alpha}{16\pi^2 4  M_W^3}\,\sum_{i,j} \mathcal{U}_{\alpha i}\, \mathcal{U}_{\alpha j}^* \bigg\{ 
C_{ij} \big[\left(m_\alpha^2 \left(m_i^2+m_j^2\right)+m_i^2 \left((D-2) M_W^2-2 m_j^2\right)+(D-2) m_j^2 M_W^2\right) C_{2} 
\nonumber\\ &\phantom{=}&
+(D-1) m_i^2 M_W^2 C_{0} +m_i^2 B_0 \big]
\nonumber\\ &\phantom{=}&
+\,C_{ij}^* m_i m_j \big[-\left(-2 m_\alpha^2+m_i^2+m_j^2-2 (D-2) M_W^2 \right) C_{2} +(D-1) M_W^2 C_{0} +B_0 \big]
 \bigg\}\,,
\end{eqnarray}
where $C_{rs}  = C_{rs}(m_\alpha^2,q^2,m_\alpha^2,M_W^2,m_i^2,m_j^2)$ and $B_0  = B_0(q^2,m_i^2,m_j^2)$.

\begin{eqnarray}
F^{(b)}_L &=& \dfrac{g^3 m_\alpha}{16\pi^2 4  M_W^3}\,\sum_i \mathcal{U}_{\alpha i}\, \mathcal{U}_{\alpha i}^* \bigg\{ 
-\left(2 M_W^2 \left(2 m_\alpha^2-m_i^2+(D-2) M_W^2\right)+M_H^2 m_i^2\right) C_{1}
\nonumber\\ &\phantom{=}&
-\left(2 m_\alpha^2 M_W^2+M_H^2 m_i^2-2 M_W^4\right) C_{0}
-m_\alpha^2 \left(M_H^2+2 M_W^2\right) C_{2}
+2 M_W^2 B_0^\alpha-M_W^2 B_0^W
 \bigg\}\,,
\end{eqnarray}

\begin{eqnarray}
F^{(b)}_R &=& \dfrac{g^3 m_\alpha}{16\pi^2 4  M_W^3}\,\sum_i \mathcal{U}_{\alpha i}\, \mathcal{U}_{\alpha i}^* \bigg\{ 
-\left(2 M_W^2 \left(2 m_\alpha^2-m_i^2+(D-2) M_W^2\right)+M_H^2 m_i^2\right) C_{2}
\nonumber\\ &\phantom{=}&
-\left(2 m_\alpha^2 M_W^2+M_H^2 m_i^2-2 M_W^4\right) C_{0}
-m_\alpha^2 \left(M_H^2+2 M_W^2\right) C_{1} +2 M_W^2 B_0^\alpha-M_W^2 B_0^W
 \bigg\}\,,
\end{eqnarray}
with $C_{rs} = C_{rs}(m_\alpha^2,q^2,m_\alpha^2,m_i^2,M_W^2,M_W^2)$, $B_0^\alpha = B_0(m_\alpha^2,m_i^2,M_W^2)$ and 
$B_0^W = B_0(q^2,M_W^2,M_W^2)$.

\mathversion{bold}
\subsection{LFC $Z\to \ell_\alpha\ell_\alpha $ form factors}
\mathversion{normal}

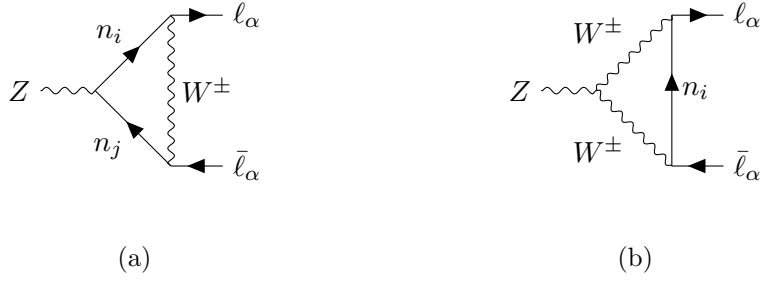
\begin{figure}[h!]
    \centering
    \begin{subfigure}[b]{0.38\textwidth}
    \centering
 \raisebox{5mm}{    \begin{tikzpicture}
    \begin{feynman}
    \vertex (a) at (0,0) {\(Z\)};
    \vertex (b) at (1,0);
    \vertex (c) at (2,1.);
    \vertex (d) at (2,-1);
    \vertex (e) at (3,1) {\( \ell_\alpha\)};
    \vertex (f) at (3,-1) {\( \bar\ell_\alpha\)};
    \diagram* {
    (a) -- [boson] (b),
    (b) -- [fermion, edge label=\(n_i\)] (c),
    (c) -- [boson, edge label=\( W^\pm\)] (d),
    (d) -- [fermion, edge label=\(n_j\)] (b),
    (c) -- [fermion] (e),
    (f) -- [fermion] (d)
    };
    \end{feynman}
    \end{tikzpicture}
    }
            \caption*{(a)}
            \label{}
    \end{subfigure}
    \begin{subfigure}[b]{0.38\textwidth}
    \centering
\raisebox{5mm}{        \begin{tikzpicture}
    \begin{feynman}
    \vertex (a) at (0,0) {\(Z\)};
    \vertex (b) at (1,0);
    \vertex (c) at (2,1.);
    \vertex (d) at (2,-1.);
    \vertex (e) at (3,1.) {\( \ell_\alpha\)};
    \vertex (f) at (3,-1.) {\( \bar\ell_\alpha\)};
    \diagram* {
    (a) -- [boson] (b),
    (b) -- [boson, edge label=\( W^\pm\)] (c),
    (c) -- [anti fermion, edge label=\(n_i\)] (d),
    (d) -- [boson, edge label=\( W^\pm\)] (b),
    (c) -- [fermion] (e),
    (f) -- [fermion] (d)
    };
    \end{feynman}
    \end{tikzpicture}
    }
            \caption*{(b)}
            \label{}
    \end{subfigure}
    \caption{Feynman diagrams contributing to LFC $Z \to \ell_\alpha \ell_\alpha$ decays (in unitary gauge).}
    \label{fig:LFCZdecays:UG}
\end{figure}

We now summarise the form factors for the flavour conserving leptonic $Z$ decays. The two contributing diagrams are presented in Fig.~\ref{fig:LFCZdecays:UG}.

\begin{eqnarray}
 F^{L\,(a)}_V &=& 
\dfrac{g^3}{16\pi^24 c_w M_W^2}\,\sum_{i,j} \mathcal{U}_{\alpha i}\, \mathcal{U}_{\alpha j}^* \bigg\{ 
 C_{ij}  \big[2 (D-2) M_W^2 C_{00} -2 M_W^2 (B_0^{i}+ B_0^j)
 \nonumber \\ &\phantom{=}&
 +m_\alpha ^2\left(2 m_\alpha ^2- m_i^2- m_j^2+8  M_W^2\right) (C_{1} + C_{2} )
 \nonumber \\ &\phantom{=}&
+\left(m_\alpha ^4-m_\alpha ^2 (m_i^2+m_j^2) +4 m_\alpha ^2 M_W^2+m_i^2 m_j^2-2 M_W^2( m_i^2+ m_j^2) -(D-6) M_W^4 +2 M_W^2 q^2\right) C_{0} 
\nonumber \\ &\phantom{=}&
+ m_\alpha ^2 \left(m_\alpha ^2+(D-2) M_W^2\right) (C_{11} +C_{22} +2 C_{12} )
-(D-6) M_W^2 B_0^{q}\big]
\nonumber \\ &\phantom{=}&
+\,C_{ij}^* m_i m_j \big[\left(3-D \right)  M_W^2 C_{0} -m_\alpha ^2  (C_{11} +C_{22} + 2  C_{12} ) -2  C_{00} + B_0^{q}\big]
 \bigg\}\,,
\end{eqnarray}

\begin{eqnarray}
F^{R\,(a)}_V  &=& \dfrac{g^3 m_\alpha^2 }{16\pi^2 4 c_w M_W^2}\,\sum_{ij} \mathcal{U}_{\alpha i}\, \mathcal{U}_{\alpha j}^* \bigg\{ 
C_{ij}  \big[(D-3) M_W^2 C_{0} + 2 (D -2) M_W^2  (C_{1} +C_{2} )
\nonumber \\ &\phantom{=}&
+\left(m_\alpha ^2 + (D-2) M_W^2\right) (C_{11}  + C_{22}  + 2C_{12} )
+2 C_{00} -B_0^{q}\big]
\nonumber \\ &\phantom{=}&
- \,C_{ij}^* m_i m_j \big[C_{11} + C_{22} +2 C_{12}  \big]
 \bigg\} \,,
\end{eqnarray}

\begin{eqnarray}
 F^{L\,(a)}_T &=& i\dfrac{g^3 m_\alpha}{16\pi^2 4 c_w M_W^2}\,\sum_{ij} \mathcal{U}_{\alpha i}\, \mathcal{U}_{\alpha j}^* \bigg\{ 
 C_{ij}  \big[\left(-m_\alpha ^2+m_j^2-2 M_W^2\right) C_{2} -D M_W^2 C_{1} 
 -\left(m_\alpha ^2+(D-2) M_W^2 \right) C_{12} 
 \nonumber \\ &\phantom{=}& 
 +\left(2 -D \right)M_W^2 C_{11} -2 M_W^2 C_{0}  -m_\alpha ^2 C_{22} \big]
 \nonumber \\ &\phantom{=}&
 +\,C_{ij}^* m_i m_j  ( C_{12} + C_{11} )
 \bigg\} \,,
\end{eqnarray}

\begin{eqnarray}
F^{R\,(a)}_T &=& i \dfrac{g^3 m_\alpha}{16\pi^2 4 c_w M_W^2}\,\sum_{ij} \mathcal{U}_{\alpha i}\, \mathcal{U}_{\alpha j}^* \bigg\{ 
 C_{ij}  \big[\left(-m_\alpha ^2+m_i^2-2 M_W^2\right) C_{1} -D M_W^2 C_{2} +\left(2 -D\right)M_W^2 C_{22} 
 \nonumber \\ &\phantom{=}&
 -\left(m_\alpha ^2+(D-2) M_W^2 \right) C_{12} -2 M_W^2 C_{0}  -m_\alpha ^2 C_{11} \big]
 \nonumber \\ &\phantom{=}&
 +\,C_{ij}^* m_i m_j  \big[ C_{22} +  C_{12} \big]
 \bigg\} \,,
\end{eqnarray}

\begin{eqnarray}
 F^{L\,(a)}_S &=& \dfrac{g^3 m_\alpha}{16\pi^2 4 c_w M_W^2}\,\sum_{ij} \mathcal{U}_{\alpha i}\, \mathcal{U}_{\alpha j}^* \bigg\{ 
C_{ij}  \big[\left(-m_\alpha ^2+m_j^2-2 M_W^2\right) C_{2} +\left(D-4 \right)M_W^2 C_{1} 
\nonumber \\ &\phantom{=}&
+\left(m_\alpha ^2-(D-2) M_W^2\right) C_{12} +\left(D -2 \right) M_W^2 C_{11}  -2 M_W^2 C_{0}  - m_\alpha ^2 C_{22} \big]
\nonumber \\ &\phantom{=}&
+\,C_{ij}^* m_i m_j \big[ C_{12} -C_{11} \big]
 \bigg\} \,,
\end{eqnarray}

\begin{eqnarray}
 F^{R\,(a)}_S &=& \dfrac{g^3 m_\alpha}{16\pi^2 4 c_w M_W^2}\,\sum_{ij} \mathcal{U}_{\alpha i}\, \mathcal{U}_{\alpha j}^* \bigg\{ 
 C_{ij}  \big[\left(m_\alpha ^2-m_i^2+2 M_W^2\right) C_{1} +(4-D) M_W^2 C_{2} +\left(2 -D \right)M_W^2 C_{22} 
 \nonumber \\ &\phantom{=}&
 +\left(-m_\alpha ^2+(D-2) M_W^2\right) C_{12} +2 M_W^2 C_{0} +m_\alpha ^2 C_{11} \big]
 \nonumber \\ &\phantom{=}&
 +\,C_{ij}^* m_i m_j  \left(C_{22} - C_{12} \right)
 \bigg\} \,,
\end{eqnarray}
with the Passarino-Veltman functions 
$C_{rs} = C_{rs}(m_\alpha^2, q^2,m_\alpha^2,M_W^2,m_i^2,m_j^2)$.

\begin{eqnarray}
 F^{L\,(b)}_V  &=& \dfrac{g^3}{16\pi^2 4 c_w M_W^2}\,\sum_i \mathcal{U}_{\alpha i}\, \mathcal{U}_{\alpha i}^* \bigg\{ 
 \left(4(D-2) c_w^2  M_W^2-2 m_i^2(s_w^2-c_w^2)\right) C_{00} 
 \nonumber \\ &\phantom{=}&
 +C_{0}  \left(-8 c_w  ^2 m_\alpha ^2 M_W^2-4 c_w  ^2 m_i ^2 M_W^2+8 c_w  ^2 M_W^4+4 c_w   m_i ^2 M_W M_Z  s_w  ^2 -2 m_\alpha ^2 m_i ^2 (s_w^2 - c_w^2)\right)
  \nonumber \\ &\phantom{=}&
+(C_{2}  +C_{1} )m_\alpha^2 \left(2(D-8) c_w  ^2  M_W^2 +4 c_w  M_W M_Z  s_w  ^2 -m_\alpha ^2 (s_w^2-c_w^2)-3  m_i ^2 (s_w^2-c_w^2)\right)
  \nonumber \\ &\phantom{=}&
+(C_{11}  + C_{22} )m_\alpha ^2 \left( 2(D-2) c_w  ^2  M_W^2 -m_\alpha ^2 (s_w^2-c_w^2)- m_i ^2 (s_w^2-c_w^2)\right)
\nonumber \\ &\phantom{=}&
+C_{12}  m_\alpha ^2 \left(4(D-2) c_w  ^2  M_W^2 -2 m_\alpha ^2(s_w^2-c_w^2)-2  m_i ^2(s_w^2-c_w^2)\right)
\nonumber \\ &\phantom{=}&-4 c_w  ^2 M_W^2 B_0^W +8 c_w  ^2 M_W^2 B_0^\alpha
 \bigg\}\,,
\end{eqnarray}

\begin{eqnarray}
 F^{R\,(b)}_V &=& \dfrac{g^3 m_\alpha^2}{16\pi^2 4 c_w M_W^2}\,\sum_i \mathcal{U}_{\alpha i}\, \mathcal{U}_{\alpha i}^* \bigg\{ 
 (C_{1} + C_{2} ) \left[2(D-2) c_w  ^2   M_W^2 -(s_w^2-c_w^2) \left(m_\alpha ^2+3 m_i ^2\right)\right]
 \nonumber \\ &\phantom{=}&
 + ( C_{11} + C_{22}   +  2 C_{12} ) \left(2(D-2) c_w  ^2  M_W^2 -(m_\alpha ^2+m_i ^2) (s_w^2-c_w^2)\right)
 \nonumber \\ &\phantom{=}&
 +2   m_i ^2 \left(c_w  ^2-s_w  ^2\right) C_{0} +  \left(2 c_w  ^2-2 s_w  ^2\right) C_{00} 
 \bigg\} \,,
\end{eqnarray}

\begin{eqnarray}
 F^{L\,(b)}_T  &=& i\dfrac{g^3 m_\alpha}{16\pi^2 4 c_w M_W^2}\,\sum_i \mathcal{U}_{\alpha i}\, \mathcal{U}_{\alpha i}^* \bigg\{ 
 C_{2}  \left(2 c_w  ^2 M_W^2 -2 c_w   M_W M_Z  s_w  ^2
 +(s_w^2-c_w^2) \left(m_\alpha ^2+m_i ^2\right)\right)
 \nonumber \\ &\phantom{=}&
 +\left(-2(D-4) c_w^2 M_W^2 +2 m_i ^2 (s_w^2-c_w^2)\right) C_{1} 
 \nonumber \\ &\phantom{=}&
 +C_{12}  \left( -2(D-2) c_w  ^2  M_W^2 + (m_\alpha ^2 +m_i ^2) (s_w^2-c_w^2)\right)
 \nonumber \\ &\phantom{=}&
 +\left(-2 (D -2)c_w  ^2  M_W^2 +m_i ^2 (s_w^2-c_w^2)\right) C_{11} 
 +m_i ^2 \left(s_w  ^2-c_w  ^2\right) C_{0} + m_\alpha ^2\left(s_w  ^2-c_w  ^2 \right) C_{22} 
 \bigg\} \,,\nonumber \\
\end{eqnarray}

\begin{eqnarray}
 F^{R\,(b)}_T &=& i \dfrac{g^3 m_\alpha }{16\pi^2 4 c_w M_W^2}\,\sum_i \mathcal{U}_{\alpha i}\, \mathcal{U}_{\alpha i}^* \bigg\{ 
C_{1}^W \left(2 c_w ^2 M_W^2-2 c_w  M_W M_Z s_w ^2+m_\alpha ^2(s_w^2-c_w^2)+m_i^2 (s_w^2-c_w^2)\right)
\nonumber \\ &\phantom{=}&
-2 \left(c_w ^2(D-4) M_W^2 -m_i^2 (s_w^2-c_w^2)\right) C_{2}^W
+\left(-2(D-2) c_w ^2  M_W^2 +m_i^2 (s_w ^2 - c_w^2)\right) C_{22}^W
\nonumber \\ &\phantom{=}&
+C_{12}^W \left(-2(D-2) c_w ^2  M_W^2 +m_\alpha ^2 (s_w ^2-c_w ^2)+m_i^2 (s_w ^2-c_w^2)\right)-m_i^2 \left(c_w ^2-s_w ^2\right) C_{0}^W
\nonumber \\ &\phantom{=}&
+m_\alpha ^2\left( s_w ^2-c_w ^2 \right) C_{11}^W
\bigg\} \,,
\end{eqnarray}

\begin{eqnarray}
 F^{L\,(b)}_S &=& \dfrac{g^3 m_\alpha}{16\pi^2 4 c_w M_W^2}\,\sum_i \mathcal{U}_{\alpha i}\, \mathcal{U}_{\alpha i}^* \bigg\{ \left(2c_w  ^2  M_W^2-2 c_w   M_W M_Z  s_w  ^2+m_i ^2 (s_w^2-c_w^2)\right) C_{2} 
 \nonumber \\ &\phantom{=}&
 +C_{12}  \left( -2 (D-2) c_w  ^2  M_W^2 -m_\alpha ^2 \left(s_w  ^2-c_w  ^2\right)+m_i ^2 \left(s_w  ^2-c_w  ^2\right)\right)
 \nonumber \\ &\phantom{=}&
 +\left(2(D-2) c_w  ^2 M_W^2 -m_i ^2 \left(s_w  ^2-c_w  ^2\right)\right) C_{11} 
 \nonumber \\ &\phantom{=}&+\left(4 c_w  ^2 M_W^2-m_i ^2 (s_w^2-c_w^2)\right) C_{1} +m_\alpha ^2 \left(s_w  ^2-c_w  ^2\right) C_{22} 
 \bigg\}\,,
\end{eqnarray}

\begin{eqnarray}
 F^{R\,(b)}_S &=& \dfrac{g^3 m_\alpha}{16\pi^2 4 c_w M_W^2}\,\sum_i \mathcal{U}_{\alpha i}\, \mathcal{U}_{\alpha i}^* \bigg\{ 
 \left(-2 c_w  ^2 M_W^2+2 c_w   M_W M_Z  s_w  ^2-m_i ^2 (s_w^2-c_w^2)\right) C_{1} 
 \nonumber \\ &\phantom{=}&
 +\left(m_i ^2 (s_w^2-c_w^2) -2c_w  ^2 (D-2) M_W^2\right) C_{22} 
 \nonumber \\ &\phantom{=}&
 +C_{12}  \left( 2(D-2) c_w  ^2  M_W^2 +m_\alpha ^2 (s_w^2-c_w^2) -m_i ^2 (s_w^2-c_w^2)\right)
 \nonumber \\ &\phantom{=}&
 +\left(m_i ^2(s_w^2-c_w^2) -4 c_w  ^2  M_W^2\right)C_{2} +m_\alpha ^2 \left(c_w^2  -s_w^2\right) C_{11} 
 \bigg\} \,,
\end{eqnarray}
with the following Passarino-Veltman functions: $C_{rs}  = C_{rs}(m_\alpha^2, q^2,m_\alpha^2,M_i^2,M_W^2,M_W^2)$, 
$B_0^q = B_0(q^2,m_i^2,m_j^2)$, 
$B_0^W = B_0(q^2,m_W^2,m_W^2)$, 
$B_0^\alpha= B_0(m_\alpha^2,m_i^2,m_W^2)$ and 
$B_0^j= B_0(m_\alpha^2,m_W^2,m_j^2)$.

\mathversion{bold}
\subsection{Invisible $Z$ decays}
\mathversion{normal}

\begin{figure}[h!]
    \centering
    \begin{subfigure}[b]{0.24\textwidth}
    \centering
 \raisebox{5mm}{    \begin{tikzpicture}
    \begin{feynman}
    \vertex (a) at (0,0) {\(Z\)};
    \vertex (b) at (1,0);
    \vertex (c) at (2,1.);
    \vertex (d) at (2,-1);
    \vertex (e) at (3,1) {\( n_a\)};
    \vertex (f) at (3,-1) {\( n_b\)};
    \diagram* {
    (a) -- [boson] (b),
    (b) -- [ edge label=\(n_i\)] (c),
    (c) -- [boson, edge label=\( Z\)] (d),
    (d) -- [edge label=\(n_j\)] (b),
    (c) --  (e),
    (f) --  (d)
    };
    \end{feynman}
    \end{tikzpicture}
    }
            \caption*{(a)}
            \label{}
    \end{subfigure}
    \hfill
    \begin{subfigure}[b]{0.24\textwidth}
    \centering
 \raisebox{5mm}{  \begin{tikzpicture}
    \begin{feynman}
    \vertex (a) at (0,0) {\(Z\)};
    \vertex (b) at (1,0);
    \vertex (c) at (2,1.);
    \vertex (d) at (2,-1);
    \vertex (e) at (3,1) {\( n_a\)};
    \vertex (f) at (3,-1) {\( n_b\)};
    \diagram* {
    (a) -- [boson] (b),
    (b) -- [ edge label=\(n_i\)] (c),
    (c) -- [scalar, edge label=\( h\)] (d),
    (d) -- [edge label=\(n_j\)] (b),
    (c) --  (e),
    (f) --  (d)
    };
    \end{feynman}
    \end{tikzpicture}
    }
            \caption*{(b)}
            \label{}
    \end{subfigure}
    \hfill
    \begin{subfigure}[b]{0.47\textwidth}
    \centering
 \raisebox{5mm}{    \begin{tikzpicture}
     \hspace*{-3.5mm}  \begin{feynman}
    \vertex (a) at (0,0) {\(Z\)};
    \vertex (b) at (1,0);
    \vertex (c) at (2,1.);
    \vertex (d) at (2,-1);
    \vertex (e) at (3,1) {\( n_a\)};
    \vertex (f) at (3,-1) {\( n_b\)};
    \diagram* {
    (a) -- [boson] (b),
    (b) -- [fermion, edge label=\(\ell\)] (c),
    (c) -- [boson, edge label=\( W^-\)] (d),
    (d) -- [fermion, edge label=\(\ell\)] (b),
    (c) --  (e),
    (f) --  (d)
    };
    \end{feynman}
    \end{tikzpicture}
    }
 \raisebox{5mm}{    \begin{tikzpicture}
    \begin{feynman}
    \vertex (a) at (0,0) {\(Z\)};
    \vertex (b) at (1,0);
    \vertex (c) at (2,1.);
    \vertex (d) at (2,-1);
    \vertex (e) at (3,1) {\( n_a\)};
    \vertex (f) at (3,-1) {\( n_b\)};
    \diagram* {
    (a) -- [boson] (b),
    (b) -- [anti fermion, edge label=\(\ell\)] (c),
    (c) -- [boson, edge label=\( W^+\)] (d),
    (d) -- [anti fermion, edge label=\(\ell\)] (b),
    (c) --  (e),
    (f) --  (d)
    };
    \end{feynman}
    \end{tikzpicture}
    }
            \caption*{(c)}
            \label{}
    \end{subfigure}
    \hfill
\\
\vspace{8mm}
    \begin{subfigure}[b]{0.49\textwidth}
    \centering
 \raisebox{5mm}{  \hspace{0.3mm}  \begin{tikzpicture}
    \begin{feynman}
    \vertex (a) at (0,0) {\(Z\)};
    \vertex (b) at (1,0);
    \vertex (c) at (2,1.);
    \vertex (d) at (2,-1.);
    \vertex (e) at (3,1.) {\( n_a\)};
    \vertex (f) at (3,-1.) {\( n_b\)};
    \diagram* {
    (a) -- [boson] (b),
    (b) -- [boson, edge label=\( W^-\)] (c),
    (c) -- [anti fermion, edge label=\(\ell\)] (d),
    (d) -- [boson, edge label=\( W^+\)] (b),
    (c) --  (e),
    (f) --  (d)
    };
    \end{feynman}
    \end{tikzpicture}
    }
    \hfill
\raisebox{5mm}{        \begin{tikzpicture}
    \begin{feynman}
    \vertex (a) at (0,0) {\(Z\)};
    \vertex (b) at (1,0);
    \vertex (c) at (2,1.);
    \vertex (d) at (2,-1.);
    \vertex (e) at (3,1.) {\( n_a\)};
    \vertex (f) at (3,-1.) {\( n_b\)};
    \diagram* {
    (a) -- [boson] (b),
    (b) -- [boson, edge label=\( W^+ \)] (c),
    (c) -- [fermion, edge label=\(\ell\)] (d),
    (d) -- [boson, edge label=\( W^-\)] (b),
    (c) --(e),
    (f) --  (d)
    };
    \end{feynman}
    \end{tikzpicture}
    }
            \caption*{(d)}
            \label{}
    \end{subfigure}
\hfill
    \begin{subfigure}[b]{0.24\textwidth}
    \centering
 \raisebox{5mm}{    \begin{tikzpicture}
    \begin{feynman}
    \vertex (a) at (0,0) {\(Z\)};
    \vertex (b) at (1,0);
    \vertex (c) at (2,1.);
    \vertex (d) at (2,-1.);
    \vertex (e) at (3,1.) {\( n_a\)};
    \vertex (f) at (3,-1.) {\( n_b\)};
    \diagram* {
    (a) -- [boson] (b),
    (b) -- [boson, edge label=\( Z\)] (c),
    (c) -- [anti fermion, edge label=\(n_i\)] (d),
    (d) -- [scalar, edge label=\( h\)] (b),
    (c) --  (e),
    (f) --  (d)
    };
    \end{feynman}
    \end{tikzpicture}
    }
            \caption*{(e)}
            \label{}
    \end{subfigure}
        \hfill
    \begin{subfigure}[b]{0.24\textwidth}
    \centering
\raisebox{5mm}{        \begin{tikzpicture}
    \begin{feynman}
    \vertex (a) at (0,0) {\(Z\)};
    \vertex (b) at (1,0);
    \vertex (c) at (2,1.);
    \vertex (d) at (2,-1.);
    \vertex (e) at (3,1.) {\( n_a\)};
    \vertex (f) at (3,-1.) {\( n_b\)};
    \diagram* {
    (a) -- [boson] (b),
    (b) -- [scalar, edge label=\( h \)] (c),
    (c) -- [anti fermion, edge label=\(n_i\)] (d),
    (d) -- [boson, edge label=\( Z\)] (b),
    (c) -- (e),
    (f) -- (d)
    };
    \end{feynman}
    \end{tikzpicture}
    }
            \caption*{(f)}
            \label{}
    \end{subfigure}
    \caption{Feynman diagrams contributing to the invisible $Z$ decays (in unitary gauge).}
    \label{fig:Zinvdecays:UG}
\end{figure}

In Fig.~\ref{fig:Zinvdecays:UG} we display the diagrams contributing to the invisible $Z$ decays, $Z \to n_a n_b$ (with $a=b$ or $a\neq b$).
Below we present the vector form factors contributing to the invisible $Z$-decays (we do not give the associated tensor and scalar form factors since the former are negligible and the latter vanish for on-shell decays); the superscript labels (a)-(f) refer to the topologies  depicted in Fig.~\ref{fig:Zinvdecays:UG}. 
\begin{eqnarray}
F_L^{V \, (a)} &=& \dfrac{g_w^3 }{16\pi^2 8 c_w^3 M_W^2 } \sum_{i,j}
\bigg\{
C_{ai}^* \bigg[C_{ij} \big[C_{jb} m_a m_i \bigg(c_w^2 m_b^2 (C_{22} +2  C_{12}+ C_{11})
+2  M_W^2 (C_{0}+C_{2}+ C_{1})
\nonumber\\ &\phantom{=}&
+ \left(m_b^2-m_j^2\right) c_w^2 (C_{2}+ C_{1})\bigg)
+C_{jb}^* m_a m_b m_i m_j c_w^2 \left( C_{22} +2  C_{12}+ C_{11}\right)\big]
\nonumber\\ &\phantom{=}&
+C_{ij}^* \big[C_{jb}^* m_a m_b\left( (D-2)c_w^2 C_{00} 
- (D-2)M_W^2 (C_{0}+ 2 C_{1} +2 C_{2} +C_{11}+C_{22} + 2 C_{12})
- q^2 c_w^2 C_{12}\right)
\nonumber\\ &\phantom{=}&
+C_{jb} m_a m_j \left((D-2)c_w^2 C_{00} 
-(D-2)  M_W^2 ( C_{0}+ C_{1}+ C_{2})- q^2 c_w^2 C_{12} 
\right)\big]\bigg]
\nonumber\\ &\phantom{=}&
+C_{ai} \bigg[C_{ij}^* \big[C_{jb}^* m_b m_i
\left((D-2)c_w^2 C_{00} 
-(D-2)M_W^2 (C_{0}+ C_{1}+ C_{2}) - q^2 c_w^2 C_{12}  \right)
\nonumber\\ &\phantom{=}&
+C_{jb} m_i m_j\left((D-2) c_w^2 C_{00} - (D-2) M_W^2 C_{0} - q^2 c_w^2 C_{12}\right)\big]
\nonumber\\ &\phantom{=}&
+C_{ij} \big[C_{jb}^* m_b m_j \left(
c_w^2  m_a^2 (C_{1} + C_{2} + C_{11} + C_{22} + 2 C_{12})
+2  M_W^2 (C_{0} +C_{1} + C_{2})
-m_i^2 c_w^2(C_{1} + C_{2}) \right)
\nonumber\\ &\phantom{=}&
+C_{jb} \bigg(\left(c_w^2 m_b^2+M_W^2 (D-2)\right) C_{11} m_a^2
-M_W^2 \left(2 B_0^{aZi}+2 B_0^{bZj}+(D-6) B_0^{qij}\right)
\nonumber\\ &\phantom{=}&
+\left(\left(m_a^2-m_i^2\right) \left(m_b^2-m_j^2\right) c_w^2+M_W^2 \left(2 m_a^2+2 m_b^2-2 m_i^2-2 m_j^2-(D+6) M_Z^2+2 q^2\right)\right) C_{0}
\nonumber\\ &\phantom{=}&
+\left(2 \left(m_a^2+3 m_b^2\right) M_W^2-c_w^2 \left(\left(m_j^2-2 m_b^2\right) m_a^2+m_b^2 m_i^2\right)\right) C_{2}
+m_b^2 \left(c_w^2 m_a^2+M_W^2 (D-2)\right) C_{22}
\nonumber\\ &\phantom{=}&
+\left(2 \left(3 m_a^2+m_b^2\right) M_W^2-c_w^2 \left(\left(m_j^2-2 m_b^2\right) m_a^2+m_b^2 m_i^2\right)\right) C_{1}
\nonumber\\ &\phantom{=}&
+\left(2 c_w^2 m_a^2 m_b^2+\left(m_a^2+m_b^2\right) M_W^2 (D-2)\right) C_{12}+2 M_W^2 (D-2) C_{00}\bigg)\big]\bigg]
\bigg\}\,,
\end{eqnarray}

\begin{eqnarray}
F_R^{V \, (a)} &=& \dfrac{g_w^3 }{16\pi^2 8 c_w^3 M_W^2 } \sum_{i,j}
\bigg\{
C_{ai} \bigg[C_{ij}^* \big[
-C_{jb} m_a m_b m_i m_j c_w^2\left(C_{11} + C_{22}+2  C_{12}\right)
\nonumber\\ &\phantom{=}&
+C_{jb}^* m_a m_i
\big(-c_w^2 m_b^2 (C_{1}+ C_{2}+ C_{22}+ C_{11}+2 C_{12} ) -2  M_W^2 (C_{0}+C_{1}+ C_{2})
+c_w^2 m_j^2(C_{1}+ C_{2})\big)
\big]
\nonumber\\ &\phantom{=}&
+C_{ij} \big[C_{jb} m_a m_b\left(q^2 c_w^2 C_{12} -(D-2)  C_{00} c_w^2
+  (D-2)M_W^2 (C_{0}+2C_{2}+2C_{1}+ C_{11} + C_{22} +2 C_{12})\right)
\nonumber\\ &\phantom{=}&
+C_{jb}^* m_a m_j\left( q^2 c_w^2 C_{12} -(D-2) c_w^2 C_{00} + (D-2) M_W^2 (C_{0}+C_{2}+ C_{1})\right)\big]\bigg]
\nonumber\\ &\phantom{=}&
+C_{ai}^* \bigg[C_{ij} \big[C_{jb} m_b m_i \left( q^2 C_{12} c_w^2-(D-2) c_w^2 C_{00} + (D-2) M_W^2 (C_{0}+ C_{2}+ C_{1})\right)
\nonumber\\ &\phantom{=}&
+C_{jb}^* m_i m_j\left( q^2 C_{12} c_w^2-(D-2)c_w^2 C_{00} +  (D-2) M_W^2C_{0}\right)\big]
\nonumber\\ &\phantom{=}&
+C_{ij}^* \big[C_{jb} m_b m_j \left(-c_w^2  m_a^2 (C_{1}+ C_{2}+ C_{11}+ C_{22} +2 C_{12}) -2  M_W^2 (C_{0}+C_{1}+ C_{2}) + c_w^2 m_i^2 (C_{1}+ C_{2}) \right)
\nonumber\\ &\phantom{=}&
+C_{jb}^* \big(M_W^2 \left(2 B_0^{aZi}+2 B_0^{bZj}+(D-6) B_0^{qij}\right) -\left(c_w^2 m_b^2+M_W^2 (D-2)\right) C_{11} m_a^2
\nonumber\\ &\phantom{=}&
+\left(M_W^2 \left(-2 m_a^2-2 m_b^2+2 m_i^2+2 m_j^2+(D-6) M_Z^2-2 q^2\right)-c_w^2 \left(m_a^2-m_i^2\right) \left(m_b^2-m_j^2\right)\right) C_{0}
\nonumber\\ &\phantom{=}&
+\left(c_w^2 \left(\left(m_j^2-2 m_b^2\right) m_a^2+m_b^2 m_i^2\right)-2 \left(m_a^2+3 m_b^2\right) M_W^2\right) C_{2}
\nonumber\\ &\phantom{=}&
-m_b^2 \left(c_w^2 m_a^2+M_W^2 (D-2)\right) C_{22}+\left(c_w^2 \left(\left(m_j^2-2 m_b^2\right) m_a^2+m_b^2 m_i^2\right)-2 \left(3 m_a^2+m_b^2\right) M_W^2\right) C_{1}
\nonumber\\ &\phantom{=}&
+\left(-2 c_w^2 m_a^2 m_b^2-(D-2) \left(m_a^2+m_b^2\right) M_W^2\right) C_{12}-2 (D-2) M_W^2 C_{00}\big)\big]\bigg]
\bigg\}\,,
\end{eqnarray}
with the Passarino-Veltman functions $C_{rs}= C_{rs}(m_a^2, q^2, m_b^2, M_Z^2, m_i^2, m_j^2)$.

\begin{eqnarray}
F_L^{V \, (b)} &=& \dfrac{g_w^3 }{16\pi^2 8 c_w^3 M_W^2 } \sum_{i,j}
\bigg\{
C_{ai}^* \bigg[C_{ij} \big[
C_{jb}^* m_a m_b m_i m_j c_w^2 \left(4 C_{0} +4  C_{1} +4  C_{2}  + C_{11}+ C_{22} +2  C_{12} \right)
\nonumber\\ &\phantom{=}&
+C_{jb} 
m_a m_i c_w^2 \left(m_j^2 (2C_{0}+ C_{1} + C_{2} )
+m_b^2 (2C_{0} +3C_{1} + 3C_{2}+ C_{11} + C_{22} +2 C_{12} )\right)\big]
\nonumber\\ &\phantom{=}& 
+C_{ij}^* \big[C_{jb}^*  m_a m_b c_w^2 \left( (D-2)C_{00}- q^2 C_{12}\right)
+C_{jb} m_a m_j c_w^2 \left( (D-2) C_{00}-q^2 C_{12}\right)\big]\bigg]
\nonumber\\ &\phantom{=}&
+C_{ai} \bigg[C_{ij} \big[C_{jb} c_w^2 \big(
m_a^2 m_b^2 (C_{0} +2C_{1} +2C_{2} + C_{11}+ C_{22} +2C_{12})
\nonumber\\ &\phantom{=}&
+ \left(m_i^2m_b^2+ m_a^2m_j^2\right) ( C_{0} + C_{1}+ C_{2} ) +m_i^2m_j^2 C_{0}\big)
\nonumber\\ &\phantom{=}&
+C_{jb}^* m_b m_j c_w^2\big( m_i^2(2 C_{0} +  C_{1}+  C_{2})    +m_a^2   (2C_{0} + 3C_{1}+3C_{2} + C_{11}+C_{22} +2C_{12}) \big)\big] 
\nonumber\\ &\phantom{=}&
+C_{ij}^* \big[C_{jb}^* m_b m_i c_w^2\left((D-2) C_{00}- q^2 C_{12}\right)
+C_{jb} m_i m_j c_w^2  \left((D-2) C_{00}- q^2 C_{12}\right)\big] \bigg]
\bigg\}\,,
\end{eqnarray}

\begin{eqnarray}
F_R^{V \, (b)} &=& \dfrac{g_w^3 }{16\pi^2 8 c_w^3 M_W^2 } \sum_{i,j}
\bigg\{
C_{ai} \bigg[C_{ij}^* \big[
-C_{jb} m_a m_b m_i m_j c_w^2 \left(4  C_{0}+4 C_{1} +4  C_{2} +C_{11} + C_{22} +2 C_{12}  \right)
\nonumber\\ &\phantom{=}&
-C_{jb}^* m_a m_i c_w^2 \left(m_j^2 (2C_{0}+C_{1}+ C_{2}) + m_b^2 (2C_{0} +3C_{1}+3C_{2} + 2 C_{12} + C_{11} +C_{22})\right)
\big]
\nonumber\\ &\phantom{=}&
+C_{ij} \big[C_{jb} m_a m_b c_w^2\left(q^2 C_{12}- (D-2)  C_{00}\right)+C_{jb}^* m_a m_j c_w^2\left( q^2 C_{12}- (D-2)  C_{00}\right)\big]\bigg]
\nonumber\\ &\phantom{=}&
+C_{ai}^* \bigg[C_{ij}^* \big[C_{jb}^* 
c_w^2\big( 
- m_a^2 m_b^2 (C_{0} +2C_{1} +2C_{2}+C_{11} +C_{22}+ 2  C_{12})
\nonumber\\ &\phantom{=}&
-\left(m_j^2m_a^2 +m_b^2 m_i^2\right) (C_{0}+ C_{1} +C_{2}) - m_i^2m_j^2  C_{0}\big)
\nonumber\\ &\phantom{=}&
+C_{jb} m_b c_w^2\left(- m_i^2m_j   (2C_{0} +C_{1} +C_{2})
- m_a^2  m_j (2C_{0}+ 3C_{1} +3C_{2}+C_{11}+C_{22} +2 C_{12})
\right)\big]
\nonumber\\ &\phantom{=}&
+C_{ij} \big[C_{jb}  m_b m_i c_w^2\left(q^2 C_{12}- (D-2)  C_{00}\right)+C_{jb}^* m_i m_j c_w^2\left( q^2 C_{12}-(D-2)  C_{00}\right)\big]\bigg]
\bigg\}\,,
\end{eqnarray}
where $ C_{rs}= C_{rs}(m_a^2, q^2, m_b^2, M_H^2, m_i^2, m_j^2)$.

\begin{eqnarray}
F_L^{V \, (c)} &=& \dfrac{g_w^3 }{16\pi^2 2 c_w M_W^2 } \sum_{\rho}
\bigg\{
m_a m_b \mathcal{U}_{\rho a} \mathcal{U}_{\rho b}^* \bigg[
-2 (g_A+g_V) C_{00} +(g_A +g_V) B_0
\nonumber\\ &\phantom{=}&
-\left(g_A \left(m_b^2-m_\rho^2+(D-2) M_W^2\right)+g_V \left(m_b^2+m_\rho^2+(D-2) M_W^2\right)\right) C_{22}
\nonumber\\ &\phantom{=}&
-\left(g_A \left(m_a^2+m_b^2-2 m_\rho^2+2 (D-2) M_W^2\right)+g_V \left(m_a^2+m_b^2+2 m_\rho^2+2(D-2) M_W^2\right)\right)C_{12} 
\nonumber\\ &\phantom{=}&
-\left(g_A \left(m_a^2-m_\rho^2+(D-2) M_W^2\right)+g_V \left(m_a^2+m_\rho^2+(D-2) M_W^2\right)\right) C_{11}
\nonumber\\ &\phantom{=}&
-(D-3) M_W^2 (g_A+g_V) C_{0}-2 (D-2) M_W^2 (g_A+g_V) (C_{1}+C_{2})
\bigg]
\nonumber\\ &\phantom{=}&
+\mathcal{U}_{\rho a}^* \mathcal{U}_{\rho b} 
\bigg[ -2 M_W^2 (g_A+g_V) (B_0^{a}+B_0^{b})
 \nonumber\\ &\phantom{=}&
+\big((g_A -g_V)m_\rho^2 - (g_A + g_V)(D-6) M_W^2 \big) B_0
 \nonumber\\ &\phantom{=}&
+ m_a^2 \left(g_A \left(m_b^2-m_\rho^2+(D-2) M_W^2\right)+g_V \left(m_b^2+m_\rho^2+(D-2) M_W^2\right)\right) C_{11}
 \nonumber\\ &\phantom{=}&
 +m_b^2 \left(g_A \left(m_a^2-m_\rho^2+(D-2) M_W^2\right)+g_V \left(m_a^2+m_\rho^2+(D-2) M_W^2\right)\right) C_{22}
 \nonumber\\ &\phantom{=}&
 +\big(g_A [m_a^2 \left(2 m_b^2-m_\rho^2+(D-2) M_W^2\right)- m_b^2 \left(m_\rho^2-(D-2) M_W^2\right)] 
 \nonumber\\ &\phantom{=}&
 + g_V [m_a^2 \left(2 m_b^2+m_\rho^2+(D-2) M_W^2\right)+m_b^2 \left(m_\rho^2+(D-2) M_W^2\right)]\big)C_{12} 
 \nonumber\\ &\phantom{=}&
 +2 \left(-g_A (m_\rho^2- (D-2) M_W^2) +g_V \left(m_\rho^2+(D-2) M_W^2\right)\right) C_{00}
 \nonumber\\ &\phantom{=}&
 +(g_A+g_V) \left(m_a^2 \left(2 m_b^2-m_\rho^2+2 M_W^2\right)-m_b^2 \left(m_\rho^2-6 M_W^2\right)\right) C_{2}
 \nonumber\\ &\phantom{=}&
 +(g_A+g_V) \left(m_a^2 \left(2 m_b^2-m_\rho^2+6 M_W^2\right)-m_b^2 \left(m_\rho^2-2 M_W^2\right)\right) C_{1}
 \nonumber\\ &\phantom{=}&
 +C_{0} \big( -g_A (D+1) m_\rho^2 M_W^2 +g_V (D-7) m_\rho^2 M_W^2 
 \nonumber\\ &\phantom{=}&
 +(g_A+g_V) \big[
 m_\rho^2(m_\rho^2-m_a^2 - m_b^2)
 + 2 M_W^2( m_a^2 + m_b^2 + q^2)
 + m_a^2 m_b^2 
 -(D-6) M_W^4 \big]\big)
\bigg]
\bigg\}\,, \nonumber \\
\end{eqnarray}

\begin{eqnarray}
F_R^{V \, (c)} &=& \dfrac{g_w^3 }{16\pi^2 2 c_w M_W^2 } \sum_{\rho}
\bigg\{
m_a m_b \mathcal{U}_{\rho a}^* \mathcal{U}_{\rho b} 
\bigg[-(g_A +g_V) B_0
\nonumber\\ &\phantom{=}&
+\left(g_A \left(m_b^2-m_\rho^2+(D-2) M_W^2\right)+g_V \left(m_b^2+m_\rho^2+(D-2) M_W^2\right)\right) C_{22}
\nonumber\\ &\phantom{=}&
+\left(g_A \left(m_a^2+m_b^2-2 m_\rho^2+2 (D-2) M_W^2\right)+g_V \left(m_a^2+m_b^2+2 m_\rho^2+2 (D-2) M_W^2\right)\right)C_{12} 
\nonumber\\ &\phantom{=}&
+\left(g_A \left(m_a^2-m_\rho^2+(D-2) M_W^2\right)+g_V \left(m_a^2+m_\rho^2+(D-2) M_W^2\right)\right) C_{11}
\nonumber\\ &\phantom{=}&
+(D-3) M_W^2 (g_A+g_V) C_{0}+2 (D-2) M_W^2 (g_A+g_V) (C_{1}+C_{2}) +2 (g_A+g_V) C_{00}
\bigg]
\nonumber\\ &\phantom{=}&
+\mathcal{U}_{\rho a} \mathcal{U}_{\rho b}^* 
\bigg[
-m_a^2 \left(g_A \left(m_b^2-m_\rho^2+(D-2) M_W^2\right)+g_V \left(m_b^2+m_\rho^2+(D-2) M_W^2\right)\right) C_{11}
\nonumber\\ &\phantom{=}&
-m_b^2 \left(g_A \left(m_a^2-m_\rho^2+(D-2) M_W^2\right)+g_V \left(m_a^2+m_\rho^2+(D-2) M_W^2\right)\right) C_{22}
\nonumber\\ &\phantom{=}&
-\big(g_A m_a^2 \left(2 m_b^2-m_\rho^2+(D-2) M_W^2\right)-g_A m_b^2 \left(m_\rho^2-(D-2) M_W^2\right)
\nonumber\\ &\phantom{=}&
+g_V m_a^2 \left(2 m_b^2+m_\rho^2+(D-2) M_W^2\right)+g_V m_b^2 \left(m_\rho^2+(D-2) M_W^2\right)\big)C_{12} 
\nonumber\\ &\phantom{=}&
+2 \left(g_A \left(m_\rho^2-(D-2) M_W^2\right)-g_V \left(m_\rho^2+(D-2) M_W^2\right)\right) C_{00}
\nonumber\\ &\phantom{=}&
-(g_A+g_V) \left(m_a^2 \left(2 m_b^2-m_\rho^2+2 M_W^2\right)-m_b^2 \left(m_\rho^2-6 M_W^2\right)\right) C_{2}
\nonumber\\ &\phantom{=}&
-(g_A+g_V) \left(m_a^2 \left(2 m_b^2-m_\rho^2+6 M_W^2\right)-m_b^2 \left(m_\rho^2-2 M_W^2\right)\right) C_{1}
\nonumber\\ &\phantom{=}&
+\big((g_A+g_V) \left(m_a^2 \left(-m_b^2+m_\rho^2-2 M_W^2\right)+m_b^2 \left(m_\rho^2-2 M_W^2\right)
-m_\rho^4 +(D-6) M_W^4 -2 M_W^2 q^2\right)
\nonumber\\ &\phantom{=}&
+g_A (D+1) m_\rho^2 M_W^2  -g_V (D-7) m_\rho^2 M_W^2
\big)C_{0} 
\nonumber\\ &\phantom{=}&
+2 M_W^2 (g_A+g_V) (B_0^{a}+ B_0^{b})- (g_A-g_V) m_\rho^2 B_0
+(g_A+g_V) (D-6)  M_W^2 B_0
\bigg]
\bigg\}\,,
\end{eqnarray}
with the following Passarino-Veltman functions $C_{rs} = C_{rs}(m_a^2, q^2, m_b^2, M_W^2, m_\rho^2, m_\rho^2)$, $B_0 = B_0(q^2,m_\rho^2,m_\rho^2)$, $B_0^a = B_0(m_a^2,M_W^2,m_\rho^2)$ and $B_0^b = B_0(m_b^2,M_W^2,m_\rho^2)$.

\begin{eqnarray}
F_L^{V \, (d)} &=& \dfrac{g_w^3 }{16\pi^2 4 c_w M_W^2 } \sum_{\rho}
\bigg\{
m_a m_b \mathcal{U}_{\rho a} \mathcal{U}_{\rho b}^* \bigg[2 c_{2w} m_\rho^2 C_{0}+2 c_{2w} C_{00}
\nonumber\\ &\phantom{=}&
+\left(c_{2w} \left(m_a^2+3 m_\rho^2\right)+2 c_w^2 (D-2) M_W^2\right) C_{1}
\nonumber\\ &\phantom{=}&
+\left(c_{2w} \left(m_b^2+3 m_\rho^2\right)+2 c_w^2 (D-2) M_W^2\right) C_{2}
\nonumber\\ &\phantom{=}&
+\left(c_{2w} \left(m_a^2+m_\rho^2\right)+2 c_w^2 (D-2) M_W^2\right) C_{11}
\nonumber\\ &\phantom{=}&
+\left(c_{2w} \left(m_b^2+m_\rho^2\right)+2 c_w^2 (D-2) M_W^2\right) C_{22}
\nonumber\\ &\phantom{=}&
+\left(c_{2w} \left(m_a^2+m_b^2+2 m_\rho^2\right)+4 c_w^2 (D-2) M_W^2\right)C_{12} \bigg]
\nonumber\\ &\phantom{=}&
+\mathcal{U}_{\rho a}^* \mathcal{U}_{\rho b} 
\bigg[ 
-m_\rho^2 \left(c_{2w} \left(m_a^2+m_b^2\right)-4 c_w \left(c_w^2-1\right) M_W M_Z\right) C_{0}
\nonumber\\ &\phantom{=}&
-m_a^2 \left(c_{2w} \left(m_b^2+m_\rho^2\right)+2 c_w^2 D M_W^2\right) C_{11}
\nonumber\\ &\phantom{=}&
-m_b^2 \left(c_{2w} \left(m_a^2+m_\rho^2\right)+2 c_w^2 D M_W^2\right) C_{22}
\nonumber\\ &\phantom{=}&
- \left(c_{2w} \left(m_a^2 \left(2 m_b^2+m_\rho^2\right)+m_b^2 m_\rho^2\right) + 2 c_w^2 M_W^2 \left(D \left(m_a^2+m_b^2\right)-2 q^2\right)\right)C_{12}
\nonumber\\ &\phantom{=}&
-\left(2 c_{2w} m_\rho^2+8 c_w^2 (D-1) M_W^2\right) C_{00}
\nonumber\\ &\phantom{=}&
+\big(2 c_w M_W \left(c_w^2 M_Z \left(m_a^2+m_b^2\right)-c_w M_W \left(m_a^2+(D-1) m_b^2 -2 q^2\right)-M_Z \left(m_a^2+m_b^2\right)\right)
\nonumber\\ &\phantom{=}&
-c_{2w} \left(m_a^2  m_b^2  +m_a^2m_\rho^2 +2 m_b^2 m_\rho^2\right)\big)C_{2} 
\nonumber\\ &\phantom{=}&
+\big(2 c_w M_W \left(c_w^2 M_Z \left(m_a^2+m_b^2\right)-c_w M_W \left((D-1) m_a^2 +m_b^2-2 q^2\right)-M_Z \left(m_a^2+m_b^2\right)\right)
\nonumber\\ &\phantom{=}&
-c_{2w} \left(m_a^2 m_b^2+2 m_a^2m_\rho^2+m_b^2 m_\rho^2\right)\big)C_{1} 
\bigg]
\bigg\}\,,
\end{eqnarray}

\begin{eqnarray}
F_R^{V \, (d)} &=& \dfrac{g_w^3 }{16\pi^2 4 c_w M_W^2 } \sum_{\rho}
\bigg\{
m_a m_b \mathcal{U}_{\rho a}^* \mathcal{U}_{\rho b} 
\bigg[-2 c_{2w} m_\rho^2 C_{0}-2 c_{2w} C_{00}
\nonumber\\ &\phantom{=}&
-\left(c_{2w} \left(m_a^2+3 m_\rho^2\right)+2 c_w^2 (D-2) M_W^2\right) C_{1}
\nonumber\\ &\phantom{=}&
-\left(c_{2w} \left(m_b^2+3 m_\rho^2\right)+2 c_w^2 (D-2) M_W^2\right) C_{2}
\nonumber\\ &\phantom{=}&
-\left(c_{2w} \left(m_a^2+m_\rho^2\right)+2 c_w^2 (D-2) M_W^2\right) C_{11}
\nonumber\\ &\phantom{=}&
-\left(c_{2w} \left(m_b^2+m_\rho^2\right)+2 c_w^2 (D-2) M_W^2\right) C_{22}
\nonumber\\ &\phantom{=}&
- \left(c_{2w} \left(m_a^2+m_b^2+2 m_\rho^2\right)+4 c_w^2 (D-2) M_W^2\right)C_{12}
\bigg]
\nonumber\\ &\phantom{=}&
+\mathcal{U}_{\rho a} \mathcal{U}_{\rho b}^* 
\bigg[m_\rho^2 \left(c_{2w} \left(m_a^2+m_b^2\right)+4 c_w \left(c_w^2-1\right) M_W M_Z\right) C_{0}
\nonumber\\ &\phantom{=}&
+m_a^2 \left(c_{2w} \left(m_b^2+m_\rho^2\right)+2 c_w^2 D M_W^2\right) C_{11}
\nonumber\\ &\phantom{=}&
+m_b^2 \left(c_{2w} \left(m_a^2+m_\rho^2\right)+2 c_w^2 D M_W^2\right) C_{22}
\nonumber\\ &\phantom{=}&
+\left(c_{2w} \left(m_a^2 \left(2 m_b^2+m_\rho^2\right)+m_b^2 m_\rho^2\right)+2 c_w^2 M_W^2 \left(D \left(m_a^2+m_b^2\right)-2 q^2\right)\right)C_{12} 
\nonumber\\ &\phantom{=}&
+\left(2 c_{2w} m_\rho^2+8 c_w^2 (D-1) M_W^2\right) C_{00}
\nonumber\\ &\phantom{=}&
+ \big(2 c_w M_W \left(c_w^2 M_Z \left(m_a^2+m_b^2\right)+c_w M_W \left(m_a^2+(D-1) m_b^2-2 q^2\right)-M_Z \left(m_a^2+m_b^2\right)\right)
\nonumber\\ &\phantom{=}&
+c_{2w} \left(m_a^2 m_b^2+m_a^2m_\rho^2+2 m_b^2 m_\rho^2\right)
\big)C_{2}
\nonumber\\ &\phantom{=}&
+ \big(2 c_w M_W \left(c_w^2 M_Z \left(m_a^2+m_b^2\right)+c_w M_W \left((D-1) m_a^2+m_b^2-2 q^2\right)-M_Z \left(m_a^2+m_b^2\right)\right)
\nonumber\\ &\phantom{=}&
+c_{2w} \left(m_a^2 m_b^2+2 m_a^2 m_\rho^2+m_b^2 m_\rho^2\right)
\big)C_{1}
\bigg]
\bigg\}\,,
\end{eqnarray}
in which $ C_{rs} = C_{rs}(m_a^2, q^2, m_b^2,  m_\rho^2, M_W^2, M_W^2)$ and $c_{2w} = \cos2\theta_w$.

\begin{eqnarray}
F_L^{V \, (e)} &=& \dfrac{g_w^3 }{16\pi^2 8 c_w^2 M_W^2 } \sum_{i}
\bigg\{
C_{ai}^*C_{ib}^* c_w m_a m_b \bigg[\left(m_i^2-m_b^2\right) (C_{2}+C_{22}) +\left(m_i^2-m_a^2\right) (C_{1}+ C_{11})
\nonumber\\ &\phantom{=}&
-\left(m_a^2+m_b^2-2 m_i^2\right) C_{12} -2 C_{00}
\bigg]
\nonumber\\ &\phantom{=}&
+C_{ai}^*C_{ib} c_w m_a m_i \bigg[\left(m_b^2-m_i^2\right) (C_{0}+  C_{2}) 
+\left(m_b^2-m_a^2\right) (C_{12}+  C_{11}) 
\nonumber\\ &\phantom{=}&
-\left(m_a^2-2 m_b^2+m_i^2\right) C_{1}
-2 C_{00}
\bigg]
\nonumber\\ &\phantom{=}&
+C_{ai}C_{ib}^* m_b m_i\bigg[
\left(c_w \left(m_i^2-m_a^2\right)+2 M_W M_Z\right) C_{0}+\left(c_w \left(m_i^2-m_b^2\right)-2 M_W M_Z\right) C_{2}
\nonumber\\ &\phantom{=}&
+\left(c_w \left(m_i^2-m_a^2\right)-2 M_W M_Z\right) C_{1}+c_w \left(m_a^2-m_b^2\right) (C_{22}+ C_{12}) -2 c_w C_{00}
\bigg]
\nonumber\\ &\phantom{=}&
+C_{ai}C_{ib}\bigg[
m_i^2 \left(c_w \left(m_b^2-m_a^2\right)+2 M_W M_Z\right) C_{0}+\left(c_w m_a^2 \left(m_b^2-m_i^2\right)-2 m_b^2 M_W M_Z\right) C_{2}
\nonumber\\ &\phantom{=}&
+\left(c_w m_a^2 \left(m_b^2-2 m_i^2\right)+c_w m_b^2 m_i^2-2 m_b^2 M_W M_Z\right) C_{1}+c_w m_a^2 \left(m_b^2-m_i^2\right) C_{11}
\nonumber\\ &\phantom{=}&
+c_w m_b^2 \left(m_a^2-m_i^2\right) C_{22}-c_w \left(m_a^2 \left(m_i^2-2 m_b^2\right)+m_b^2 m_i^2\right) C_{12}-2 c_w m_i^2 C_{00}
\bigg]
\bigg\}\,,
\end{eqnarray}

\begin{eqnarray}
F_R^{V \, (e)} &=& \dfrac{g_w^3 }{16\pi^2 8 c_w^2 M_W^2 } \sum_{i}
\bigg\{
C_{ai} C_{ib} c_w m_a m_b\bigg[
\left(m_b^2-m_i^2\right) (C_{2}+ C_{22})+\left(m_a^2-m_i^2\right) (C_{1} + C_{11})
\nonumber\\ &\phantom{=}&
+\left(m_a^2+m_b^2-2 m_i^2\right) C_{12}+2 C_{00}
\bigg]
\nonumber\\ &\phantom{=}&
+ C_{ai} C_{ib}^* c_w m_a m_i\bigg[
\left(m_i^2-m_b^2\right) (C_{0} + C_{2}) +\left(m_a^2-m_b^2\right) (C_{12}+ C_{11}) 
\nonumber\\ &\phantom{=}&
+\left(m_a^2-2 m_b^2+m_i^2\right) C_{1}+2 C_{00}
\bigg]
\nonumber\\ &\phantom{=}&
+ C_{ai}^* C_{ib} m_b m_i\bigg[
\left(c_w \left(m_a^2-m_i^2\right)-2 M_W M_Z\right) C_{0}+\left(c_w \left(m_b^2-m_i^2\right)+2 M_W M_Z\right) C_{2}
\nonumber\\ &\phantom{=}&
+\left(c_w \left(m_a^2-m_i^2\right)+2 M_W M_Z\right) C_{1}+c_w \left(m_b^2-m_a^2\right) (C_{22}+ C_{12}) +2 c_w C_{00}
\bigg]
\nonumber\\ &\phantom{=}&
+ C_{ai}^* C_{ib}^* \bigg[
m_i^2 \left(c_w \left(m_a^2-m_b^2\right)-2 M_W M_Z\right) C_{0}+\left(c_w m_a^2 \left(m_i^2-m_b^2\right)+2 m_b^2 M_W M_Z\right) C_{2}
\nonumber\\ &\phantom{=}&
+\left(2 m_b^2 M_W M_Z-c_w \left(m_a^2 \left(m_b^2-2 m_i^2\right)+m_b^2 m_i^2\right)\right) C_{1}+c_w m_a^2 \left(m_i^2-m_b^2\right) C_{11}
\nonumber\\ &\phantom{=}&
+c_w m_b^2 \left(m_i^2-m_a^2\right) C_{22}+c_w \left(m_a^2 \left(m_i^2-2 m_b^2\right)+m_b^2 m_i^2\right) C_{12}+2 c_w m_i^2 C_{00}
\bigg]
\bigg\}\,,
\end{eqnarray}
with $ C_{rs}=C_{rs}(m_a^2, q^2, m_b^2,  m_i^2, M_Z^2, M_H^2)$.

\begin{eqnarray}
F_L^{V \, (f)} &=& \dfrac{g_w^3 }{16\pi^2 8 c_w^2 M_W^2 } \sum_{i}
\bigg\{
C_{ai}^* C_{ib}^* c_w m_a m_b \bigg[
\left(m_i^2-m_b^2\right) (C_{2}+  C_{22}) +\left(m_i^2-m_a^2\right) (C_{1}+ C_{11})
\nonumber\\ &\phantom{=}&
-\left(m_a^2+m_b^2-2 m_i^2\right) C_{12} -2 C_{00}
\bigg]
\nonumber\\ &\phantom{=}&
+ C_{ai}^* C_{ib} m_a m_i \bigg[
\left(c_w \left(m_i^2-m_b^2\right)+2 M_W M_Z\right) C_{0}+\left(c_w \left(m_i^2-m_b^2\right)-2 M_W M_Z\right) C_{2}
\nonumber\\ &\phantom{=}&
+\left(c_w \left(m_i^2-m_a^2\right)-2 M_W M_Z\right) C_{1}+c_w \left(m_b^2-m_a^2\right) (C_{12} + C_{11}) -2 c_w C_{00}
\bigg]
\nonumber\\ &\phantom{=}&
+ C_{ai} C_{ib}^* c_w m_b m_i \bigg[
\left(m_a^2-m_i^2\right) (C_{0} + C_{1})  +\left(m_a^2-m_b^2\right) (C_{22}+C_{12})
\nonumber\\ &\phantom{=}&
-\left(-2 m_a^2+m_b^2+m_i^2\right) C_{2} -2 C_{00}
\bigg]
\nonumber\\ &\phantom{=}&
+ C_{ai} C_{ib}\bigg[
m_i^2 \left(c_w \left(m_a^2-m_b^2\right)+2 M_W M_Z\right) C_{0}+\left(c_w m_b^2 \left(m_a^2-m_i^2\right)-2 m_a^2 M_W M_Z\right) C_{1}
\nonumber\\ &\phantom{=}&
+\left(c_w m_a^2 \left(m_b^2+m_i^2\right)-2 c_w m_b^2 m_i^2-2 m_a^2 M_W M_Z\right) C_{2}+c_w m_b^2 \left(m_a^2-m_i^2\right) C_{22}
\nonumber\\ &\phantom{=}&+c_w m_a^2 \left(m_b^2-m_i^2\right) C_{11}-c_w \left(m_a^2 \left(m_i^2-2 m_b^2\right)+m_b^2 m_i^2\right) C_{12}-2 c_w m_i^2 C_{00}
\bigg]
\bigg\}\,,
\end{eqnarray}

\begin{eqnarray}
F_R^{V \, (f)} &=& \dfrac{g_w^3 }{16\pi^2 8 c_w^2 M_W^2 } \sum_{i}
\bigg\{
C_{ai} C_{ib} c_w m_a m_b\bigg[
\left(m_b^2-m_i^2\right) (C_{2}+ C_{22}) +\left(m_a^2-m_i^2\right) (C_{1}+C_{11})
\nonumber\\ &\phantom{=}&
+\left(m_a^2+m_b^2-2 m_i^2\right) C_{12}  +2 C_{00}
\bigg]
\nonumber\\ &\phantom{=}&
+ C_{ai} C_{ib}^* m_a m_i\bigg[
\left(c_w \left(m_b^2-m_i^2\right)-2 M_W M_Z\right) C_{0}+\left(c_w \left(m_b^2-m_i^2\right)+2 M_W M_Z\right) C_{2}
\nonumber\\ &\phantom{=}&
+\left(c_w \left(m_a^2-m_i^2\right)+2 M_W M_Z\right) C_{1}+c_w \left(m_a^2-m_b^2\right) (C_{12}+ C_{11})+2 c_w C_{00}
\bigg]
\nonumber\\ &\phantom{=}&
+ C_{ai}^* C_{ib} c_w m_b m_i \bigg[
\left(m_i^2-m_a^2\right) (C_{0}+ C_{1})  +\left(m_b^2-m_a^2\right) (C_{22}+ C_{12})
\nonumber\\ &\phantom{=}&
+\left(-2 m_a^2+m_b^2+m_i^2\right) C_{2}+2 C_{00}
\bigg]
\nonumber\\ &\phantom{=}&
+ C_{ai}^* C_{ib}^*\bigg[
-m_i^2 \left(c_w \left(m_a^2-m_b^2\right)+2 M_W M_Z\right) C_{0}+\left(c_w m_b^2 \left(m_i^2-m_a^2\right)+2 m_a^2 M_W M_Z\right) C_{1}
\nonumber\\ &\phantom{=}&
+\left(-c_w m_a^2 \left(m_b^2+m_i^2\right)+2 c_w m_b^2 m_i^2+2 m_a^2 M_W M_Z\right) C_{2}+c_w m_b^2 \left(m_i^2-m_a^2\right) C_{22}
\nonumber\\ &\phantom{=}&
+c_w m_a^2 \left(m_i^2-m_b^2\right) C_{11}+c_w \left(m_a^2 \left(m_i^2-2 m_b^2\right)+m_b^2 m_i^2\right) C_{12}+2 c_w m_i^2 C_{00} 
\bigg]
\bigg\}\,,
\end{eqnarray}
where $ C_{rs}=  C_{rs}(m_a^2, q^2, m_b^2, m_i^2, M_H^2, M_Z^2)$.

\section{SM extensions via sterile fermions and neutrino mass generation: framework and constraints}
\label{app:ISS}

In this section we describe the underlying framework 
of our study: a well-motivated, low-scale mechanism of neutrino mass generation (relying on a minimal SM extension via sterile fermions).

\subsection{The Inverse Seesaw}
The type I seesaw (and its variants, as is the case of the Inverse Seesaw) are among the most minimal yet successful extensions of the SM accounting for a mechanism of neutrino mass generation. Moreover, it allows for a ''natural" (in the sense of 't~Hooft~\cite{tHooft:1979rat}) explanation of the smallness of the the observed neutrino masses, as the latter vanish in the limit of lepton number symmetry restoration.  

In the case of the ISS, two distinct species of sterile  fermions, $X$ and $\nu_R$, are added to the SM content; the Lagrangian encoding the mass terms for the neutral lepton sector can generically be written as 
\begin{equation}\label{eq:ISS:lagrangian}
    \mathcal L_\text{ISS}\, =\, -Y^D_{ij} \,\overline{L_i^c}\,\widetilde H \,\nu_{Rj}^c - M_R^{ij}\, \overline{\nu_{Ri}}\, X_j - \frac{1}{2}\mu_R^{ij}\, \overline{\nu_{Ri}^c}\,\nu_{Rj} - \frac{1}{2} \mu_X^{ij}\, \overline{X_i^c}\, X_j + \text{H.c.}\,,
\end{equation}
in which $\mu_X$ and $\mu_R$ are the only source of lepton number violation. After the Higgs boson acquires its vacuum expectation value, $v$, a Dirac mass term is generated, given by $m_D\equiv Y_D v$. 
In the limit of vanishing $\mu_{X,R}$, lepton number is restored, and thus light neutrinos masses vanish. In this sense, assuming the hierarchy of scales $\mu_{X,R}\ll m_D\ll M_R$ is natural in the 't Hooft sense~\cite{tHooft:1980xss,Hettmansperger:2011bt}. 
Large deviations from unitarity of the PMNS can still be present, even in the massless neutrino limit, as $\eta$ is given by
\begin{equation}
    \eta = \frac{1}{2}m_D^* \left(M_R^{-1}\right)^{\dagger}\left(M_R^{-1}\right)m_D^T\,,
    \label{eq:def_eta_ISS}
\end{equation}
and thus it does not depend on $\mu_{X,R}$. Moreover, in the limit of approximate lepton number conservation, the term $\mu_R$ only contributes to light neutrino masses through loop effects~\cite{Dev:2012sg}. Therefore, in the following we will not consider its contribution.

Although more minimal ISS realisations exist~\cite{Abada:2014vea}, here we will work in the (3,3) realisation, corresponding to the addition of 
$n_R = n_X = 3$ generations of heavy neutral states. 
The diagonalisation of the $9\times9$ mass matrix allows obtaining the full neutrino spectrum. In the approximate lepton number conserving limit in which $\mu_X \ll m_D\ll M_R$, one can derive an approximate expression for the masses of the light 
(mostly active) neutrinos, given by
\begin{equation}
    m_\nu \simeq  m_D \, \left( M_{R}^{-1} \right)^{T} \, \mu_X \, M_{R}^{-1} \, m_D^T\, \equiv U_\text{PMNS}^\ast\, m_\nu^\text{diag}\, U_\text{PMNS}^\dagger\,.
    \label{eq:ISS:lightmasses}
\end{equation}
A detailed description of the relevant charged and neutral Lagrangian terms is provided in Appendix~\ref{app:Lagrangian}, and the associated vertices 
in~\ref{app:FeynmanRules}.

\subsection{Relevant Lagrangian terms}\label{app:Lagrangian}
In what follows we collect the most relevant terms in the Lagragian of the lepton sector (interactions with neutral and charged gauge bosons, Higgs and Goldstone bosons). The terms are presented in the physical lepton bases, and reflect the Majorana nature of the neutral leptons. (Some of the terms have been individually discussed in the main body of the manuscript, see Eqs.~(\ref{eq:lagrangian:W}, \ref{eq:lagrangian:HZ}).)
\begin{align}\label{eq:lagrangian:WGHZ}
& \mathcal{L}_{W^\pm}\, =\, -\frac{g_w}{\sqrt{2}} \, W^-_\mu \,
\sum_{\alpha=1}^{3} \sum_{j=1}^{3 + n_S} \mathcal{U}_{\alpha j} \bar \ell_\alpha 
\gamma^\mu P_L \nu_j \, + \, \text{H.c.}\,, \nonumber \\
& \mathcal{L}_{Z^0}^{\nu}\, = \,-\frac{g_w}{4 \cos \theta_w} \, Z_\mu \,
\sum_{i,j=1}^{3 + n_S} \bar \nu_i \gamma ^\mu \left(
P_L {C}_{ij} - P_R {C}_{ij}^* \right) \nu_j\,, \nonumber \\
& \mathcal{L}_{Z^0}^{\ell}\, = \,-\frac{g_w}{2 \cos \theta_w} \, Z_\mu \,
\sum_{\alpha=1}^{3}  \bar \ell_\alpha \gamma ^\mu \left(
{\bf C}_{V} - {\bf C}_{A} \gamma_5 \right) \ell_\alpha\,, \nonumber \\
& \mathcal{L}_{H^0}\, = \, -\frac{g_w}{4 M_W} \, H  \,
\sum_{i\ne j= 1}^{3 + n_S}    \bar \nu_i\,\left[{C}_{ij}\,\left(
P_L m_i + P_R m_j \right) +{C}_{ij}^\ast\left(
P_R m_i + P_L m_j \right) \right] \nu_j\ , \nonumber \\
& \mathcal{L}_{G^0}\, =\,\frac{i g_w}{4 M_W} \, G^0 \,
\sum_{i,j=1}^{3 + n_S}  \bar \nu_i \left[ {C}_{ij}
\left(P_R m_j  - P_L m_i  \right) + {C}_{ij}^\ast
\left(P_R m_i  - P_L m_j  \right)\right] \nu_j\,, \nonumber  \\
& \mathcal{L}_{G^\pm}\, =\, -\frac{g_w}{\sqrt{2} M_W} \, G^- \,
\sum_{\alpha=1}^{3}\sum_{j=1}^{3 + n_S} \mathcal{U}_{\alpha j}
\bar \ell_\alpha\left(
m_\alpha P_L - m_j P_R \right) \nu_j\, + \, \text{H.c.}\,.
\end{align}
Finally, we recall that ${C}_{ij} $ are defined as in Eq.~(\ref{eq:cij}): 
\begin{equation}
    {C}_{ij} = \sum_{\rho = 1}^3
  \mathcal{U}_{i\rho}^\dagger \,\mathcal{U}_{\rho j}^{\phantom{\dagger}}\:. \nonumber
\end{equation}

\subsection{Feynman rules}\label{app:FeynmanRules}
Following the presentation of the Lagrangian terms, see Eq.~(\ref{eq:lagrangian:WGHZ}), we list in Table~\ref{table:feynrules} the Feynman rules for the vertices which were used in the computations carried out in this manuscript. Again notice that the neutral leptons are assumed to be of Majorana nature.

In the $Z$ and Higgs vertices, the arrows denote the momentum flow. 
We note here that diagrams including at least one $Z n_i n_j$ or $H n_i n_j$ vertex have to be symmetrised (factor 2) due to the Majorana nature of the physical neutrinos.

\hspace*{-2mm}
\begin{table}[h!]
\begin{tabular}{m{2.4cm}m{5.35cm}m{2.4cm}m{5.35cm}}
        \begin{tikzpicture}
    \begin{feynman}
    \vertex (a) at (0,0) {\(Z_\mu\)};
    \vertex (b) at (1,0);
    \vertex (c) at (2,1.){\(n_i\)};
    \vertex (d) at (2,-1.){\(n_j\)};
    \diagram* {
    (a) -- [boson] (b),
    (c) -- [momentum'=\( \)] (b),
    (b) -- [momentum'=\( \)] (d),
    };
    \end{feynman}
    \end{tikzpicture} 
    & $= \, -i \dfrac{g_w}{4 c_w} \gamma_\mu \left(C_{ij}^* \,P_L -C_{ij} \,P_R \right)$ 
    & & 
\\
\\
        \begin{tikzpicture}
    \begin{feynman}
    \vertex (a) at (0,0) {\(H\)};
    \vertex (b) at (1,0);
    \vertex (c) at (2,1.){\(n_i\)};
    \vertex (d) at (2,-1.){\(n_j\)};
    \diagram* {
    (a) -- [scalar] (b),
    (c) -- [momentum'=\( \)] (b),
    (b) -- [momentum'=\( \)] (d),
    };
    \end{feynman}
    \end{tikzpicture}
    & $=\, -i\dfrac{g_w}{4 M_W} \left[C_{ij} \left( m_i\, P_L + m_j \,P_R \right)
    + C_{ij}^* \left( m_i\, P_R + m_j \,P_L \right) \right]$
    \\
    \\
    \begin{tikzpicture}
    \begin{feynman}
    \vertex (a) at (0,0) {\(W_\mu^-\)};
    \vertex (b) at (1,0);
    \vertex (c) at (2,1.){\(n_i\)};
    \vertex (d) at (2,-1.){\(\ell_\alpha^-\)};
    \diagram* {
    (a) -- [boson] (b),
    (c) -- [fermion] (b),
    (b) -- [fermion] (d),
    };
    \end{feynman}
    \end{tikzpicture}
    & $=\, -i\dfrac{g_w}{\sqrt{2}} \,\mathcal{U}_{\alpha i}\, \gamma_\mu\, P_L$
    &
    \begin{tikzpicture}
    \begin{feynman}
    \vertex (a) at (0,0) {\(W_\mu^+\)};
    \vertex (b) at (1,0);
    \vertex (c) at (2,1.){\(\ell_\alpha^+\)};
    \vertex (d) at (2,-1.){\(n_i\)};
    \diagram* {
    (a) -- [boson] (b),
    (c) -- [fermion] (b),
    (b) -- [fermion] (d),
    };
    \end{feynman}
    \end{tikzpicture}
    &
    $=\, -i\dfrac{g_w}{\sqrt{2}} \,\mathcal{U}_{\alpha i}^*\, \gamma_\mu\, P_L$
    \\
    \\
    \begin{tikzpicture}
    \begin{feynman}
    \vertex (a) at (0,0) {\(G^-\)};
    \vertex (b) at (1,0);
    \vertex (c) at (2,1.){\(n_i\)};
    \vertex (d) at (2,-1.){\(\ell_\alpha^-\)};
    \diagram* {
    (a) -- [scalar] (b),
    (c) -- [fermion] (b),
    (b) -- [fermion] (d),
    };
    \end{feynman}
    \end{tikzpicture}  
    &
    $=\, i\dfrac{g_w}{\sqrt{2}M_W} \,\mathcal{U}_{\alpha i} \,\left(m_i\, P_R - m_\alpha\, P_L\right)$
&
    \begin{tikzpicture}
    \begin{feynman}
    \vertex (a) at (0,0) {\(G^+\)};
    \vertex (b) at (1,0);
    \vertex (c) at (2,1.){\(\ell_\alpha^+\)};
    \vertex (d) at (2,-1.){\(n_i\)};
    \diagram* {
    (a) -- [scalar] (b),
    (c) -- [fermion] (b),
    (b) -- [fermion] (d),
    };
    \end{feynman}
    \end{tikzpicture}
    &
    $=\, i\dfrac{g_w}{\sqrt{2}M_W} \,\mathcal{U}_{\alpha i}^* \,\left(m_i \,P_L - m_\alpha \,P_R\right)$
\end{tabular}
\caption{Feynman rules for $W$, $Z$ and Higgs interactions (and associated Goldstone bosons) in SM extensions via Majorana sterile fermions. }\label{table:feynrules}
\end{table}

\subsection{Constraints on HNL extensions of the SM}\label{app:constraints}
In the phenomenological analysis whose results are summarised in Sections~\ref{sec:NumericalResults} and~\ref{sec:ComplementarityProbing}, we have taken into account numerous constraints, which were applied to the present ISS (3,3) realisation.
In addition to ensuring that such a mechanism of neutrino mass generation does comply with neutrino oscillation data, we have further imposed several experimental limits, including
EW precision observables, universality bounds from tau and meson decays, cLFV bounds, among others. 
Finally, we also took into account perturbative unitarity constraints for the heavy sterile states (including 
tree-level decay widths of heavy $N$ from the channels $N_i\to \ell_\alpha W$, $N_i\to N_j Z$ and $N_i \to N_j H$ where $N_j$ can also be light).

\paragraph{Oscillation data} The first constrain on any SM extension aiming at accounting for neutrino masses and mixings (in particular via the addition of HNL) is that of reproducing the measured neutrino oscillation parameters. From the latest NuFIT 5.1 global fit results~\cite{Esteban:2020cvm}, these parameters are given in Table~\ref{tab:nufit}. As already stated in the main text (see Section~\ref{sec:NumericalResults}), we vary the light neutrino mass in the range $m_0\in \left[ 10^{-10},\, 10^{-3}\right]$~eV and fix oscillation data to their best-fit values as summarised in Table~\ref{tab:nufit}, assuming normal ordering for the light neutrino spectrum, while varying the CP phase within its range. Note that none of these assumptions have any impact in our results nor on the constraints.
\renewcommand{\arraystretch}{1.3}
\begin{table}[h!]
    \centering
    \begin{tabular}{|c|c|c|}
        \hline
        & Normal ordering & Inverted ordering \\
        \hline
        \hline
      $\sin^2\theta_{12}$  & $0.304^{+0.013}_{-0.012}$ & $0.304^{+0.012}_{-0.012}$ \\
      \hline
      $\sin^2\theta_{23}$ & $0.573^{+0.018}_{-0.023}$ & $0.578^{+0.017}_{-0.021}$\\
      \hline
      $\sin^2\theta_{13}$ & $0.02220^{+0.00068}_{-0.00062}$ & $0.02238^{+0.00064}_{-0.00062}$\\
      \hline
      $\Delta m_{21}^2/10^{-5}\,\mathrm{eV}$ & $7.42^{+0.21}_{-0.20}$ & $7.42^{+0.21}_{-0.20}$\\
      \hline
      $\Delta m_{3\ell}^2/10^{-3}\,\mathrm{eV}$ & $2.515^{+0.028}_{-0.028}$ & $-2.498^{+0.028}_{-0.029}$\\
      \hline
    \end{tabular}
    \caption{Global fit results obtained by NuFIT 5.1~\cite{Esteban:2020cvm} for neutrino mixing data, without the inclusion of the atmospheric sample from Super-Kamiokande. For normal ordering we have $\Delta m_{3\ell}^2\equiv\Delta m_{31}^2>0$, while in the case of inverted ordering one finds $\Delta m_{3\ell}^2\equiv \Delta m_{32}^2<0$.}
    \label{tab:nufit}
\end{table}
\renewcommand{\arraystretch}{1.}

\paragraph{Universality bounds from tau-lepton and light meson decays} 
The non-unitarity of the PMNS mixing matrix impacts several low-energy observables, which can then place strong constraints on the active-sterile mixings.
Among them, we have the universality ratios from kaon and pion leptonic decays, defined as~\cite{Abada:2013aba,Abada:2012mc}
\begin{equation}
    \Delta r_P\equiv \frac{\Gamma\left(P^+\rightarrow e\nu\right)}{\Gamma\left(P^+\rightarrow \mu\nu\right)}-1\quad \left(P=\pi,K\right),
\end{equation}
as well as $\tau$ decays (also sensitive to the modification of the $W\ell \nu$ vertex), with the associated observable, $R_{\tau}$, given by
\begin{equation}
    R_{\tau}\equiv \frac{\Gamma\left(\tau \rightarrow \mu \nu \nu\right)}{\Gamma\left(\tau \rightarrow e \nu \nu\right)}.
\end{equation}

\paragraph{Charged lepton flavour violation}
There are several cLFV transitions and decays placing strong constraints on neutrino mass models upon the appearance of flavour violation. Although here we will mostly focus on LFUV (while aiming at exploring regimes with suppressed cLFV), it is interesting to consider the interplay between the two sets of observables in order to discuss future prospects. Particularly relevant are cLFV three-body  decays, $\mu - e$ conversion in atoms, as well as cLFV $Z$ boson decays. 
The expressions for the cLFV observables (in the context of SM extensions via HNL) can be found, for instance, in~\cite{Ilakovac:1994kj,Alonso:2012ji,Abada:2018nio,Abada:2022asx,Riemann:1982rq,Riemann:1999ab,Illana:1999ww,Mann:1983dv,Illana:2000ic,Ma:1979px,Gronau:1984ct,Deppisch:2004fa,Deppisch:2005zm,Dinh:2012bp,Abada:2014kba,Abada:2015oba,Abada:2015zea,Abada:2016vzu,Arganda:2014dta} 
The current bounds and future sensitivities are collected in Table~\ref{tab:cLFV_lep}.

\renewcommand{\arraystretch}{1.3}
\begin{table}[h!]
    \centering
    \hspace*{-2mm}{\small\begin{tabular}{|c|c|c|}
    \hline
    Observable & Current bound & Future sensitivity  \\
    \hline\hline
    $\text{BR}(\mu\to e \gamma)$    &
    \quad $<4.2\times 10^{-13}$ \quad (MEG~\cite{TheMEG:2016wtm})   &
    \quad $6\times 10^{-14}$ \quad (MEG II~\cite{Baldini:2018nnn}) \\
    $\text{BR}(\tau \to e \gamma)$  &
    \quad $<3.3\times 10^{-8}$ \quad (BaBar~\cite{Aubert:2009ag})    &
    \quad $3\times10^{-9}$ \quad (Belle II~\cite{Kou:2018nap})      \\
    $\text{BR}(\tau \to \mu \gamma)$    &
     \quad $ <4.4\times 10^{-8}$ \quad (BaBar~\cite{Aubert:2009ag})  &
    \quad $10^{-9}$ \quad (Belle II~\cite{Kou:2018nap})     \\
    \hline
    $\text{BR}(\mu \to 3 e)$    &
     \quad $<1.0\times 10^{-12}$ \quad (SINDRUM~\cite{Bellgardt:1987du})    &
     \quad $10^{-15(-16)}$ \quad (Mu3e~\cite{Blondel:2013ia})   \\
    $\text{BR}(\tau \to 3 e)$   &
    \quad $<2.7\times 10^{-8}$ \quad (Belle~\cite{Hayasaka:2010np})&
    \quad $5\times10^{-10}$ \quad (Belle II~\cite{Kou:2018nap})     \\
    $\text{BR}(\tau \to 3 \mu )$    &
    \quad $<3.3\times 10^{-8}$ \quad (Belle~\cite{Hayasaka:2010np})  &
    \quad $5\times10^{-10}$ \quad (Belle II~\cite{Kou:2018nap})     \\
    & & \quad$5\times 10^{-11}$\quad (FCC-ee~\cite{Abada:2019lih})\\
        $\text{BR}(\tau^- \to e^-\mu^+\mu^-)$   &
    \quad $<2.7\times 10^{-8}$ \quad (Belle~\cite{Hayasaka:2010np})&
    \quad $5\times10^{-10}$ \quad (Belle II~\cite{Kou:2018nap})     \\
    $\text{BR}(\tau^- \to \mu^-e^+e^-)$ &
    \quad $<1.8\times 10^{-8}$ \quad (Belle~\cite{Hayasaka:2010np})&
    \quad $5\times10^{-10}$ \quad (Belle II~\cite{Kou:2018nap})     \\
    $\text{BR}(\tau^- \to e^-\mu^+e^-)$ &
    \quad $<1.5\times 10^{-8}$ \quad (Belle~\cite{Hayasaka:2010np})&
    \quad $3\times10^{-10}$ \quad (Belle II~\cite{Kou:2018nap})     \\
    $\text{BR}(\tau^- \to \mu^-e^+\mu^-)$   &
    \quad $<1.7\times 10^{-8}$ \quad (Belle~\cite{Hayasaka:2010np})&
    \quad $4\times10^{-10}$ \quad (Belle II~\cite{Kou:2018nap})     \\
    \hline
    $\text{CR}(\mu- e, \text{N})$ &
     \quad $<7 \times 10^{-13}$ \quad  (Au, SINDRUM~\cite{Bertl:2006up}) &
    \quad $10^{-14}$  \quad (SiC, DeeMe~\cite{Nguyen:2015vkk})    \\
    & &  \quad $2.6\times 10^{-17}$  \quad (Al, COMET~\cite{Krikler:2015msn,COMET:2018auw,Moritsu:2022lem})  \\
    & &  \quad $8 \times 10^{-17}$  \quad (Al, Mu2e~\cite{Bartoszek:2014mya})\\
    \hline
    $\mathrm{BR}(Z\to e^\pm\mu^\mp)$ & \quad$< 4.2\times 10^{-7}$\quad (ATLAS~\cite{Aad:2014bca}) & \quad$\mathcal O (10^{-10})$\quad (FCC-ee~\cite{Abada:2019lih})\\
    $\mathrm{BR}(Z\to e^\pm\tau^\mp)$ & \quad$< 4.1\times 10^{-6}$\quad (ATLAS~\cite{ATLAS:2021bdj}) & \quad$\mathcal O (10^{-10})$\quad (FCC-ee~\cite{Abada:2019lih})\\
    $\mathrm{BR}(Z\to \mu^\pm\tau^\mp)$ & \quad$< 5.3\times 10^{-6}$\quad (ATLAS~\cite{ATLAS:2021bdj}) & \quad $\mathcal O (10^{-10})$\quad (FCC-ee~\cite{Abada:2019lih})\\
    \hline
    \end{tabular}}
    \caption{Current experimental bounds and future sensitivities on relevant cLFV observables. The quoted limits are given at $90\%\:\mathrm{C.L.}$ (Belle II sensitivities correspond to an integrated luminosity of $50\:\mathrm{ab}^{-1}$.)}
    \label{tab:cLFV_lep}
\end{table}
\renewcommand{\arraystretch}{1.}

\paragraph{EW precision tests}
The presence of the HNL leads to a shift on the EW oblique parameters $S$, $T$ and $U$~\cite{Peskin:1991sw} from the SM predictions. The dominant contribution from heavy neutrinos is found for the $T$ parameter~\cite{Fernandez-Martinez:2015hxa}. Even if present bounds on $T$ are not currently competitive to constrain the explored parameter space, notice that prospective sensitivities of FCC-ee (improving current constraints by about two orders of magnitude~\cite{FCC:2018byv}) will place very strong bounds on the size of the deviations of $T$ from its SM expectation.

Moreover, muon decays are affected by the presence of the heavy sterile states. Given that we use $G_{\mu}$ - the measurement of the Fermi constant - as an input parameter, we must take into account the non-unitarity effect of HNL already at tree level. In particular, the muon decay rate is now given by
\begin{equation}
    \Gamma_{\mu}= \frac{m_{\mu}^5 G_\mu^2}{192\pi^3} =\frac{m_{\mu}^5 G_F^2}{192\pi^3}\sum_{i,j=1}^3|\mathcal{U}_{\mu i}|^2|\mathcal{U}_{ej}|^2\simeq \frac{m_{\mu}^5G_F^2}{192\pi^3}\left(1-2\eta_{\mu\mu}-2\eta_{ee}\right)\,,
\end{equation}
from which we identify $G_{\mu}\equiv G_F\left(1-\eta_{\mu\mu}-\eta_{ee}\right)$, assuming $\eta_{\alpha\alpha}\ll1$.

\bigskip
\paragraph{Perturbative unitarity}
We can also impose bounds on the HNL parameter space by taking into account perturbative unitarity constraints~\cite{Chanowitz:1978mv,Durand:1989zs,Bernabeu:1993up,Fajfer:1998px,Ilakovac:1999md}, which restricts the heavy states decay width to comply with
\begin{equation}
    \frac{\Gamma(N_i)}{m_{N_i}}<\frac{1}{2},\quad \text{for}\quad i\geq4\,.
\end{equation}
For heavy neutrinos as considered here, the main contributions to the decay rate will be given by the two-body decays into a SM boson and a lepton. At tree level, these are given by
\begin{eqnarray}
    \Gamma\left(N_i\rightarrow Wl_{\alpha}\right)&=&\frac{g_w^2}{64\pi}\left|\mathcal{U}_{\alpha i}\right|^2\frac{\lambda^{1/2}(m_i^2,M_W^2,m_{\ell_{\alpha}}^2)}{m_i}\left\lbrace 1+\frac{m_{\ell_{\alpha}}-2M_W^2}{m_i^2}+\frac{(m_i^2-m_{\ell_{\alpha}}^2)^2}{m_i^2M_W^2}\right\rbrace,\\
    \Gamma\left(N_i\rightarrow ZN_j\right)&=&\frac{g_w^2}{128c_w^2\pi}\left|C_{ij}\right|^2\frac{\lambda^{1/2}(m_i^2,M_Z^2,m_{j}^2)}{m_i}\left\lbrace 1+\frac{m_{j}-2M_Z^2}{m_i^2}+\frac{(m_i^2-m_{j}^2)^2}{m_i^2M_Z^2}\right\rbrace,\\
    \Gamma\left(N_i\rightarrow HN_j\right)&=&\frac{g_w^2}{128\pi}\left|C_{ij}\right|^2\frac{m_i^2}{M_W^2}\frac{\lambda^{1/2}(m_i^2,M_H^2,m_j^2)}{m_i}\left\lbrace\left(1+\frac{m_j^2-M_H^2}{m_i^2}\right)\left(1+\frac{m_j^2}{m_i^2}\right)+4\frac{m_j^2}{m_i^2}\right\rbrace, \nonumber
    \\
\end{eqnarray}
where $\lambda(a,b,c)=\left[a-(\sqrt{b}-\sqrt{c})^2\right]\left[a-(\sqrt{b}+\sqrt{c})^2\right]$ is the K\"all\'en function.

\end{document}